\documentclass[aps,prb,twocolumn,longbibliography]{revtex4-2}
\usepackage{amsmath,amssymb}
\usepackage{hyperref,graphicx}
\hypersetup{colorlinks = true, urlcolor = blue, linkcolor = blue, citecolor = blue}
\usepackage{physics}
\usepackage{xcolor}
\usepackage{bm}
\usepackage[section]{placeins}

\begin{document}

	\title{Crossover between trivial zero modes in Majorana nanowires}
	\author{Haining Pan}
	\affiliation{Condensed Matter Theory Center and Joint Quantum Institute, Department of Physics, University of Maryland, College Park, Maryland 20742, USA}
	\author{Sankar Das Sarma}
	\affiliation{Condensed Matter Theory Center and Joint Quantum Institute, Department of Physics, University of Maryland, College Park, Maryland 20742, USA}

\begin{abstract}
	We consider the superconductor-semiconductor nanowire hybrid Majorana platform (``Majorana nanowire'') in the presence of a deterministic spatially slowly varying inhomogeneous chemical potential and a random spatial quenched potential disorder, both of which are known to produce nontopological almost-zero-energy modes mimicking the theoretically predicted topological Majorana zero modes.   We study the crossover among these mechanisms by calculating the tunnel conductance while varying the relative strength between inhomogeneous potential and random disorder in a controlled manner.  We find that the entire crossover region manifests abundant trivial zero modes, many of which showing the apparent ``quantization'' of the zero-bias conductance peak at $ 2e^2/h $, with occasional disorder-dominated peaks exceeding $ 2e^2/h $.  We present animations of the simulated crossover behavior and discuss experimental implications. {Additionally, in order to simulate the realistic disorder in experimental nanowires, we also study in depth the case of disorder arising from random individual static impurities along the wire, and consider crossover associated with such impurity effects.} Our results, when compared qualitatively with existing Majorana nanowire experimental results, indicate the dominant role of random disorder in the experiments. {It turns out that all three mechanisms may produce trivial zero-bias peaks in the tunnel conductance, and the crossover among these physical mechanisms (i.e., when more than one mechanism is present in the system) is smooth and continuous, making it difficult \textit{a priori} to conclude which mechanism is dominant in a particular sample just by a casual inspection of the zero-bias conductance peaks.}

\end{abstract}
\maketitle

\section{Introduction}\label{sec:intro}
Following the predicted possible realization of non-Abelian Majorana zero modes (MZMs) in superconductor-semiconductor hybrid platforms in the presence of the superconducting proximity effect, spin-orbit coupling, and spin splitting~\cite{lutchyn2010majorana,sau2010generic,sau2010nonabelian,oreg2010helical}, a large number of experiments from many different groups reported the observation of zero-bias conductance peaks (ZBCPs) in tunneling spectroscopy of InAs and InSb nanowires, presumably as evidence for the predicted topological MZM~\cite{mourik2012signatures,das2012zerobias,deng2012anomalous,churchill2013superconductornanowire,finck2013anomalous, deng2016majorana,nichele2017scaling,zhang2017ballistic,vaitiekenas2018effective,moor2018electric,zhang2021large,zhang2018quantizeda,bommer2019spinorbit,grivnin2019concomitant,anselmetti2019endtoend,menard2020conductancematrix,puglia2021closing}. Recent experiments~\cite{nichele2017scaling,zhang2021large,zhang2018quantizeda} have even reported ZBCPs with the approximate conductance value of $ 2e^2/h $, which is the predicted topologically quantized value for MZM conductance~\cite{sengupta2001midgap,law2009majorana,flensberg2010tunneling,wimmer2011quantum}. This created considerable excitement in the community that perhaps the elusive non-Abelian MZM has finally been observed.

It was, however, quickly realized that topological MZMs are unlikely to have been observed in these tunneling measurements.  First, the current nanowires may be simply too short, most likely shorter than the superconducting coherence length, and thus, the system is most likely not in the topological regime.  Second, there is no evidence for a bulk gap opening (or more generally, no sign for a topological quantum phase transition to the MZM-carrying topological superconducting phase) when the ZBCP shows up, which is a necessary topological requirement by virtue of the bulk-boundary correspondence.  Third, no signature of the predicted MZM oscillations~\cite{dassarma2012splitting}, associated with the overlap of the Majorana wavefunctions from the two wire ends, has ever been reported.  Fourth, no nonlocal experimental feature, for example, correlated ZBCPs from tunneling at both ends~\cite{lai2019presence,pan2021threeterminal,rosdahl2018andreev}, has ever been observed, casting doubt on the nonlocal topological nature of the observed ZBCPs just from one end of the nanowire.  Fifth, the observed ZBCPs are typically not stable as a function of system parameters such as applied tunnel barrier, magnetic field, and gate voltages, casting doubt on their robust topological nature.  Although these features indicate serious difficulties with the MZM interpretation of the observed ZBCPs, perhaps the most compelling argument against the MZM interpretation of the experimentally observed ZBCPs is that two persuasive non-MZM physical mechanisms have been theoretically identified which produce nontopological (i.e., trivial) ZBCPs generically in nanowires, and these trivial ZBCPs appear consistent with all the observed features in the tunneling measurements, leading to a consensus that the reported ZBCPs so far are most likely trivial and not topological.

These two trivial ZBCP mechanisms of nontopological origin, which we have recently dubbed ``bad'' and ``ugly''~\cite{pan2020physical}, are, respectively, a slowly varying chemical potential due to the presence of an inhomogeneous potential and random spatial disorder arising from unknown impurities and defects in the system.  The possibility that an inhomogeneous chemical potential could give rise to subgap fermionic states was pointed out early in the Majorana nanowire literature~\cite{kells2012nearzeroenergy,prada2012transport,liu2017andreev}, but its importance in determining the tunnel conductance measurements was not immediately appreciated.  Following the experiment by Deng \textit{et al}~\cite{deng2016majorana}, where a claim was made for the observation of Majorana bound states (a different name for MZMs) from coalescing Andreev bound states (ABSs) in the InAs quantum dot-nanowire-Al hybrid system, it was pointed out that the observations are more consistent with almost-zero-energy trivial ZBCPs arising from nontopological fermionic subgap states induced by the inhomogeneous potential associated with the quantum dot.  This is the ``bad'' scenario for ZBCPs, where trivial ABSs produce rather stable zero-energy states in nanowires, often giving rise to ZBCPs with values close to $ 2e^2/h $ value~\cite{stanescu2019robust,moore2018quantized,moore2018twoterminal}. Similar trivial ZBCPs arise from a smooth slowly spatially varying potential along the wire also. We will refer to these inhomogeneous potential-induced ZBCPs as ``bad'' zero modes for notational convenience. (The truly topological MZMs will be referred to as ``good'' following the nomenclature introduced in Ref.~\onlinecite{pan2020physical}.)

The fact that random disorder by itself could produce ZBCPs in the nanowire mimicking MZM behavior was also pointed out early~\cite{mi2014xshaped,sau2013density,pikulin2012zerovoltage,bagrets2012class,liu2012zerobias}, but its relevance to the experimental tunneling spectroscopy has only been appreciated recently~\cite{pan2020generic,pan2020physical}. In particular, we recently established that random disorder in the nanowire by itself can produce relatively stable trivial ZBCPs with a high probability of achieving $\sim  2e^2/h  $ conductance value.  We will refer to these disorder-induced ZBCPs as ``ugly'' for descriptive brevity as in our recent publications~\cite{pan2020physical,pan2021threeterminal}. They are also sometimes referred to as ``class D'' peaks alluding to their connection to antilocalization effects in systems breaking time-reversal invariance and spin rotational symmetry~\cite{pan2020generic}.

Since the superconductor-nanowire hybrid systems are likely to have both inhomogeneous potentials and random disorder, neither of which is intentional and therefore not controllable, it is important to consider their interplay by taking into account both mechanisms together.  This is precisely what we do in this work by calculating the tunnel conductance and the local density of states (LDOS) of the Majorana nanowire including both bad and ugly mechanisms and using a tuning parameter to study the crossover between the two, going from the completely bad situation (with only potential inhomogeneity) to the completely ugly situation (with only random disorder) in a controlled manner.  We find the ubiquitous presence of trivial zero modes throughout the crossover region, often with ZBCP values $ \sim2e^2/h $, thus considerably complicating the interpretation of experimental results where ZBCPs, particularly with conductance  $\sim 2e^2/h $ value, are assumed to be synonymous with the existence of topological good MZMs.  We present, for the sake of a direct comparison, results for the pristine ``good'' situation also, where neither disorder nor inhomogeneous potential is present in the nanowire leading, therefore, to the presence of topological MZMs in the system.  
{In addition, we present the calculated tunnel conductance and the LDOS of the nanowire in the crossover between two ugly cases arising from different random disorder configurations, which also manifests the ubiquitous trivial ZBCPs throughout the crossover region, for a complete story of crossover physics.}
We also present, in the Supplemental Material~\cite{crossover_SM}, animations showing the simulated conductance {and LDOS} results with varying amounts of inhomogeneous potential and random disorder. These crossover results should help better understand experimental results, where both inhomogeneous chemical potential and random disorder are invariably present.

{In addition to the crossover with the potential disorder, we also study the crossover with the local impurity disorder since it is more likely to happen in real experiments~\cite{woods2021charge}. In this case, the chemical potential is modified by a single impurity or a bunch of impurities located randomly in the nanowire. We find that although the underlying mechanism of impurity disorder is different from potential disorder, the results are qualitatively similar.}

{The rest of this paper is organized as follows: In Sec.~\ref{sec:model}, we describe the crossover models we use.  In Sec.~\ref{sec:theory}, we briefly describe the underlying theory and the calculational details, presenting and discussing our detailed results for the calculated tunnel conductance as a function of bias voltage and applied magnetic field in Sec.~\ref{sec:results}. We conclude in Sec.~\ref{sec:conclusion}. Appendix~\ref{app:A} provides the conductance of a different set of realizations of potential disorder. Appendix~\ref{app:B} provides the calculated LDOS at the two ends of the nanowire as well as in the middle of the wire for a comparison with the corresponding conductance results presented in Sec.~\ref{sec:results} and Appendix~\ref{app:A}. The Supplemental Material contains detailed animations for the crossover conductance behavior as a continuous function of the tuning parameter controlling the crossover~\cite{crossover_SM}.
}

\section{Crossover Model}\label{sec:model}

\begin{figure}[htbp]
	\centering
	\includegraphics[width=3.4in]{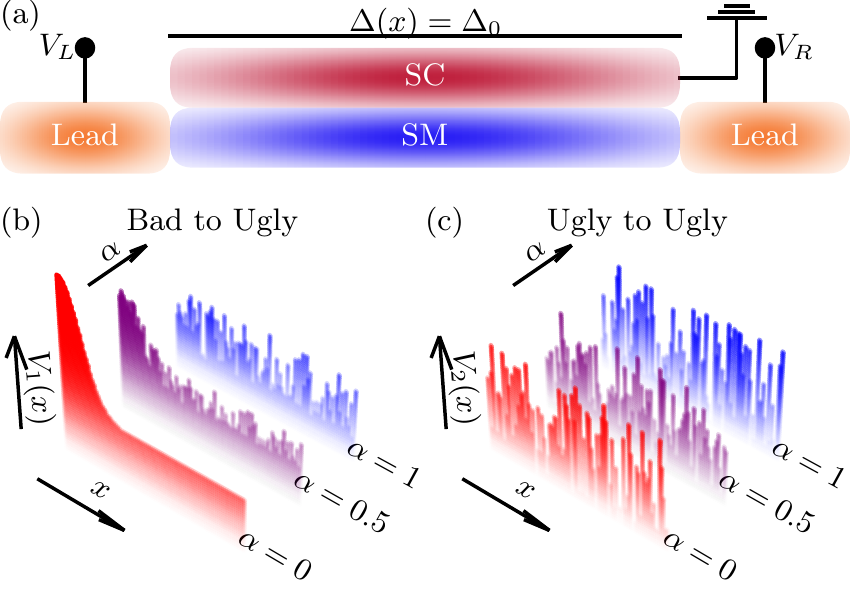}
	\caption{(a) The schematic for the three-terminal device of the SC-SM hybrid nanowire. (b) The crossover between the bad ZBCP and the ugly ZBCP controlled by $ \alpha $. $ V_1(x) $ transforms from an inhomogeneous potential at $ \alpha=0 $ to a random disorder potential at $ \alpha=1 $. (c) The crossover between two distinct ugly ZBCPs controlled by $ \alpha $. $ V_2(x) $ transforms from a random disorder potential at $ \alpha=0 $ to another random disorder potential at $ \alpha=1 $.}
	\label{fig:schematic}
\end{figure}

In this section, we briefly revisit three types of ZBCPs--- the good, the bad, and the ugly--- as first introduced in Ref.~\onlinecite{pan2020physical} by presenting their Hamiltonians and the corresponding mechanisms. The experimental device that we are theoretically simulating is a three-terminal superconductor-semiconductor (SC-SM) hybrid nanowire as shown in Fig.~\ref{fig:schematic}(a). The semiconductor which is covered by a grounded $ s $-wave pairing superconductor is attached with two normal leads with the bias voltages ($ V_L $ and $ V_R $) applying to the left and right end, respectively. We assume the nanowire has a spatially constant proximitized superconductivity induced by the parent SC gap $ \Delta_0 $ but may have an inhomogeneous chemical potential $ V(x) $ as we will discuss later.

{We note that this single-band 1D nanowire model is the minimal \emph{realistic} model that can represent the essential physics, and is already sufficient to describe all the crossover physics. Using other more detailed models, e.g., a multi band model~\cite{woods2020subband,liu2019conductance}, do not change the physics of the topological superconductivity qualitatively, but just unnecessarily complicates the calculation. In fact, given that any realistic information about the actual experimental Majorana nanowire situation (e.g., the carrier density, the number of occupied subbands, or the \textit{in situ} spin-orbit coupling strength, or the amount of disorder, or the applicable $g$ factor in the hybrid structure) is unknown at this stage, it is more reasonable to obtain general and generic results based on the parametrized minimal model used in the current paper, particularly since recent work has shown that detailed realistic simulations provide results similar to that obtained from this minimal model, which already includes the essential physics of spin-orbit coupling, proximity superconductivity, Zeeman splitting, finite wire length, and disorder~\cite{pan2021quantized,woods2021charge}. We emphasize that, unlike the Kitaev 1D chain model, our minimal model includes all the relevant physical ingredients of the experimental superconductor-semiconductor hybrid structure.}
\subsection{Good ZBCP}
 The good ZBCP is the ideal case of the pristine nanowire without any inhomogeneous potential or disorder. This pristine limit can be described by the standard Bogoliubov-de Gennes (BdG) Hamiltonian of the superconductor-semiconductor hybrid nanowire~\cite{sau2010generic,sau2010nonabelian,oreg2010helical,lutchyn2010majorana},
\begin{eqnarray}\label{eq:nanowire}
	H_{\text{nanowire}}&=&\frac12\int_{0}^{L} dx ~\hat{\Psi}^\dagger(x) \left[\left(-\frac{\partial^2_x}{2m^*}  -i \alpha_R \partial_x \sigma_y - \mu \right)\tau_z \right. \nonumber\\
	&+& \left. V_Z\sigma_x + \Sigma(\omega) \vphantom{\frac12} \right] \hat{\Psi}(x),
\end{eqnarray}
where $\hat{\Psi}(x)=\left(\hat{\psi}_{\uparrow}(x),\hat{\psi}_{\downarrow}(x),\hat{\psi}_{\downarrow}^\dagger(x),-\hat{\psi}_{\uparrow}^\dagger(x)\right)^{\intercal}$ represents a position-dependent Nambu spinor, and $\vec{\bm{\sigma}}$ and $\vec{\bm{\tau}}$ are vectors of Pauli matrices that act on the spin space and particle-hole space, respectively. $ L $ is the wire length, $ m^* $ is the effective mass of the conduction band, $ \alpha_R $ is the strength of Rashba-type spin-orbit coupling, and $ V_Z $ is the Zeeman field applied along the nanowire. 

The self-energy term $ \Sigma(\omega) $, which accounts for the proximitized superconductivity in the semiconductor by integrating out the degrees of freedom in the parent superconductor~\cite{stanescu2010proximity,stanescu2017proximityinduced}, is
\begin{equation}\label{eq:selfenergy}
	\Sigma(\omega)=-\gamma\frac{\omega+\Delta_0 \tau_x}{\sqrt{\Delta_0^2-\omega^2}},
\end{equation}
where $ \omega $ is the energy in the retarded Green's function for the BdG Hamiltonian, $ \Delta_0 $ is the pairing energy of the parent superconductor, and $ \gamma $ is the effective SC-SM coupling (tunneling) strength producing the SC proximity effect in the SM nanowire. {The simpler version without the self-energy of SC is obtained by replacing the $ \Sigma(\omega) $ in Hamiltonian~\eqref{eq:nanowire} by a constant $ s $-wave pairing $ \Delta $ to represent the proximitized superconductivity in the semiconductor.  This simpler version will be useful if we only focus on the states near zero energy .}

Without loss of generality, we choose the set of parameters corresponding to the experimental platform of the InSb-Al hybrid system~\cite{lutchyn2018majorana} {as a representative example, (which does not qualitatively affect the essential physics, and the parameters for the alternate InAs-Al structure are very similar)}: the effective mass $ m^*=0.015m_e $ with $ m_e $ being the electron rest mass, the parent SC gap $ \Delta_0=0.2 $ meV, the effective SC-SM coupling strength $ \gamma=0.2 $ meV, the Rashba-type spin-orbit coupling strength $ \alpha_R=0.5 $ eV\AA, and the chemical potential $ \mu=1 $ meV. These are typical numbers for Majorana nanowire systems in real units. 

{We focus on the long-wire limit with $ L=3~\mu $m to avoid the misleading finite-size effect as topological effects cannot manifest at all in short wires. (We do warn, however, that the current experimental nanowires may actually be in the short-wire limit.)
We emphasize that for short wires, $\sim$ 1 $\mu$m , no topological physics is expected in the system since the Majorana separation between the two ends of the wire then is shorter than the coherence length. To evaluate the effective length of the nanowire, we estimate the coherence length at zero magnetic field from the BCS theory~\cite{annett2004superconductivity} by
\begin{equation}
	\xi=\frac{\hbar v_F}{\pi\Delta},
\end{equation}
where $v_F=\alpha_R+\sqrt{\alpha_R^2+\frac{2\mu}{m^*}}$~\cite{oreg2010helical} is the Fermi velocity, and $\Delta\sim\frac{\gamma\Delta_0}{\gamma+\Delta_0}$~\cite{stanescu2017proximityinduced} is the proximity-induced SC gap. In our particular choice of parameters, Fermi velocity is around $2.4\times 10^5 $ m/s and the proximity-induced gap is roughly $0.1$ meV. Thus, the coherence length is roughly $0.5~\mu$m. Therefore, a one-micron nanowire has an effective length of $2\xi$, which is short. Our 3-micron wire length choice makes the dimensionless wire length to be 6 in units of the coherence length.  We mention that the actual coherence length in experimental nanowires is unknown at finite magnetic field of topological interest since it has never been directly measured.}

We calculate our theoretical results at zero temperature under the aforementioned parameters unless otherwise stated. Inclusion of temperature is straightforward in the theory~\cite{setiawan2017electron} and is an unnecessary complication--- typical experiments are carried out at 20-25 mK which should be well represented by our zero-temperature theory.

\subsection{Bad ZBCP}
We omit the effect of the electrostatic potential arising from various gate voltages in the pristine ``good MZM'' limit. However, this is too ideal in the realistic scenario: The charge impurity and various gate voltages may induce an inhomogeneous smooth confining potential in the semiconductor~\cite{kells2012nearzeroenergy,liu2017andreev,liu2018distinguishing,stanescu2019robust}. This inhomogeneous potential is a common mechanism that gives rise to the bad ZBCP~\cite{pan2020physical} in the tunneling spectroscopy, which could easily be mistaken as the topological Majorana zero mode.

Since the details of this inhomogeneous chemical potential are usually not precisely known in experiments, without loss of generality~\cite{liu2017andreev}, we model the inhomogeneous chemical potential $ V_{\text{bad}}(x) $ in the form of the Gaussian function
\begin{equation}\label{eq:inhomogeneous}
V_{\text{bad}}(x)=V_{\text{max}}\exp(-\frac{x^2}{2\sigma^2}),
\end{equation}
where $ \sigma $ and $ V_{\text{max}} $ define the linewidth and height of the peak of the inhomogeneous potential. In the following calculations, we choose $ \sigma=0.4~\mu $m and $ V_{\text{max}}=1.2 $ meV. Therefore, the inhomogeneous potential can be construed to give an effective chemical potential $ \mu-V_{\text{bad}}(x) $ in  Hamiltonian~\eqref{eq:nanowire}. We show an example of $ V_{\text{bad}}(x) $ in Fig.~\ref{fig:schematic}(b) at $ \alpha=0 $.

\subsection{Ugly ZBCP}\label{sec:ugly}
Besides the bad ZBCP, another type of trivial ZBCP is the ugly ZBCP arising from random disorder in the chemical potential of the nanowire. Potential disorder is unintentional and unavoidable due to the imperfect sample quality, subband occupation~\cite{woods2020subband}, etc. Thus, we model potential disorder as the randomness in the chemical potential  $ V_{\text{ugly}}(x) $  which is drawn from an uncorrelated Gaussian distribution with the mean value of zero and variance of $ \sigma_\mu^2 $. Spatially dependent random potential disorder also gives an effective chemical potential $ \mu-V_{\text{ugly}}(x) $ in Hamiltonian~\eqref{eq:nanowire}. We sketch a representative random disorder $ V_{\text{ugly}}(x) $ in Fig.~\ref{fig:schematic}(b) at $ \alpha=1 $. Note that the inhomogeneous potential producing bad ZBCP is deterministic, varying smoothly spatially, whereas potential disorder, producing ugly ZBCP, is spatially random; although impurities may give rise to both effects, their physical origins are qualitatively different as are their effects on the Majorana physics.  We also emphasize that neither is ``noise'' in the usual sense,  which is the random temporal variation of current and voltage in electronic systems.

Because weak potential disorder preserves the topological properties of the nanowire which does not induce the trivial ZBCP, and strong potential disorder completely destroys the hybrid nanowire model, which should be instead described by a random matrix approach in a class D ensemble~\cite{beenakker1997randommatrix,guhr1998randommatrix,brouwer1999distribution,beenakker2015randommatrix,mi2014xshaped,pan2020generic}, {and leaves almost no signature in the conductance spectrum}, we focus on intermediate disorder with $ \sigma_\mu/\mu\sim1 $. We clarify that all following results are presented just for one particular random realization, without taking the ensemble average, which is the appropriate theory for low-temperature nanowire experiments.

We also mention that there have been several theoretical studies in the literature already, focusing on the cases with just inhomogeneous potential or just disorder, but ours studies both appearing together in the system.

\subsection{Crossover between the bad and the ugly ZBCP}
We introduce the first type of crossover: the crossover between the bad ZBCP and the ugly ZBCP [Fig.~\ref{fig:schematic}(b)], and the second type of crossover between one ugly ZBCP and another ugly ZBCP with different potential disorder realizations [Fig.~\ref{fig:schematic}(c)]. In Figs.~\ref{fig:schematic}(b) and~\ref{fig:schematic}(c), we use the tuning parameter $ \alpha $ to control the crossover. 

The first type of crossover [Fig.~\ref{fig:schematic}(b)] is the case where the nanowire starts with showing the bad ZBCP arising from the inhomogeneous potential denoted by $ V_{\text{bad}}(x) $ and ends up with showing the ugly ZBCP arising from random potential disorder denoted by $ V_{\text{ugly}}(x) $. We interpolate the two limits linearly such that the nanowire experiences an effective potential $ V_1(x;\alpha) $ continuously during the crossover as per
\begin{equation}\label{eq:badtougly}
	V_1(x;\alpha)=(1-\alpha) V_{\text{bad}}(x)+ \alpha V_{\text{ugly}}(x).
\end{equation} 
Therefore, $ \alpha$=0 corresponds to the inhomogeneous potential that gives rise to the strictly bad ZBCP, and $ \alpha $=1 corresponds to random disorder in the chemical potential that gives rise to the strictly ugly ZBCP. When $ \alpha\in\qty(0,1) $, the nanowire lies in an intermediate state [e.g., Fig.~\ref{fig:schematic}(b) at $\alpha=0.5  $] which is of our interest. In this crossover regime, which is the likely generic experimental situation, the ZBCP is neither strictly bad nor strictly ugly but is some kind of a complicated combined bad-ugly mixture depending on the value of $ \alpha $. We present representative results at several $ \alpha $  along with two extreme cases of $ \alpha=0 $ and 1 in Sec.~\ref{sec:results}. For a complete process of the crossover, we refer to an animation in the Supplemental Material to show the crossover continuously~\cite{crossover_SM}.

\subsection{Crossover between two distinct ugly ZBCPs}
The second type of crossover [Fig.~\ref{fig:schematic}(c)] involves two ugly ZBCPs arising from two distinct potential disorder realizations denoted by $ V_{\text{ugly}}^{(1)}(x) $ and $ V_{\text{ugly}}^{(2)}(x) $. Similar to the first type of crossover in Eq.~\eqref{eq:badtougly} between the bad ZBCP and the ugly ZBCP, we also interpolate the two limits of the ugly ZBCPs linearly. Namely, the nanowire is in the presence of an effective potential
\begin{equation}\label{eq:ugly2ugly}
	V_2(x;\alpha)=(1-\alpha)  V_{\text{ugly}}^{(1)}(x) +\alpha  V_{\text{ugly}}^{(2)}(x)
\end{equation}
during the crossover.
Therefore, $ \alpha $=0 or 1 corresponds to either one of the predetermined random realizations, and $ \alpha\in\qty(0,1) $ corresponds to an intermediate state. We also present representative results at several $ \alpha $  along with two extreme cases of $ \alpha $=0 and 1 in Sec.~\ref{sec:results}. The continuous process of the crossover is shown in the form of an animation in the Supplemental Material as well~\cite{crossover_SM}.

In fact, the linear interpolation of two predetermined disorder realizations actually decreases the variance of the original potential disorder from 1 [e.g., Fig.~\ref{fig:schematic}(c) at $ \alpha=0 $ and 1] to $ 1-2\alpha+2\alpha^2 $ [e.g., Fig.~\ref{fig:schematic}(c) at $ \alpha=0.5 $]. However, this is not an issue because such disorder inside the nanowire is, in principle, susceptible to many sources, and, thus, the chemical potential is unknown \textit{a priori}. There is no particular reason why the disorder variance is conserved as system parameters change to vary the disorder from one to another realization. For example, the electrostatic potential in the nanowire may be different even when the device undergoes a charge jump with all gate voltages returning to the initial state~\cite{zhang2021large,zhang2018quantizeda}. 
Theoretically, we are simulating a nanowire that was initially in the presence of one particular random potential disorder and is now in the presence of another distinct random potential disorder. Thus, it is not guaranteed that disorder will conserve its strength throughout the process of the voltage switch in the laboratory. 

Nevertheless, we have also carried out the variance-conserving crossover calculations between two ugly scenarios {following the interpolation as per
\begin{equation}\label{eq:ugly2ugly_sqrt}
	V_2(x;\alpha)=\sqrt{1-\alpha}  V_{\text{ugly}}^{(1)}(x) +\sqrt{\alpha}  V_{\text{ugly}}^{(2)}(x),
\end{equation}
} {but the results are qualitatively the same as that obtained from the model of Eq.~\eqref{eq:ugly2ugly}.
There is no current experimental information on the detailed forms of available $ V_\text{bad} $ and $ V_\text{ugly} $ and therefore, of $ V_1 $ [Eq.~\eqref{eq:badtougly}] and $ V_2 $ [Eqs.~\eqref{eq:ugly2ugly} and~\eqref{eq:ugly2ugly_sqrt}]. Our results are therefore of qualitative validity in the generic system.}
\subsection{Crossover with impurity disorder}

Besides the aforementioned potential disorder, where a Gaussian spatially random potential in the Hamiltonian represents the disorder, which is present throughout the nanowire, we also consider the effect of the random localized impurities, which we call the impurity disorder model in this paper as opposed to potential disorder defined by a random potential. 

We first start with the simplest case--- single-impurity disorder--- where only one impurity resides in the nanowire, and it is located in the middle of the nanowire, i.e., $ 1.5~\mu $m of a $ 3~\mu $m-nanowire as we stated before. Therefore, only the chemical potential at the middle point in the nanowire is changed by
\begin{equation}\label{eq:Vx}
	V(x)=V\delta_{x,x_0},
\end{equation}
where $ x_0=L/2 $, and $ \delta_{i,j} $ is Kronecker's delta function. We tune the strength of impurity disorder $ V $ from a large negative value to a large positive value to study the crossover physics. Note that although we refer to this case as ``disorder'' because it arises from an impurity, there is nothing random about Eq.~\eqref{eq:Vx}, which simply denotes an onsite potential of strength $V$.

Similarly, we also consider double-impurity disorder, where only two impurities exist in the nanowire. For simplicity, we make the two impurities of equal strength and place them spatially symmetrically at one-third and two-thirds of the nanowire length, i.e., at $ 1$ and $ 2~\mu $m of a $ 3~\mu $m-nanowire. Thus, the chemical potential is changed by 
\begin{equation}
	V(x)=V(\delta_{x,x_0}+\delta_{x,x_1}),
\end{equation}
where $ x_0=L/3 $ and $ x_1=2L/3 $. Again, the disorder strength $ V $ is the tuning parameter during the crossover, and there is nothing random about this ``disorder''.

Beyond the simple single-impurity and double-impurity disorder, which are completely deterministic, we also study a more complicated case where a small portion of the nanowire is occupied by many impurities at random different sites. These impurities are randomly spatially distributed in the nanowire with equal strength of disorder but random signs. In this case, the randomness lies in the positions of impurities and their signs, but not in their strength in contrast to our Gaussian random potential disorder of Sec.~\ref{sec:ugly}. In Sec.~\ref{sec:results}, we study the cases where 10\% and 30\% sites of the nanowire are randomly occupied by impurities of random sign, respectively.

Finally, we consider a mixed case with both potential disorder and impurity disorder: impurities are placed throughout the nanowire. This is a different type of disorder from the Gaussian potential disorder which induces the ugly ZBCP; however, we will show later in Sec.~\ref{sec:results} that by changing the strength of the onsite impurity potential disorder, we obtain the same qualitative results as in creating the ugly ZBCP case arising from a Gaussian random disorder potential. Thus, our disorder results for the ugly ZBCP are universal as long as the underlying potential is random independently of the specific random model we use.

\color{black}
\section{Theory}\label{sec:theory}
\subsection{Local density of states}
To explicitly show the presence or absence of the gap closing and reopening features in the bulk region, and the end-to-end correlation (or lack thereof) at two ends in the good, bad, and ugly cases, we resort to the local density of states in the middle of the nanowire and at both ends of the nanowire. We first discretize the BdG Hamiltonian using a fictitious lattice constant $ a_0 =10$ nm and replacing the derivatives with the finite differences to construct a tight-binding model~\cite{dassarma2016how}. Thus, the LDOS corresponding to the tight-binding model at the energy $ \omega $ and position $ x_i $ is defined as 
\begin{equation}\label{eq:LDOS}
	\text{LDOS}(\omega,x_i)=-\frac{1}{\pi} \Im\qty[\tr_{\sigma,\tau}\qty(\frac{1}{\omega+\eta-H})]_{i,i},
\end{equation}
where $ \tr_{\sigma,\tau} $ is a partial trace over the spin and particle-hole space, $ \Im\qty[\dots] $ takes the imaginary part, $ H $ is the Hamiltonian of the nanowire, the subscript $ i,i $ takes the $ i $-th diagonal term in the matrix, and $ \eta $ is a standard positive infinitesimal required to ensure the causality. 

We show the LDOS of the Hamiltonian in the pristine limit as well as the aforementioned crossovers in Appendix~\ref{app:B}. For distances larger than $ a_0 $ our calculated LDOS corresponds to the continuum system of interest, and the details of the tight-binding prescription become coarse-grained.

\subsection{Tunnel conductance}
To simulate the experimental measurements of the tunneling spectroscopies, we also calculate the tunnel conductance based on the {Blonder-Tinkham-Klapwijk }formalism~\cite{blonder1982transition,datta1995electronic,anantram1996current}. We first attach two semi-infinite normal leads on both ends of the nanowire [Fig.~\ref{fig:schematic}(a)], where the Hamiltonian of the lead takes the same form as that of the nanowire except for the superconducting term (i.e., no proximitized superconductivity in normal leads). The chemical potential of the lead is 25 meV, and the tunnel barrier height at the normal-superconductor (NS) junction interface is 10 meV following the choice in Ref.~\onlinecite{pan2021threeterminal}. Then we assume a propagating wave in the normal lead and calculate the $ S $ matrix at the NS interface. The calculation of the $ S $ matrix is done with the help of the Python scattering matrix package KWANT~\cite{groth2014kwant}. Details of the tunneling conductance calculation are not provided since they are standard and can be found in the literature~\cite{prada2012transport,dassarma2016how,liu2017andreev,setiawan2015conductance,rosdahl2018andreev}.

\section{Results}\label{sec:results}

\subsection{Crossover between bad and ugly ZBCPs}
\begin{figure}[htbp]
	\centering
	\includegraphics[width=3.4in]{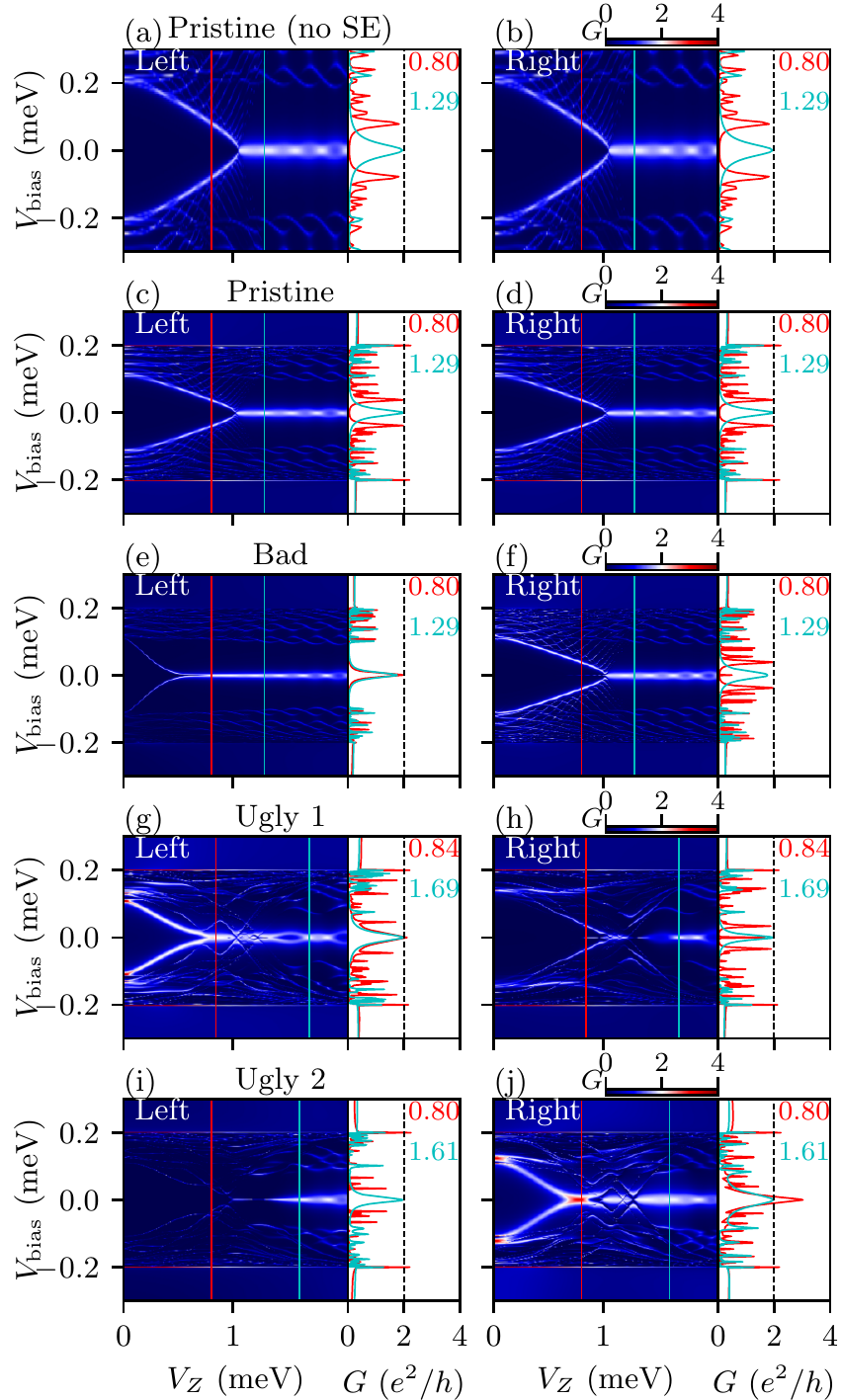}
	\caption{The differential conductance as a function of Zeeman field $ V_Z $ and bias voltage $ V_{\text{bias}} $ from the left end (left panels) and right end (right panels) for different static cases. The common parameters are as follows: chemical potential $ \mu=1 $ meV, parent SC gap $ \Delta_0=0.2 $ meV, spin-orbit coupling $ \alpha=0.5 $ eV\AA, wire length $ L=3~\mu $m, dissipation is $ 10^{-4} $ meV, and zero temperature. The TQPT is at $ V_Z=1.02 $ meV and the corresponding line cuts in the trivial regime (red) and the topological regime (cyan) are shown right to the color plot of the conductance.
	(a),(b) The pristine wire without the self-energy of SC. The proximitized $ \Delta =0.2$ meV.    
	(c),(d) The pristine wire with the self-energy of the SC. The SC-SM coupling strength $ \gamma=0.2 $ meV.
	(e),(f) The bad ZBCP in the presence of the inhomogeneous potential with $ V_{\text{max}}=1.2 $ meV and $ \sigma=0.4~\mu $m.
	(g),(h) The ugly ZBCP in the presence of random potential disorder with $ \sigma_\mu=1 $ meV. 
	(i),(j) The ugly ZBCP in the presence of another random potential disorder with $ \sigma_\mu=1 $ meV.
	The corresponding LDOSs are shown in Fig.~\ref{fig:static_LDOS}.
	}
	\label{fig:static}
\end{figure}
In this section, we first present the results for the calculated tunnel conductance as a function of the Zeeman splitting and bias voltage in the pristine wire limit corresponding to Hamiltonian~\eqref{eq:nanowire} as shown in Figs.~\ref{fig:static}(c) and~\ref{fig:static}(d). These results correspond to the good (i.e., topologically protected) Majoranas. Figures~\ref{fig:static}(c) and~\ref{fig:static}(d) show the tunnel conductances measured from the left and right end, respectively. On the right of the color plot of conductance, we show two line cuts in the trivial regime at $ V_Z=0.8 $ meV (red) and the topological regime $ V_Z=1.29 $ meV (cyan). The conductances measured from two ends in this ideal case show a perfect end-to-end correlation because of the nonlocal topological nature of MZMs. The ZBCPs shown here are all topological good ZBCPs with a robust quantized plateau of $ 2e^2/h $ above the topological quantum phase transition (TQPT) at $ V_Z=1.02 $ meV where the bulk gap closes. The good ZBCP also manifests an increasing Majorana oscillation as the Zeeman field increases above the TQPT because the dimensionless separation between the two end Majoranas decreases with increasing Zeeman field because of the decrease in the topological gap. In addition, we present the corresponding LDOSs at two ends and in the middle of the nanowire in Figs.~\ref{fig:static_LDOS}(d),~\ref{fig:static_LDOS}(e), and~\ref{fig:static_LDOS}(f) in Appendix~\ref{app:B}. We find the bulk gap closing and reopening features are very prominent showing up in the middle of the nanowire at the TQPT, but not at the wire ends, as shown in Fig.~\ref{fig:static_LDOS}(e). 

\begin{figure}[htbp]
	\centering
	\includegraphics[width=3.4in]{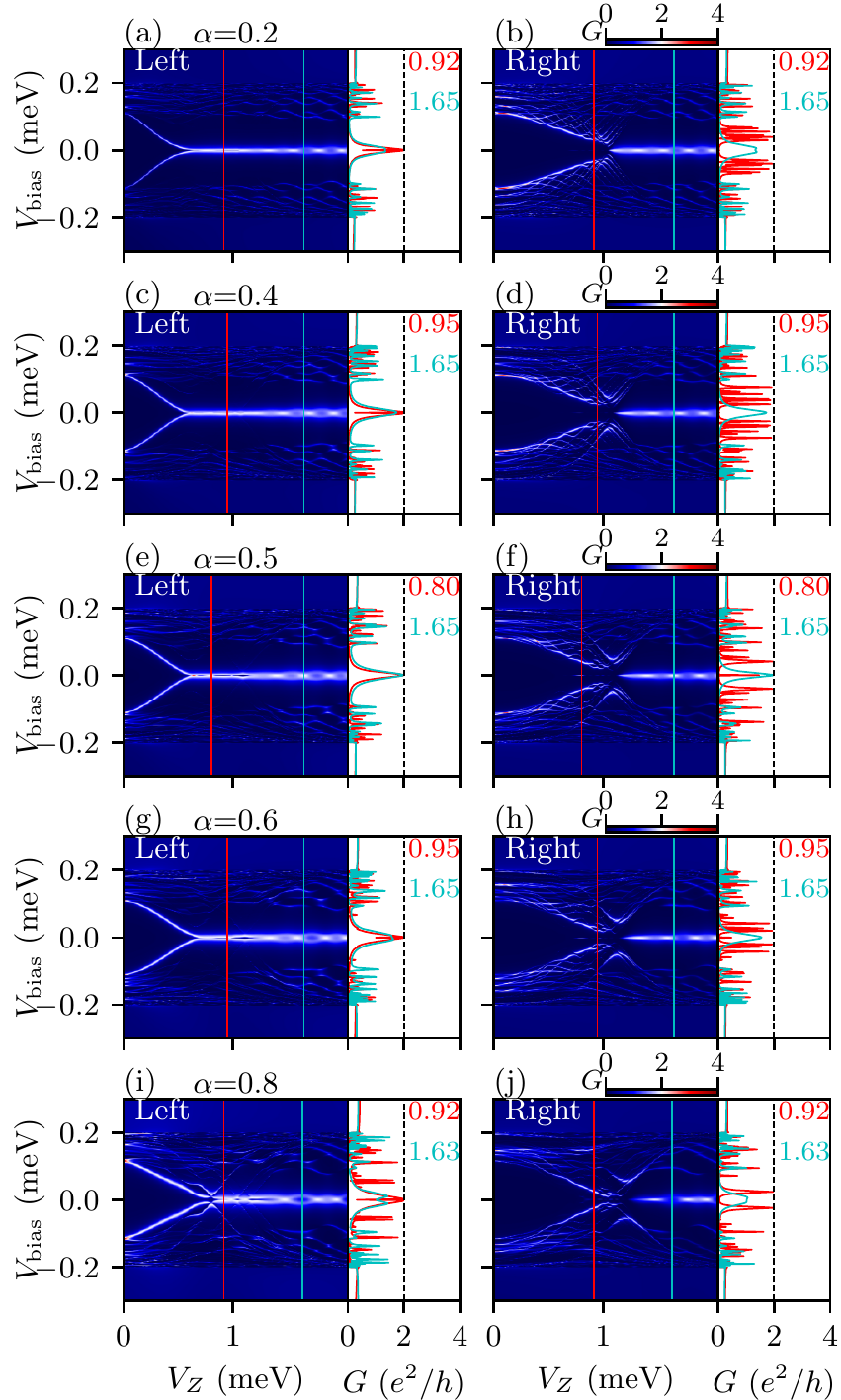}
	\caption{The tunnel conductance measured from the left end (in the left panels) and right end (in the right panels) of the nanowire in the crossover between the bad ZBCP [Figs.~\ref{fig:static}(a) and~\ref{fig:static}(b) for $ \alpha=0 $] arising from an inhomogeneous potential and ugly ZBCP [Figs.~\ref{fig:static}(g) and~\ref{fig:static}(h) for $ \alpha=1 $] arising from random disorder. 
	From the top to the bottom panels, more disorder is blended with the inhomogeneous chemical potential. 
	The corresponding line cuts in the trivial regime (red) and the topological regime (cyan) are shown right to the color plot of the conductance. Refer to Fig.~\ref{fig:static} for other parameters. The corresponding LDOSs are shown in Fig.~\ref{fig:bad_ugly_LDOS}.
	}
	\label{fig:bad_ugly}
\end{figure}

However, the pristine wire limit rarely applies in the laboratory; therefore, a more realistic scenario is the nanowire in the presence of an inhomogeneous potential and/or disorder. In Fig.~\ref{fig:bad_ugly}, we show a representative result of the tunnel conductance in the crossover between the bad ZBCP [Figs.~\ref{fig:static}(e) and~\ref{fig:static}(f)] arising from the inhomogeneous potential and the ugly ZBCP [Figs.~\ref{fig:static}(g) and~\ref{fig:static}(h)] arising from random potential disorder. In Fig.~\ref{fig:bad_ugly}, the left (right) panels show the tunnel conductances measured from the left (right) end of the nanowire with the corresponding line cuts in the trivial regime (red) and topological regime (cyan), respectively. 
At $ \alpha=0 $ [Figs.~\ref{fig:static}(e) and~\ref{fig:static}(f)], when there is no disorder by definition, the nanowire shows a bad ZBCP with a quantized plateau at the left end [Fig.~\ref{fig:static}(e)] below the TQPT, which is the quasi-Majorana arising from the inhomogeneous potential~\cite{vuik2019reproducing}. The conductance at the right end [Fig.~\ref{fig:static}(f)] does not show any trivial ZBCPs because the right end of the nanowire does not have inhomogeneity to confine a quasi-Majorana. 
When disorder is introduced by increasing $\alpha$, for example for $ \alpha=0.5 $ [Figs.~\ref{fig:bad_ugly}(e) and~\ref{fig:bad_ugly}(f)], the system is in the intermediate crossover between the bad ZBCP and the ugly ZBCP, where the conductance spectrum at the left end shows the remnants of the bad ZBCP with an almost-quantized plateau in the trivial regime, while the conductance spectrum at the right end manifesting a disappearing segment of ZBCP ($1.02< V_Z<1.4 $ meV) is the remnant of the good topological Majorana, but suppressed strongly by disorder. The topological ZBCP reappears above around $ V_Z=1.4 $ meV because of an effective disorder-induced shift in the TQPT to a higher Zeeman field. 
At $ \alpha=1 $ [Figs.~\ref{fig:static}(g) and~\ref{fig:bad_ugly}(h)], where disorder is strong, the nanowire manifests the trivial ugly ZBCP with the conductance peak above $ 2e^2/h $ at the left end. At the right end, there are only sporadic ZBCPs with arbitrary values of conductances below $ 2e^2/h $, unlike the bad ZBCPs at $ \alpha=0 $ whose conductances are almost quantized at $ 2e^2/h $, which explains the terminology of ``quasi-Majorana'' often used to describe this nontopological ZBCP. The whole crossover regime, $alpha$ =0 to 1, is thus smooth in going from purely inhomogeneous potential induced quasi-Majorana to trivial disorder-induced zero bias peaks, and in between both effects may manifest, complicating interpretation.

A generic feature throughout the crossover region from $ \alpha $=0 to 1 is the absence of any end-to-end correlation (Fig.~\ref{fig:bad_ugly}) below the TQPT because the trivial ABS is a fermionic state which does not have the nonlocal property of the topological MZM. Near the ugly region [$ \alpha >0.8$ in Figs.~\ref{fig:bad_ugly}(i) and~\ref{fig:bad_ugly}(j)], the end-to-end correlation even disappears above the putative TQPT ($ V_Z=1.02 $ meV) since the topological regime is suppressed by strong potential disorder.  However, near the bad region [$ \alpha=0.2 $ in Figs.~\ref{fig:bad_ugly}(a) and~\ref{fig:bad_ugly}(b)], the end-to-end correlation is restored above the TQPT along with the appearance of Majorana oscillation, which indicates the metamorphosis from the trivial ABS to the topological MZM as the Zeeman field increases in the inhomogeneous potential without the presence of very strong disorder. We also note that the bad ZBCP arising from the inhomogeneous potential shows considerable stability with the Zeeman field even in the presence of some disorder (i.e., $ \alpha=0.5 $). Thus, it may be possible to observe topological MZMs in nanowires in the crossover regime with both inhomogeneous potential and random disorder present in the system, as long as the disorder is not particularly strong, and the experiment is probing the topological regime above TQPT.  For strong disorder, however, only ugly trivial ZBCPs survive with all MZM effects completely suppressed as already emphasized in the literature~\cite{dassarma2021disorderinduced}.

We also calculate the crossover LDOS between the bad ZBCP and the ugly ZBCP in Fig.~\ref{fig:bad_ugly_LDOS} of Appendix~\ref{app:B}. Similarly to the local conductance in Fig.~\ref{fig:bad_ugly}, we see the lack of the end-to-end correlation below the TQPT. In particular, the LDOS in the middle of the nanowire clearly shows the gap closing and reopening features. At $ \alpha=0 $ [Fig.~\ref{fig:static}(h)], we see sharp gap closing and reopening features at the TQPT ($ V_Z=1.02 $ meV), which indicates the same transition from the trivial ABS to the topological Majorana bound states as the reappearance of the end-to-end correlation in the local conductance in Figs.~\ref{fig:static}(e) and~\ref{fig:static}(f) does. However, as $ \alpha $ increases [from the top to bottom panels of Fig.~\ref{fig:bad_ugly_LDOS}], more potential disorder is blended with the inhomogeneous chemical potential, suppressing the topological Majorana physics. Thus, we notice the gap closing and reopening features gradually becoming ambiguous, i.e., the gap closing and reopening do not happen simultaneously. This ambiguity becomes worse in the ugly region at $ \alpha=1 $ [Fig.~\ref{fig:static_LDOS}(k)] as the gap closes and reopens at different Zeeman fields, which indicates that the TQPT is strongly renormalized by strong disorder.

\begin{figure}[htbp]
	\centering
	\includegraphics[width=3.4in]{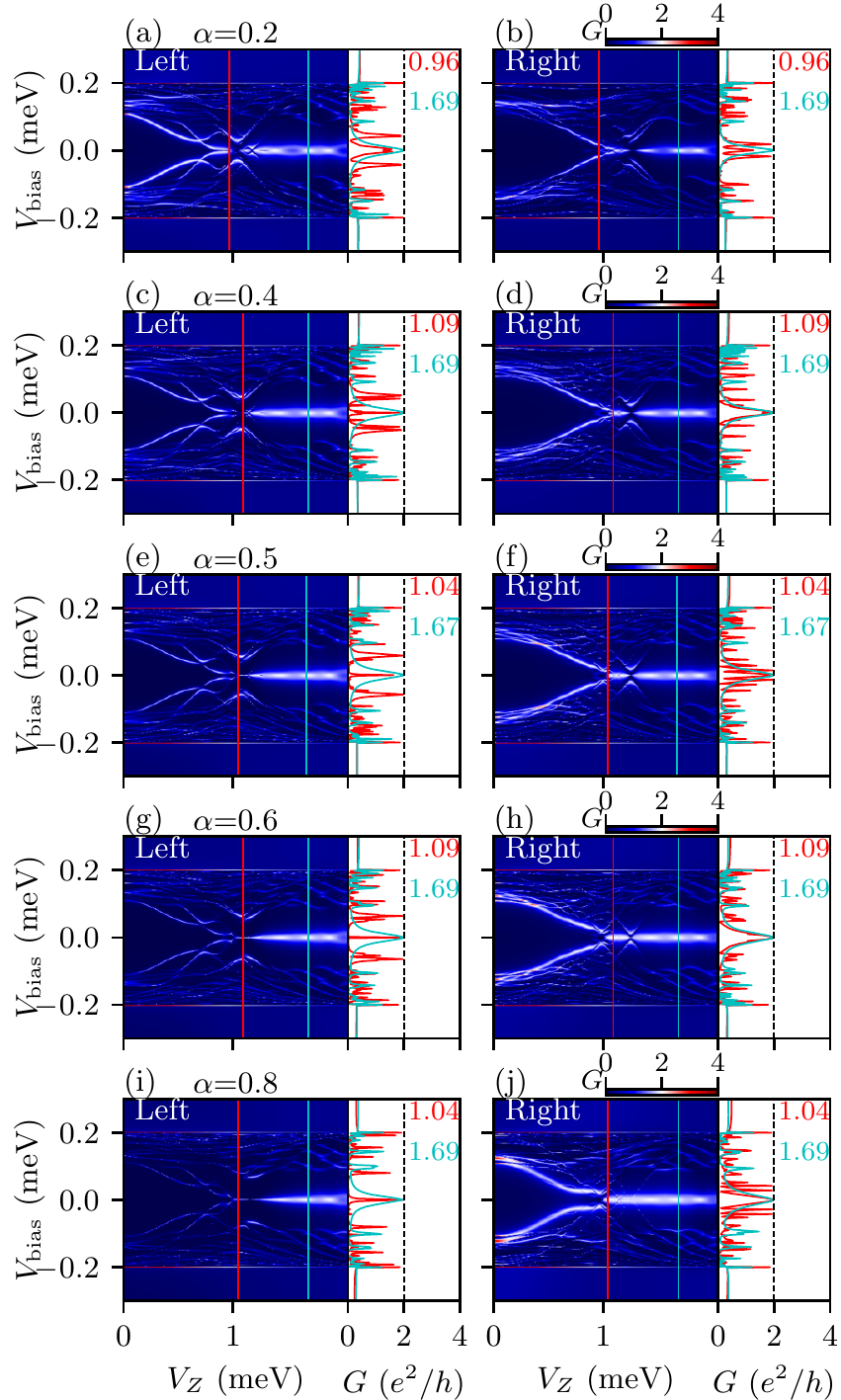}
	\caption{The tunnel conductance measured from the left end (in the left panels) and right end (in the right panels) of the nanowire in the crossover between two ugly ZBCPs [Figs.~\ref{fig:static}(g) and~\ref{fig:static}(h) for $ \alpha=0 $ and Figs.~\ref{fig:static}(i) and~\ref{fig:static}(j) for $ \alpha=1 $] using the simple linear interpolation of Eq.~\eqref{eq:ugly2ugly}. 
	From the top to the bottom panels, potential disorder becomes closer to the realization of Figs.~\ref{fig:static}(i) and~\ref{fig:static}(j).
	The corresponding line cuts in the trivial regime (red) and the topological regime (cyan) are shown right to the color plot of the conductance. Refer to Fig.~\ref{fig:static} for other parameters. The corresponding LDOSs are shown in Fig.~\ref{fig:ugly_ugly_linear_LDOS}.
	}
	\label{fig:ugly_ugly_linear}
\end{figure}
\subsection{Crossover between ugly ZBCPs}

In Fig.~\ref{fig:ugly_ugly_linear}, we present the tunnel conductance in the crossover between two ugly ZBCPs, arising from distinct disorder configurations, using the simple linear interpolation of Eq.~\eqref{eq:ugly2ugly}. Namely, we start with the random potential disorder configuration in Figs.~\ref{fig:static}(g) and~\ref{fig:static}(h) corresponding to $ \alpha=0 $ , and let it transform into another set of realizations of potential disorder in Figs.~\ref{fig:static}(i) and~\ref{fig:static}(j) corresponding to $ \alpha=1 $. At these two limits, they are both ugly ZBCPs without any end-to-end correlation, which is similar to the previous crossover between the bad ZBCP and the ugly ZBCP. The crossover now is between different disorder configurations, which may actually happen experimentally as various gate voltages are tuned in the nanowire during the experiment, causing arbitrary disorder annealing parametrized by our crossover parameter $\alpha$. However, during the crossover, one additional noteworthy feature is the ubiquitous presence of the trivial ZBCPs: The conductance peaks show arbitrary values between 0 and $ 4e^2/h $ at random ranges of Zeeman fields in the trivial regime at different $ \alpha $'s. The ubiquitous trivial ZBCPs in the crossover from $ \alpha $ =0 to 1 resemble the experimentally observed ZBCPs during a cycle of voltage switch or a cycle of heat-up and cool-down~\cite{zhang2018quantizeda,zhang2021large}, which are arising from the slowly varying disorder inside the sample as the impurities move around during voltage and field annealing. 
We present an animation that shows the random appearance and disappearance of the ugly ZBCPs in the Supplemental Material~\cite{crossover_SM} corresponding to Fig.~\ref{fig:ugly_ugly_linear}.

In Appendix~\ref{app:B}, we also present the LDOS in Fig.~\ref{fig:ugly_ugly_linear_LDOS} corresponding to the local conductance in Fig.~\ref{fig:ugly_ugly_linear}. Similarly to the crossover between the bad ZBCP and the ugly ZBCP, we find the absence of the end-to-end correlation at two ends, and the gap closing and reopening features are also not sharp (i.e., the bulk gap does not reopen immediately after it closes) throughout the crossover [which can be seen from LDOS in the middle of the nanowire as shown in in the middle panels of Fig.~\ref{fig:ugly_ugly_linear_LDOS}] because of the ubiquitous zero-energy Andreev modes in the trivial regime in the presence of potential disorder.

\begin{figure}[htbp]
	\centering
	\includegraphics[width=3.4in]{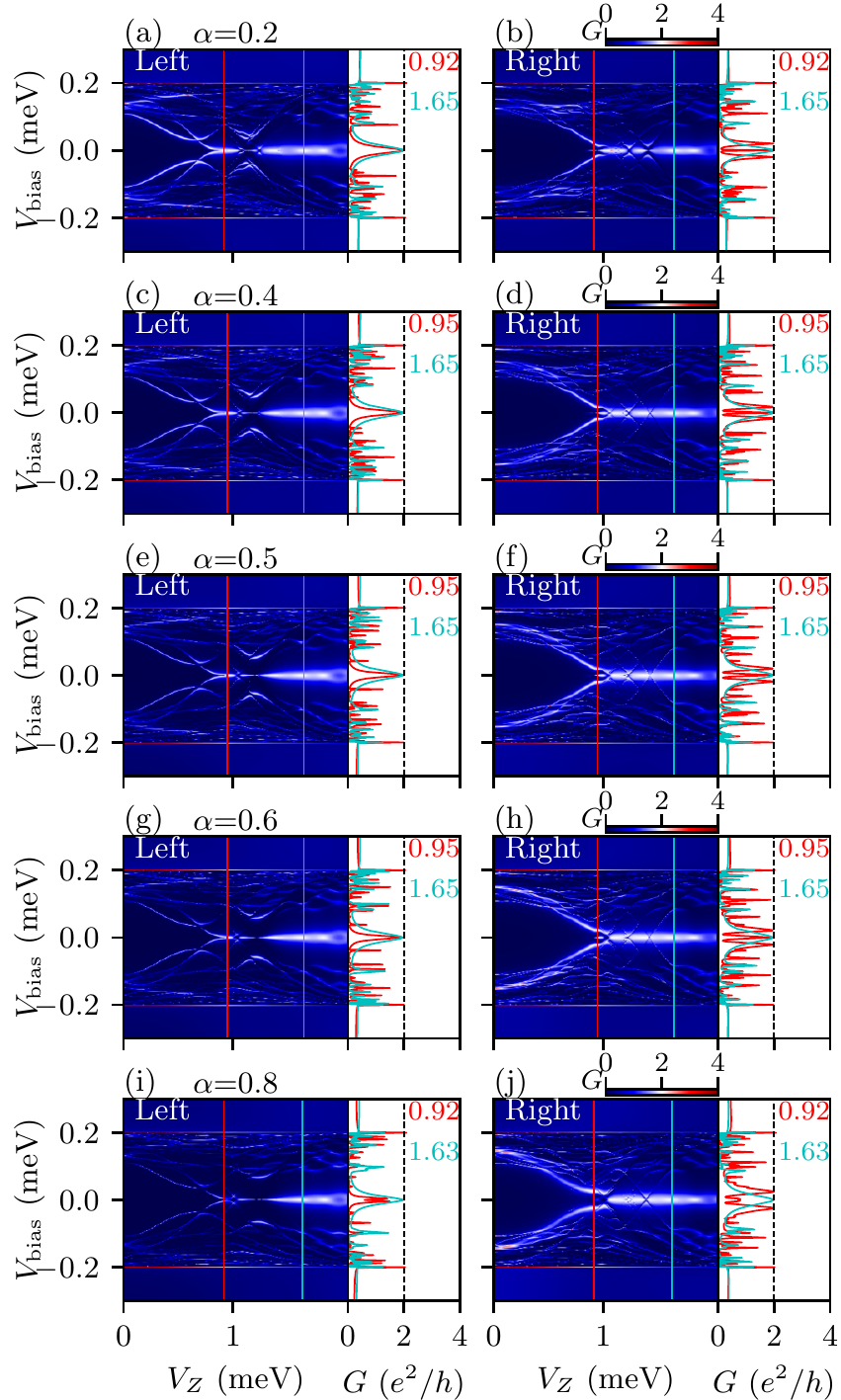}
	\caption{The tunnel conductance measured from the left end (in the left panels) and right end (in the right panels) of the nanowire in the crossover between two ugly ZBCPs [Figs.~\ref{fig:static}(g) and~\ref{fig:static}(h) for $ \alpha=0 $ and Figs.~\ref{fig:static}(i) and~\ref{fig:static}(j) for $ \alpha=1 $] using the variance-conserving interpolation of Eq.~\eqref{eq:ugly2ugly_sqrt}. 
	From the top to the bottom panels, the potential disorder becomes closer to the realization of Figs.~\ref{fig:static}(i) and~\ref{fig:static}(j).
	The corresponding line cuts in the trivial regime (red) and the topological regime (cyan) are shown right to the color plot of the conductance. Refer to Fig.~\ref{fig:static} for other parameters. The corresponding LDOSs are shown in Fig.~\ref{fig:ugly_ugly_var_LDOS}.
		}
	\label{fig:ugly_ugly_var}
\end{figure}

In addition to the linear interpolation, we also use the variance-conserving interpolation and present the tunnel conductance in the crossover between the same two ugly ZBCPs in Fig.~\ref{fig:ugly_ugly_var}. We show several values of $ \alpha $  between 0.2 and 0.8, and we find that the key feature of the ubiquitous ZBCPs does not change in this new crossover model. We also provide the crossover in the presence of another set of potential disorder realization in Figs.~\ref{fig:ugly_ugly_linear_2} and~\ref{fig:ugly_ugly_var_2} to show the generality and ubiquity of the trivial ZBCPs in Appendix~\ref{app:A} along with their corresponding LDOS in Figs.~\ref{fig:ugly_ugly_linear_2_LDOS} and~\ref{fig:ugly_ugly_var_2_LDOS} in Appendix~\ref{app:B}.

The key generic messages of Figs.~\ref{fig:ugly_ugly_linear} and~\ref{fig:ugly_ugly_var} are that (1) disorder could randomly produce ZBCPs with conductance values at, below, above $ 2e^2/h $; (2) disorder-induced ZBCPs are unstable, without manifesting any robustness in the Zeeman field; (3) disorder-induced ZBCPs are uncorrelated for tunneling from both ends; (4) disorder-induced ZBCPs could occasionally suddenly disappear with small parameter variations, reflecting the so-called ``charge jumps'' or ``voltage switches'' associated with random traps in electronic materials; (5) on a qualitative level, most existing Majorana nanowire experimental results appear to be consistent with the system manifesting crossovers between different disorder configurations as gate voltages are tuned with the observed ZBCPs being the ``ugly'' ones.

\subsection{Crossover with the strength of disorder}

\begin{figure}[htbp]
	\centering
	\includegraphics[width=3.4in]{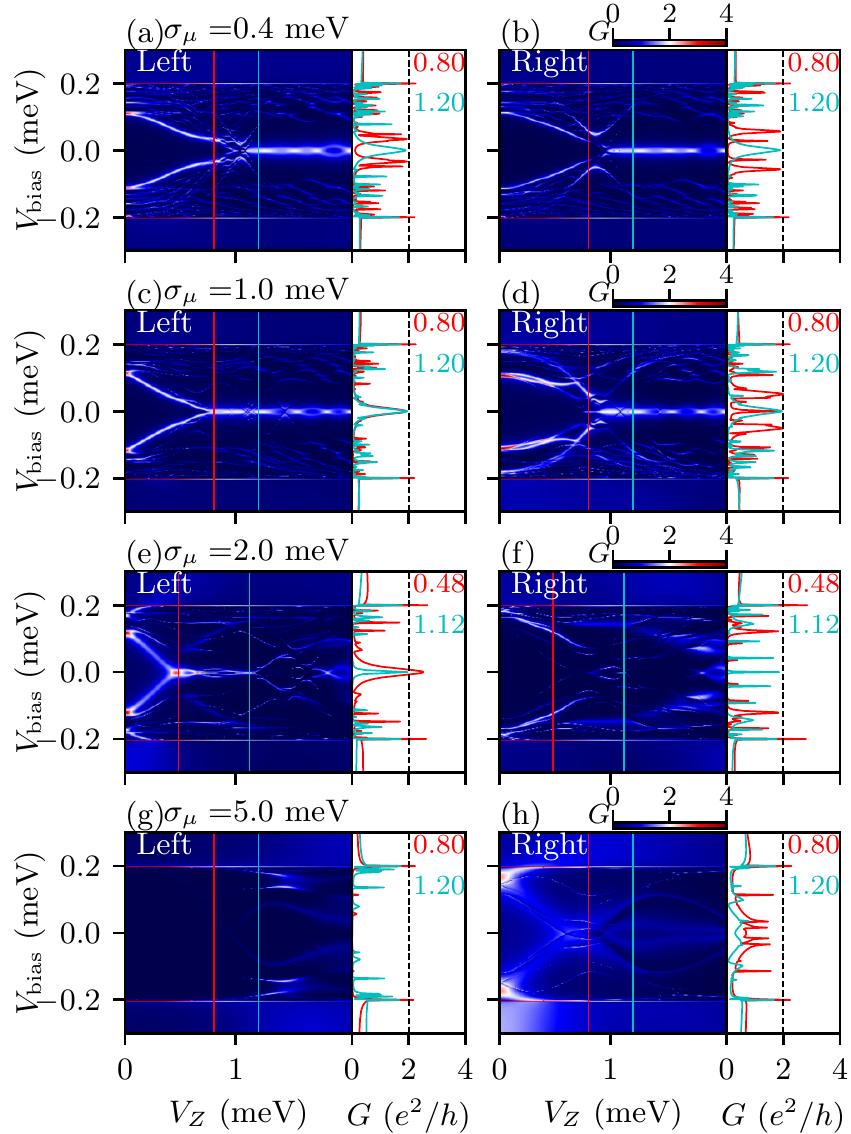}
	\caption{The tunnel conductance measured from the left end (in the left panels) and right end (in the right panels) of the nanowire as the potential disorder increases from a weak regime [(a),(b) with $ \sigma_\mu=0.4 $ meV] to a very strong regime [(g),(h) with  $ \sigma_\mu=5 $ meV]. 
	The corresponding line cuts in the trivial regime (red) and the (nominal) topological regime (cyan) are shown right to the color plot of the conductance.
	Refer to Fig.~\ref{fig:static} for other parameters.
	}
	\label{fig:ugly}
\end{figure}

Besides the crossover between two static potentials (an inhomogeneous potential to a quenched random potential disorder or two distinct quenched potential disorder realizations), we also present the tunnel conductance spectrum as the potential disorder strength increases in Fig.~\ref{fig:ugly}, where the realization of potential disorder in each $ \sigma_\mu $ is chosen independently, and the crossover is not a continuous process from one to another, as opposed to aforementioned crossovers. Qualitatively, the changing disorder in each realization represents discrete charge switching events known to be present in semiconductor nanowires~\cite{zhang2021large}. 

In Figs.~\ref{fig:ugly}(a) and~\ref{fig:ugly}(b), where the disorder is in a weak regime with $ \sigma_\mu=0.4 $ meV, we find the topological properties are protected: The topological ZBCP with Majorana oscillations appears beyond TQPT ($ V_Z>1.02 $ meV) and no trivial ABS is present in the trivial regime ($ V_Z < 1.02$ meV ). This is a case of good ZBCP surviving in the presence of weak disorder because of topological immunity.

When the disorder increases further to an intermediate regime, $ \sigma_\mu = 1$ meV as shown in Figs.~\ref{fig:ugly}(c) and~\ref{fig:ugly}(d), we find an example of the ugly ZBCP as the trivial ABS emerging in the trivial regime while the topological MZM is still present, being protected by a bulk gap beyond TQPT. Thus, both trivial and topological zero modes may be present in the disordered system at different Zeeman energies, with the topological MZMs arising always at higher magnetic field values.

When the disorder continues to increase to the strong regime, $ \sigma_\mu = 2$ meV as shown in Figs.~\ref{fig:ugly}(e) and~\ref{fig:ugly}(f), we find not only that trivial ABS is induced by strong potential disorder below TQPT but also the topological MZM is destroyed. In this case, the bulk gap closes before the nominal TQPT and does not reopen anymore. Therefore, the concept of the topological superconductivity breaks down in such a strong-disorder regime. This is the strong-disorder-induced Anderson localized fixed point of the system where topological superconductivity is completely suppressed.

Finally, if the disorder increases to a very strong regime, $ \sigma_\mu = 5$ meV as shown in Figs.~\ref{fig:ugly}(g) and~\ref{fig:ugly}(h), we find almost nothing in the tunnel conductances from both ends of the wire. In this case, very strong disorder just destroys the whole SC-SM nanowire model and leaves nothing in the tunnel conductance spectrum. This is a scenario that should be avoided in experiments; however, we believe this is what is happening in many samples in the current stage of experiments, where very few ($\sim$1\%) of the samples produce any subgap conductance signatures indicating the dominance of strong disorder. The only solution to this problem is improved materials science leading to cleaner samples with less disorder.

\subsection{Crossover with impurity disorder}
\subsubsection{Single-impurity disorder}
\begin{figure}[htbp]
	\centering
	\includegraphics[width=3.4in]{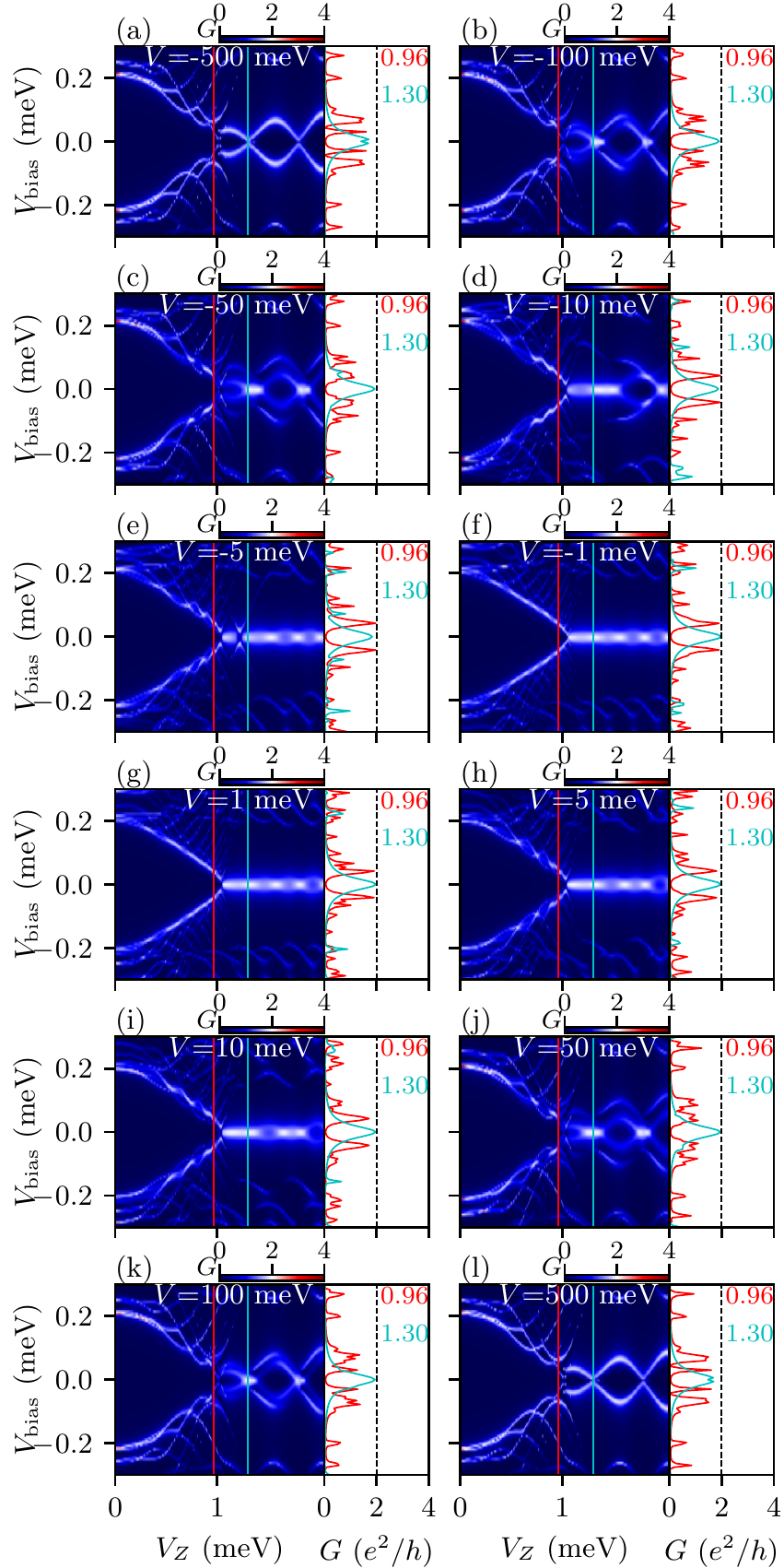}
	\caption{The tunnel conductance measured from the left end of the nanowire in the presence of single-impurity disorder located in the middle of the nanowire ($ 1.5~\mu $m). 	
	From (a) to (l), the strength of impurity $ V $  increases from -500 meV to 500 meV. For a comparison, the pristine case ($ V=0 $ ) is shown in Fig.~\ref{fig:static}(a). 
	The corresponding line cuts in the trivial regime (red) and the topological regime (cyan) are shown right to the color plot of the conductance. Refer to Fig.~\ref{fig:static} for other parameters.
	}
	\label{fig:impurity_1}
\end{figure}

In this section, we study impurity disorder in contrast to the random Gaussian potential disorder studied in the last section. 
We first present the results of single-impurity disorder in Fig.~\ref{fig:impurity_1}. The impurity is located in the middle of the wire. Therefore, we only show the conductance from the left end since the right end shows an identical conductance because of the inversion symmetry in the impurity disorder. 
Here we do not include the self-energy in the nanowire because we are only concerned about the effect of impurities, and the self-energy itself does not play a role in zero-bias states. 

From Fig.~\ref{fig:impurity_1}(a) to Fig.~\ref{fig:impurity_1}(l), the strength of impurity disorder $ V $ increases from -500 to 500 meV.
When the disorder is weak, e.g., $ \abs{V}=1 $ meV in Figs.~\ref{fig:impurity_1}(f) and~\ref{fig:impurity_1}(g), the conductances show a good ZBCP similar to a pristine wire in Figs.~\ref{fig:static}(a) and~\ref{fig:static}(b) with the topological MZM emerging beyond the TQPT and no trivial ABS in the trivial regime. On the right of the color plot of the conductance, we show the corresponding line cuts in the trivial regime (red), where no subgap states appear, and line cuts in the topological regime (cyan) showing a quantized peak at $ 2e^2/h $. Thus, the weak impurity is invisible in the zero energy states.

As disorder increases, however, we find the emergence of trivial ABSs induced by impurity disorder in the trivial regime, e.g., $ {V}=5 $ meV in Fig.~\ref{fig:impurity_1}(e). But the topological MZM is still protected by a bulk gap, although the impurity potential is more than an order of magnitude larger than the induced SC gap.

When impurity disorder increases to a very large value, e.g., $ \abs{V}=50 $ meV in Figs.~\ref{fig:impurity_1}(c) and~\ref{fig:impurity_1}(j), the topological MZM is partially destroyed by impurity disorder, where we can only find a small segment near $ V_Z\sim $ 1.3 meV that shows a quantized peak. It is, however, quite amazing that even an impurity potential almost three orders of magnitude larger than the SC gap still does not destroy the MZM completely, indicating the fundamentally robust nature of the topological immunity.

Finally, in the limits of $ \abs{V}=500 $ meV [Figs.~\ref{fig:impurity_1}(a) and~\ref{fig:impurity_1}(l)], the wire is effectively halved because the left and right partitions are separated from each other by a high potential barrier or well. Each part serves as an isolated nanowire with a halved wire length. Therefore, we see a very large Majorana oscillation beyond TQPT as a result of a shorter effective wire length. So, only in the limit of an unphysically large impurity potential, the system is divided into two effective wires, both with their own end MZMs, leading to enhanced Majorana oscillations.
\subsubsection{Double-impurity disorder}
\begin{figure}[htbp]
	\centering
	\includegraphics[width=3.4in]{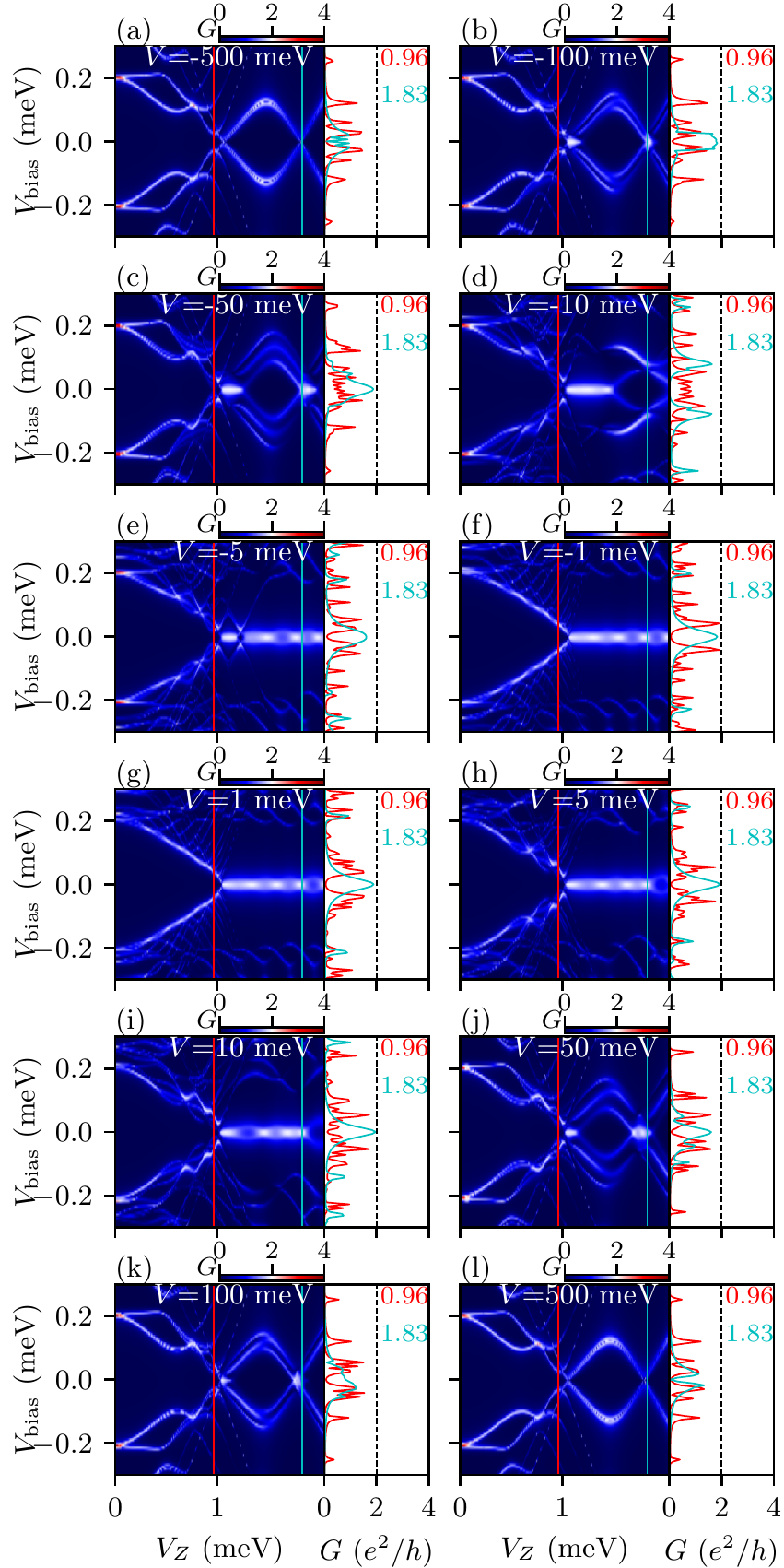}
	\caption{The tunnel conductance measured from the left end of the nanowire in the presence of double-impurity disorder located at 1 and 2 $ \mu $m of the nanowire. 
	From (a) to (l), the strength of both impurities $ V $  increases from -500 meV to 500 meV. For a comparison, the pristine case ($ V=0 $ ) is shown in Fig.~\ref{fig:static}(a). 
	The corresponding line cuts in the trivial regime (red) and the topological regime (cyan) are shown right to the color plot of the conductance. Refer to Fig.~\ref{fig:static} for other parameters.
	}
	\label{fig:impurity_2}
\end{figure}

In addition to single-impurity disorder, we also show double-impurity disorder results in Fig.~\ref{fig:impurity_2}, where the two impurities are located at $ 1 $ and $ 2~\mu $m of the nanowire. Thus, the inversion symmetry of the nanowire is also preserved and we only need to show the conductance from one end as we did before in the case of single-impurity since the conductance from the other end is identical. 

In the weak-disorder regime, e.g., $ \abs{V}=1 $ meV in Figs.~\ref{fig:impurity_2}(f) and~\ref{fig:impurity_2}(g), we find the good ZBCPs similar to the pristine case in Figs.~\ref{fig:static}(a) and~\ref{fig:static}(b). Again, the zero-energy subgap spectrum is unaffected by the presence of multiple weak impurities.

As the double-impurity disorder increases, we find subgap conductance features similar to those in the previous results in the presence of single-impurity disorder: the trivial ABS emerges as shown in Fig.~\ref{fig:impurity_2}(e) and the topological MZM is only partially destroyed by impurities as shown in Figs.~\ref{fig:impurity_2}(c) and~\ref{fig:impurity_2}(j). In the limits of $ \abs{V}=500 $ meV, we find very strong Majorana oscillations because the wire is effectively cut into three segments by the two impurities with strong disorder $ V $, leading to six effective strongly overlapping MZMs at the ends of the three segmented pieces.

The examples above of one and two spatially fixed impurities are deterministic (similar in some sense to the inhomogeneous potential situation) and do not involve any random disorder, which is what we consider now by considering a certain number of randomly spatially located impurities along the wire.  We consider a certain percentage of the sites along the nanowire to be randomly occupied by impurities of equal potential with random signs.

\subsubsection{10\%-impurity disorder}

\begin{figure}[htbp]
	\centering
	\includegraphics[width=3.4in]{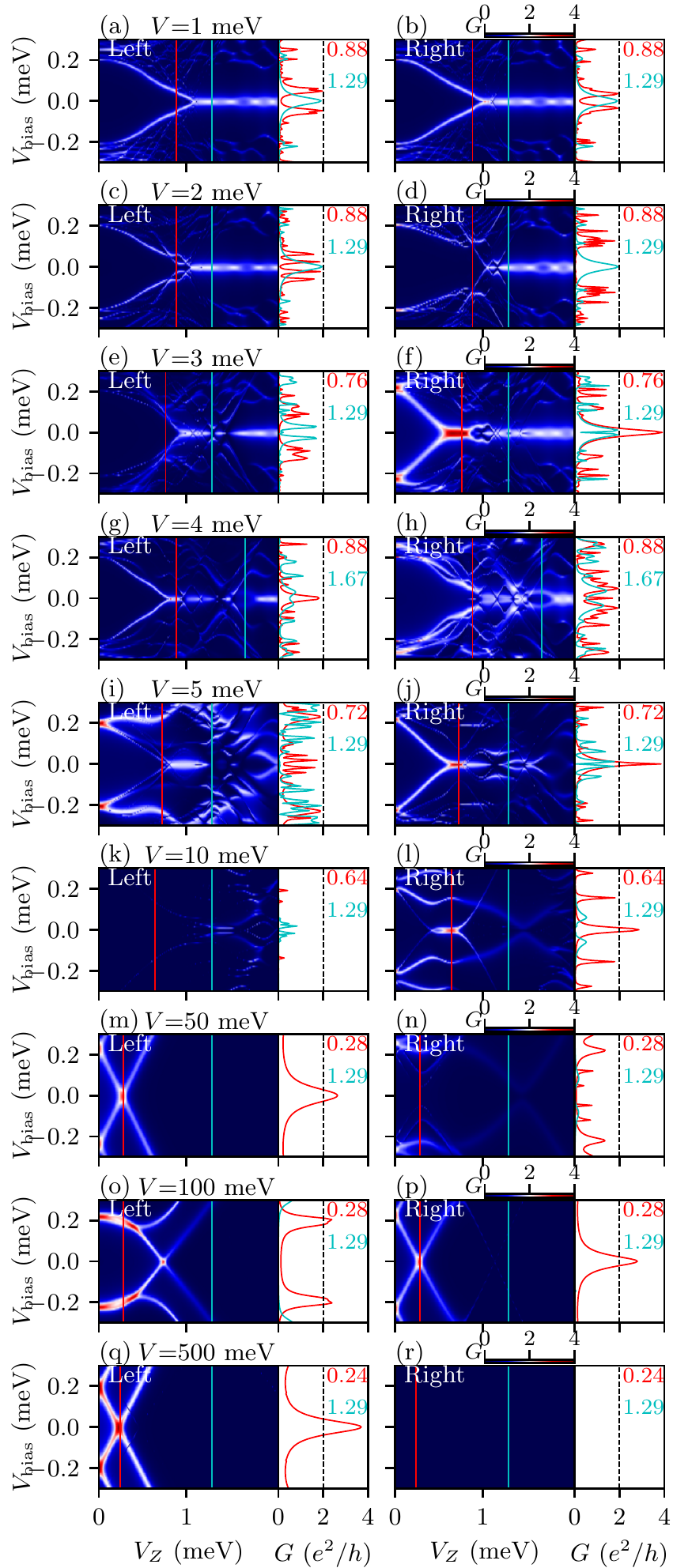}
	\caption{The tunnel conductance measured from the left end (in the left panels) and right end (in the right panels) of the nanowire in the presence of 10\%-impurity disorder.
	From the top to the bottom panels, the impurity disorder increases from a weak regime [(a),(b) with $ \sigma_\mu =$1 meV] to a very strong regime [(q),(r) with $ \sigma_\mu =$500 meV]. 
	The corresponding line cuts in the trivial regime (red) and the (nominal) topological regime (cyan) are shown right to the color plot of the conductance.
	Refer to Fig.~\ref{fig:static} for other parameters.
	}
	\label{fig:impurity_10p}
\end{figure}

In Fig.~\ref{fig:impurity_10p} we study the case where 10\% sites of the nanowire are randomly occupied by impurities. In the weak-disorder regime, e.g., $ V=1 $ meV in Figs.~\ref{fig:impurity_10p}(a) and~\ref{fig:impurity_10p}(b), the conductance resembles the good ZBCP in Figs.~\ref{fig:static}(c) and~\ref{fig:static}(d) and topological ZBCP emerges beyond the TQPT. The line cuts in the trivial regime (red) and those in the topological regime (cyan) on the right of the color plot show the absence of the in-gap fermionic state and a quantized peak of $ 2e^2/h $, respectively. The system is clearly topological and protected against the weak impurity disorder.

As impurity disorder becomes stronger, e.g., $V=2$ meV in Figs.~\ref{fig:impurity_10p}(e) and~\ref{fig:impurity_10p}(f), we find the trivial ABS emerging below the TQPT. Especially, it sometimes even creates a very stable plateau of trivial ZBCP by accident, e.g., Fig.~\ref{fig:impurity_10p}(f), which shows a peak of $4e^2/h$ in the red line cut. 

With further increase in disorder, the topological MZM in the nanowire is destroyed as shown in Figs.~\ref{fig:impurity_10p}(g), ~\ref{fig:impurity_10p}(h),~\ref{fig:impurity_10p}(i), and~\ref{fig:impurity_10p}(j). The ZBCPs shown in the tunnel conductance spectra in these figures are all trivial ABSs induced by impurity disorder. Note that, not surprisingly, the situation here is qualitatively similar to the random Gaussian potential disorder case, with the emergence of trivial ZBCPs for $V > 2$ meV, and eventually for $V> 5$  meV, all features of the topological MZMs disappear.  For very large $V$ , all subgap conductance features are suppressed.

At very large impurity disorder, we find there is no stable ZBCP at all; only crossings of the subgap states appear in the tunnel conductance spectrum as shown in Figs.~\ref{fig:impurity_10p}(m), ~\ref{fig:impurity_10p}(o),~\ref{fig:impurity_10p}(p), and~\ref{fig:impurity_10p}(q). In addition, we verify that most of the realizations of such very large disorder do not produce any conductance signal as shown in Fig.~\ref{fig:impurity_10p}(r), which are consistent with the results of Gaussian potential disorder as shown in Figs.~\ref{fig:ugly}(g) and~\ref{fig:ugly}(h). 

\subsubsection{30\%-impurity disorder}

\begin{figure}[htbp]
	\centering
	\includegraphics[width=3.4in]{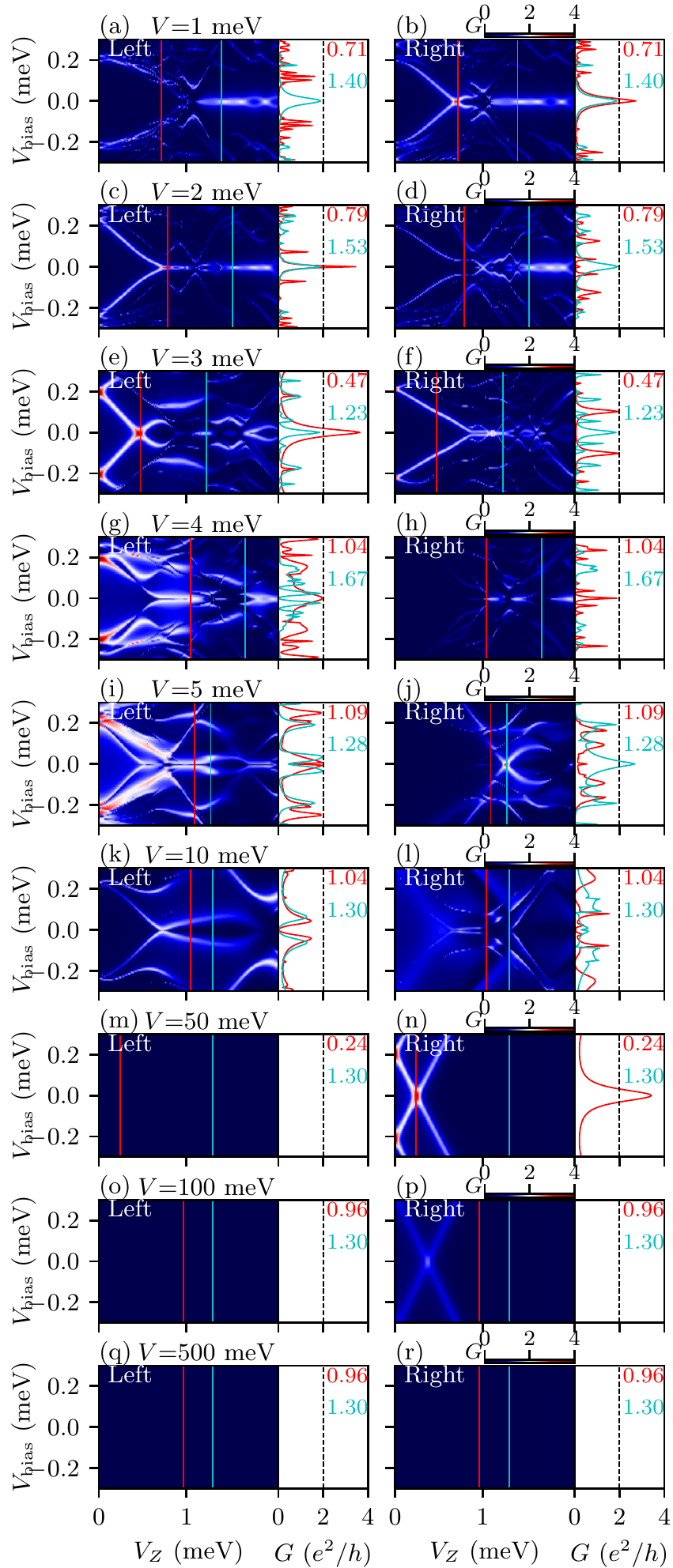}
	\caption{The tunnel conductance measured from the left end (in the left panels) and right end (in the right panels) of the nanowire in the presence of 30\%-impurity disorder.
	From the top to the bottom panels, the impurity disorder increases from a weak regime [(a),(b) with $ \sigma_\mu =$1 meV] to a very strong regime [(q),(r) with $ \sigma_\mu =$500 meV]. 
	The corresponding line cuts in the trivial regime (red) and the (nominal) topological regime (cyan) are shown right to the color plot of the conductance.
	Refer to Fig.~\ref{fig:static} for other parameters.
	}
	\label{fig:impurity_30p}
\end{figure}

We also present the tunnel conductance of the nanowire in the presence of 30\%-impurity disorder in Fig.~\ref{fig:impurity_30p}, where 30\% sites of the nanowire are randomly occupied by impurities.

In contrast to the case of 10\%-impurity disorder where $V=1$ meV still gives the good ZBCP, here $V=1$ meV already induces the ugly ZBCP with the trivial ABS emerging below the TQPT as shown in Fig.~\ref{fig:impurity_30p}(b). We find that the features of 30\%-impurity disorder are qualitatively similar to that of the 10\% disorder, with the difference being that the threshold of the disorder determining the ``weak'' and ``strong'' regime becomes smaller. Namely, in 30\%-impurity disorder,  disorder with $V=1$ meV [Figs.~\ref{fig:impurity_30p}(a) and~\ref{fig:impurity_30p}(b)] is already strong enough to create the ugly ZBCP, and disorder with $V=50$ meV is already a very large disorder that suppresses all subgap conductance features, whereas the corresponding thresholds for these two regimes in the case of 10\%-impurity disorder are $V=3$ and $V=500$ meV, respectively. Thus, both the spatial impurity distribution and the impurity potential strength are relevant in determining the topological immunity; a few randomly placed strongly coupled charged impurities are as effective in destroying the topological protection as many weakly interacting impurities.

\subsubsection{Impurity disorder maintaining charge neutrality}

\begin{figure}[htbp]
	\centering
	\includegraphics[width=3.4in]{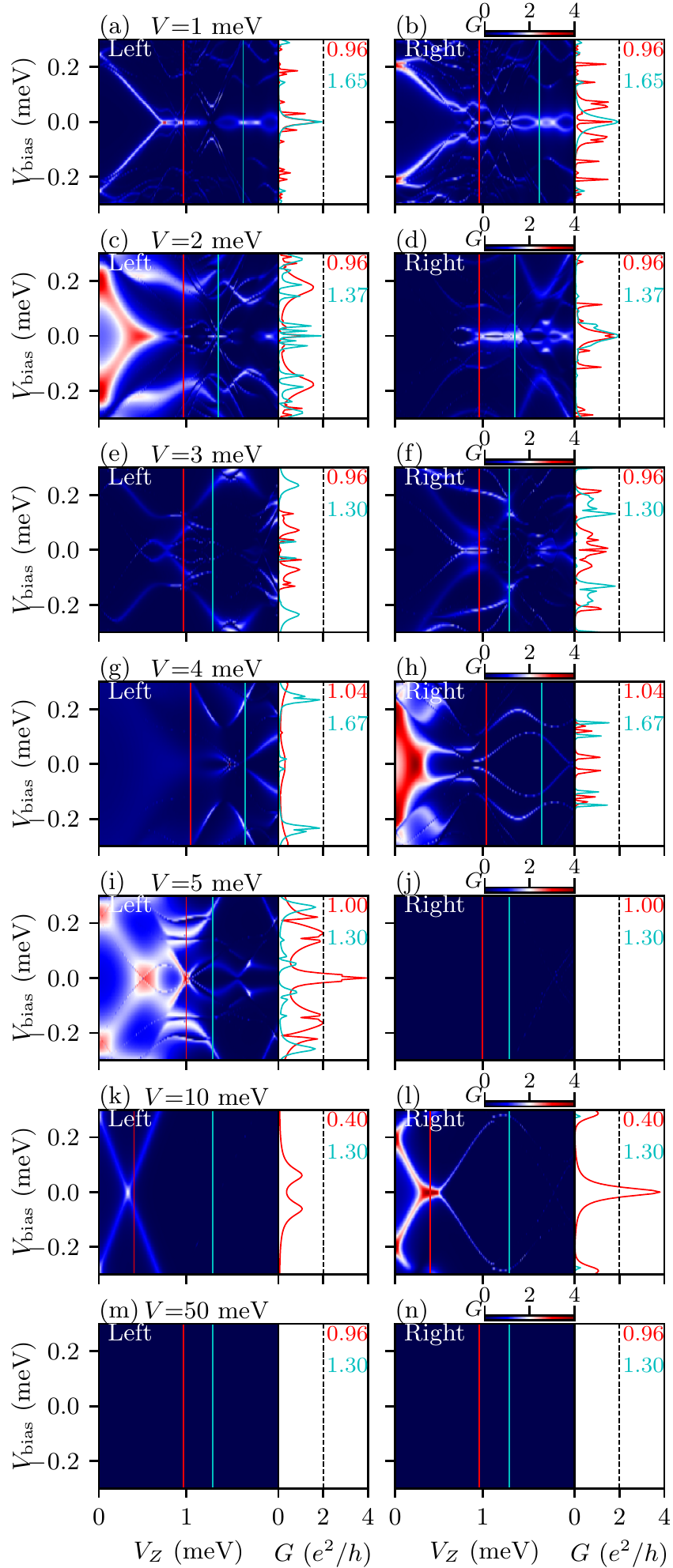}
	\caption{The tunnel conductance measured from the left end (in the left panels) and right end (in the right panels) of the nanowire in the presence of impurity disorder maintaining charge neutrality.
	From the top to the bottom panels, the impurity disorder increase from a weak regime [(a),(b) with $ \sigma_\mu =$1 meV] to a very strong regime [(q),(r) with $ \sigma_\mu =$500 meV]. 
	The corresponding line cuts in the trivial regime (red) and the (nominal) topological regime (cyan) are shown right to the color plot of the conductance.
	Refer to Fig.~\ref{fig:static} for other parameters.
	}
	\label{fig:impurity_neutral}
\end{figure}

In Fig.~\ref{fig:impurity_neutral}, we show the tunnel conductance spectrum in the presence of impurity disorder maintaining charge neutrality. Although this is a different model of disorder from the Gaussian potential disorder, we find the features of the tunnel conductance spectrum are qualitatively the same as the Gaussian potential disorder shown in Fig.~\ref{fig:ugly}. For $V=1$ meV in Figs.~\ref{fig:ugly}(a) and~\ref{fig:ugly}(b), trivial ABSs are induced manifesting as ugly ZBCPs. As disorder increases, as shown in Fig.~\ref{fig:ugly}(c)-Fig.~\ref{fig:ugly}(f), the topological MZMs are suppressed by impurity disorder. For very large disorder, as shown in Fig.~\ref{fig:ugly}(e)-Fig.~\ref{fig:ugly}(h), there are only crossings of the subgap states with no stable ZBCP, not even any ugly ZBCPs. For most cases of the large disorder, the tunnel conductance spectrum shows nothing, which is consistent with the large Gaussian potential disorder results as shown in Figs.~\ref{fig:ugly}(g) and~\ref{fig:ugly}(h).

\color{black}
\section{Conclusion}\label{sec:conclusion}
We theoretically study Majorana nanowires in superconductor-semiconductor hybrid platforms by focusing on the crossover behavior of tunnel conductance and local density of states arising from the interplay among inhomogeneous potential, random potential disorder, {and local impurity disorder}. The main qualitative finding is that the crossover behavior is dominated by trivial zero modes, some of which manifest conductance peaks with values $\sim2e^2/h $, mimicking the predicted topological MZM behavior. {A main finding of this work is that the crossover behavior is smooth, allowing no clear-cut conclusion about which physical mechanism may be the dominant one in preventing the emergence of topological superconductivity and end Majorana zero modes.}

For an inhomogeneous potential, as has already been pointed out in the theoretical literature~\cite{pan2021quantized}, the $ 2e^2/h $ trivial ZBCP may be stable as a function of system parameters even in the presence of some finite disorder since the emergent Andreev bound states, sometimes called quasi-Majoranas in this context, may remain pinned near zero energy for finite ranges of parameters (e.g., Zeeman field). The difference between these crossover trivial zero modes and their topological MZM counterparts in pristine nanowires is not necessarily the value of conductance or the existence of zero-bias peaks, which are ubiquitous in the trivial situation, even in the presence of some random disorder, but the facts that the topological MZMs manifest generic end-to-end correlations (i.e., the observed tunnel conductance peak is similar from both ends) and the topological MZMs manifest generic Majorana oscillations (i.e., the zero-bias conductance shows oscillatory behavior with increasing magnetic field), and that the topological MZMs are necessarily accompanied by the opening of a superconducting gap in the bulk.  

Another qualitative finding of the current work is that disorder typically may produce zero-bias conductance values slightly higher than $ 2e^2/h $ (as observed recently~\cite{zhang2021large,zhang2018quantizeda,nichele2017scaling}) whereas inhomogeneous smooth potential tends to mostly produce zero-bias peaks with conductance pinned at $ 2e^2/h $ or below~\cite{pan2021quantized} even in the presence of some random disorder.

{Since the realistic disorder in experiments arises from the presence of random impurities~\cite{woods2021charge}, we also study the crossover with impurity disorder and find qualitatively similar results for the Gaussian potential disorder.}

Based on our results and its qualitative agreement with much of the existing Majorana nanowire measurements in the literature, we suggest the following five criteria as the minimal conditions for any future experimental claim of the possible observation of topological Majorana zero modes in nanowires based just on the tunneling spectroscopy: (1) There must be stable (both in gate voltage and in magnetic field) zero-bias conductance peak {on a generally low subgap conductance background} with a value close to (but not above) $ 2e^2/h $ at the lowest experimental temperature; (2) the conductance value must saturate with lowering temperature and varying tunnel barrier to a value close to $ 2e^2/h $ (but not above—in fact, the expected Majorana conductance under experimental conditions should be slightly below $ 2e^2/h $); (3) similar ZBCPs with ``quantized conductance'' must be observed in tunneling from both ends of the wire (the whole tunneling spectra need not be identical from both ends, but the ZBCPs must be); (4) there should be some signatures for Majorana oscillations as the magnetic field increases above the field value where the ZBCP appears; (5) there should be some signatures of a gap reopening when the ZBCP shows up.  If these five criteria are not satisfied, chances are very high, as shown explicitly in the current work, that the system is most likely manifesting nontopological Andreev bound state induced zero modes in some complicated crossover behavior between inhomogeneous chemical potential and random disorder dominated trivial regimes (or more likely, simply crossover among distinct random disorder configurations as system parameters vary, changing the impurity states in the environment). It is important to emphasize that neither any reported experimental observation nor any of our trivial zero-mode simulations satisfy these topological criteria although both often manifest zero-bias conductance peaks with approximate $ 2e^2/h $ value over narrow fine-tuned parameter ranges. The observation of a ZBCP, even with a value close to $ 2e^2/h $, is at best a necessary condition for the existence of topological Majorana zero modes, satisfying the sufficient conditions, as discussed above, and requires more work, particularly in much cleaner disorder-free samples.

This work is supported by the Microsoft Corporation, the Laboratory for Physical Sciences, and the University of Maryland High-Performance Computing Cluster~\cite{hpcc}.

\bibliography{Paper_crossover}

\begin{thebibliography}{66}%
\makeatletter
\providecommand \@ifxundefined [1]{%
 \@ifx{#1\undefined}
}%
\providecommand \@ifnum [1]{%
 \ifnum #1\expandafter \@firstoftwo
 \else \expandafter \@secondoftwo
 \fi
}%
\providecommand \@ifx [1]{%
 \ifx #1\expandafter \@firstoftwo
 \else \expandafter \@secondoftwo
 \fi
}%
\providecommand \natexlab [1]{#1}%
\providecommand \enquote  [1]{``#1''}%
\providecommand \bibnamefont  [1]{#1}%
\providecommand \bibfnamefont [1]{#1}%
\providecommand \citenamefont [1]{#1}%
\providecommand \href@noop [0]{\@secondoftwo}%
\providecommand \href [0]{\begingroup \@sanitize@url \@href}%
\providecommand \@href[1]{\@@startlink{#1}\@@href}%
\providecommand \@@href[1]{\endgroup#1\@@endlink}%
\providecommand \@sanitize@url [0]{\catcode `\\12\catcode `\$12\catcode
  `\&12\catcode `\#12\catcode `\^12\catcode `\_12\catcode `\%12\relax}%
\providecommand \@@startlink[1]{}%
\providecommand \@@endlink[0]{}%
\providecommand \url  [0]{\begingroup\@sanitize@url \@url }%
\providecommand \@url [1]{\endgroup\@href {#1}{\urlprefix }}%
\providecommand \urlprefix  [0]{URL }%
\providecommand \Eprint [0]{\href }%
\providecommand \doibase [0]{https://doi.org/}%
\providecommand \selectlanguage [0]{\@gobble}%
\providecommand \bibinfo  [0]{\@secondoftwo}%
\providecommand \bibfield  [0]{\@secondoftwo}%
\providecommand \translation [1]{[#1]}%
\providecommand \BibitemOpen [0]{}%
\providecommand \bibitemStop [0]{}%
\providecommand \bibitemNoStop [0]{.\EOS\space}%
\providecommand \EOS [0]{\spacefactor3000\relax}%
\providecommand \BibitemShut  [1]{\csname bibitem#1\endcsname}%
\let\auto@bib@innerbib\@empty
\bibitem [{\citenamefont {Lutchyn}\ \emph {et~al.}(2010)\citenamefont
  {Lutchyn}, \citenamefont {Sau},\ and\ \citenamefont
  {Das~Sarma}}]{lutchyn2010majorana}%
  \BibitemOpen
  \bibfield  {author} {\bibinfo {author} {\bibfnamefont {R.~M.}\ \bibnamefont
  {Lutchyn}}, \bibinfo {author} {\bibfnamefont {J.~D.}\ \bibnamefont {Sau}},\
  and\ \bibinfo {author} {\bibfnamefont {S.}~\bibnamefont {Das~Sarma}},\
  }\bibfield  {title} {\bibinfo {title} {Majorana {{Fermions}} and a
  {{Topological Phase Transition}} in {{Semiconductor}}-{{Superconductor
  Heterostructures}}},\ }\href {https://doi.org/10.1103/PhysRevLett.105.077001}
  {\bibfield  {journal} {\bibinfo  {journal} {Phys. Rev. Lett.}\ }\textbf
  {\bibinfo {volume} {105}},\ \bibinfo {pages} {077001} (\bibinfo {year}
  {2010})}\BibitemShut {NoStop}%
\bibitem [{\citenamefont {Sau}\ \emph {et~al.}(2010{\natexlab{a}})\citenamefont
  {Sau}, \citenamefont {Lutchyn}, \citenamefont {Tewari},\ and\ \citenamefont
  {Das~Sarma}}]{sau2010generic}%
  \BibitemOpen
  \bibfield  {author} {\bibinfo {author} {\bibfnamefont {J.~D.}\ \bibnamefont
  {Sau}}, \bibinfo {author} {\bibfnamefont {R.~M.}\ \bibnamefont {Lutchyn}},
  \bibinfo {author} {\bibfnamefont {S.}~\bibnamefont {Tewari}},\ and\ \bibinfo
  {author} {\bibfnamefont {S.}~\bibnamefont {Das~Sarma}},\ }\bibfield  {title}
  {\bibinfo {title} {Generic {{New Platform}} for {{Topological Quantum
  Computation Using Semiconductor Heterostructures}}},\ }\href
  {https://doi.org/10.1103/PhysRevLett.104.040502} {\bibfield  {journal}
  {\bibinfo  {journal} {Phys. Rev. Lett.}\ }\textbf {\bibinfo {volume} {104}},\
  \bibinfo {pages} {040502} (\bibinfo {year} {2010}{\natexlab{a}})}\BibitemShut
  {NoStop}%
\bibitem [{\citenamefont {Sau}\ \emph {et~al.}(2010{\natexlab{b}})\citenamefont
  {Sau}, \citenamefont {Tewari}, \citenamefont {Lutchyn}, \citenamefont
  {Stanescu},\ and\ \citenamefont {Das~Sarma}}]{sau2010nonabelian}%
  \BibitemOpen
  \bibfield  {author} {\bibinfo {author} {\bibfnamefont {J.~D.}\ \bibnamefont
  {Sau}}, \bibinfo {author} {\bibfnamefont {S.}~\bibnamefont {Tewari}},
  \bibinfo {author} {\bibfnamefont {R.~M.}\ \bibnamefont {Lutchyn}}, \bibinfo
  {author} {\bibfnamefont {T.~D.}\ \bibnamefont {Stanescu}},\ and\ \bibinfo
  {author} {\bibfnamefont {S.}~\bibnamefont {Das~Sarma}},\ }\bibfield  {title}
  {\bibinfo {title} {Non-{{Abelian}} quantum order in spin-orbit-coupled
  semiconductors: {{Search}} for topological {{Majorana}} particles in
  solid-state systems},\ }\href {https://doi.org/10.1103/PhysRevB.82.214509}
  {\bibfield  {journal} {\bibinfo  {journal} {Phys. Rev. B}\ }\textbf {\bibinfo
  {volume} {82}},\ \bibinfo {pages} {214509} (\bibinfo {year}
  {2010}{\natexlab{b}})}\BibitemShut {NoStop}%
\bibitem [{\citenamefont {Oreg}\ \emph {et~al.}(2010)\citenamefont {Oreg},
  \citenamefont {Refael},\ and\ \citenamefont {{von Oppen}}}]{oreg2010helical}%
  \BibitemOpen
  \bibfield  {author} {\bibinfo {author} {\bibfnamefont {Y.}~\bibnamefont
  {Oreg}}, \bibinfo {author} {\bibfnamefont {G.}~\bibnamefont {Refael}},\ and\
  \bibinfo {author} {\bibfnamefont {F.}~\bibnamefont {{von Oppen}}},\
  }\bibfield  {title} {\bibinfo {title} {Helical {{Liquids}} and {{Majorana
  Bound States}} in {{Quantum Wires}}},\ }\href
  {https://doi.org/10.1103/PhysRevLett.105.177002} {\bibfield  {journal}
  {\bibinfo  {journal} {Phys. Rev. Lett.}\ }\textbf {\bibinfo {volume} {105}},\
  \bibinfo {pages} {177002} (\bibinfo {year} {2010})}\BibitemShut {NoStop}%
\bibitem [{\citenamefont {Mourik}\ \emph {et~al.}(2012)\citenamefont {Mourik},
  \citenamefont {Zuo}, \citenamefont {Frolov}, \citenamefont {Plissard},
  \citenamefont {Bakkers},\ and\ \citenamefont
  {Kouwenhoven}}]{mourik2012signatures}%
  \BibitemOpen
  \bibfield  {author} {\bibinfo {author} {\bibfnamefont {V.}~\bibnamefont
  {Mourik}}, \bibinfo {author} {\bibfnamefont {K.}~\bibnamefont {Zuo}},
  \bibinfo {author} {\bibfnamefont {S.~M.}\ \bibnamefont {Frolov}}, \bibinfo
  {author} {\bibfnamefont {S.~R.}\ \bibnamefont {Plissard}}, \bibinfo {author}
  {\bibfnamefont {E.~P. A.~M.}\ \bibnamefont {Bakkers}},\ and\ \bibinfo
  {author} {\bibfnamefont {L.~P.}\ \bibnamefont {Kouwenhoven}},\ }\bibfield
  {title} {\bibinfo {title} {Signatures of {{Majorana Fermions}} in {{Hybrid
  Superconductor}}-{{Semiconductor Nanowire Devices}}},\ }\href
  {https://doi.org/10.1126/science.1222360} {\bibfield  {journal} {\bibinfo
  {journal} {Science}\ }\textbf {\bibinfo {volume} {336}},\ \bibinfo {pages}
  {1003} (\bibinfo {year} {2012})}\BibitemShut {NoStop}%
\bibitem [{\citenamefont {Das}\ \emph {et~al.}(2012)\citenamefont {Das},
  \citenamefont {Ronen}, \citenamefont {Most}, \citenamefont {Oreg},
  \citenamefont {Heiblum},\ and\ \citenamefont {Shtrikman}}]{das2012zerobias}%
  \BibitemOpen
  \bibfield  {author} {\bibinfo {author} {\bibfnamefont {A.}~\bibnamefont
  {Das}}, \bibinfo {author} {\bibfnamefont {Y.}~\bibnamefont {Ronen}}, \bibinfo
  {author} {\bibfnamefont {Y.}~\bibnamefont {Most}}, \bibinfo {author}
  {\bibfnamefont {Y.}~\bibnamefont {Oreg}}, \bibinfo {author} {\bibfnamefont
  {M.}~\bibnamefont {Heiblum}},\ and\ \bibinfo {author} {\bibfnamefont
  {H.}~\bibnamefont {Shtrikman}},\ }\bibfield  {title} {\bibinfo {title}
  {Zero-bias peaks and splitting in an {{Al}}\textendash{{InAs}} nanowire
  topological superconductor as a signature of {{Majorana}} fermions},\ }\href
  {https://www.nature.com/articles/nphys2479} {\bibfield  {journal} {\bibinfo
  {journal} {Nature Physics}\ }\textbf {\bibinfo {volume} {8}},\ \bibinfo
  {pages} {887} (\bibinfo {year} {2012})}\BibitemShut {NoStop}%
\bibitem [{\citenamefont {Deng}\ \emph {et~al.}(2012)\citenamefont {Deng},
  \citenamefont {Yu}, \citenamefont {Huang}, \citenamefont {Larsson},
  \citenamefont {Caroff},\ and\ \citenamefont {Xu}}]{deng2012anomalous}%
  \BibitemOpen
  \bibfield  {author} {\bibinfo {author} {\bibfnamefont {M.~T.}\ \bibnamefont
  {Deng}}, \bibinfo {author} {\bibfnamefont {C.~L.}\ \bibnamefont {Yu}},
  \bibinfo {author} {\bibfnamefont {G.~Y.}\ \bibnamefont {Huang}}, \bibinfo
  {author} {\bibfnamefont {M.}~\bibnamefont {Larsson}}, \bibinfo {author}
  {\bibfnamefont {P.}~\bibnamefont {Caroff}},\ and\ \bibinfo {author}
  {\bibfnamefont {H.~Q.}\ \bibnamefont {Xu}},\ }\bibfield  {title} {\bibinfo
  {title} {Anomalous {{Zero}}-{{Bias Conductance Peak}} in a
  {{Nb}}\textendash{{InSb Nanowire}}\textendash{{Nb Hybrid Device}}},\ }\href
  {https://doi.org/10.1021/nl303758w} {\bibfield  {journal} {\bibinfo
  {journal} {Nano Letters}\ }\textbf {\bibinfo {volume} {12}},\ \bibinfo
  {pages} {6414} (\bibinfo {year} {2012})}\BibitemShut {NoStop}%
\bibitem [{\citenamefont {Churchill}\ \emph {et~al.}(2013)\citenamefont
  {Churchill}, \citenamefont {Fatemi}, \citenamefont {{Grove-Rasmussen}},
  \citenamefont {Deng}, \citenamefont {Caroff}, \citenamefont {Xu},\ and\
  \citenamefont {Marcus}}]{churchill2013superconductornanowire}%
  \BibitemOpen
  \bibfield  {author} {\bibinfo {author} {\bibfnamefont {H.~O.~H.}\
  \bibnamefont {Churchill}}, \bibinfo {author} {\bibfnamefont {V.}~\bibnamefont
  {Fatemi}}, \bibinfo {author} {\bibfnamefont {K.}~\bibnamefont
  {{Grove-Rasmussen}}}, \bibinfo {author} {\bibfnamefont {M.~T.}\ \bibnamefont
  {Deng}}, \bibinfo {author} {\bibfnamefont {P.}~\bibnamefont {Caroff}},
  \bibinfo {author} {\bibfnamefont {H.~Q.}\ \bibnamefont {Xu}},\ and\ \bibinfo
  {author} {\bibfnamefont {C.~M.}\ \bibnamefont {Marcus}},\ }\bibfield  {title}
  {\bibinfo {title} {Superconductor-nanowire devices from tunneling to the
  multichannel regime: {{Zero}}-bias oscillations and magnetoconductance
  crossover},\ }\href {https://doi.org/10.1103/PhysRevB.87.241401} {\bibfield
  {journal} {\bibinfo  {journal} {Phys. Rev. B}\ }\textbf {\bibinfo {volume}
  {87}},\ \bibinfo {pages} {241401} (\bibinfo {year} {2013})}\BibitemShut
  {NoStop}%
\bibitem [{\citenamefont {Finck}\ \emph {et~al.}(2013)\citenamefont {Finck},
  \citenamefont {Van~Harlingen}, \citenamefont {Mohseni}, \citenamefont
  {Jung},\ and\ \citenamefont {Li}}]{finck2013anomalous}%
  \BibitemOpen
  \bibfield  {author} {\bibinfo {author} {\bibfnamefont {A.~D.~K.}\
  \bibnamefont {Finck}}, \bibinfo {author} {\bibfnamefont {D.~J.}\ \bibnamefont
  {Van~Harlingen}}, \bibinfo {author} {\bibfnamefont {P.~K.}\ \bibnamefont
  {Mohseni}}, \bibinfo {author} {\bibfnamefont {K.}~\bibnamefont {Jung}},\ and\
  \bibinfo {author} {\bibfnamefont {X.}~\bibnamefont {Li}},\ }\bibfield
  {title} {\bibinfo {title} {Anomalous {{Modulation}} of a {{Zero}}-{{Bias
  Peak}} in a {{Hybrid Nanowire}}-{{Superconductor Device}}},\ }\href
  {https://doi.org/10.1103/PhysRevLett.110.126406} {\bibfield  {journal}
  {\bibinfo  {journal} {Phys. Rev. Lett.}\ }\textbf {\bibinfo {volume} {110}},\
  \bibinfo {pages} {126406} (\bibinfo {year} {2013})}\BibitemShut {NoStop}%
\bibitem [{\citenamefont {Deng}\ \emph {et~al.}(2016)\citenamefont {Deng},
  \citenamefont {Vaitiek{\.e}nas}, \citenamefont {Hansen}, \citenamefont
  {Danon}, \citenamefont {Leijnse}, \citenamefont {Flensberg}, \citenamefont
  {Nyg{\aa}rd}, \citenamefont {Krogstrup},\ and\ \citenamefont
  {Marcus}}]{deng2016majorana}%
  \BibitemOpen
  \bibfield  {author} {\bibinfo {author} {\bibfnamefont {M.}~\bibnamefont
  {Deng}}, \bibinfo {author} {\bibfnamefont {S.}~\bibnamefont
  {Vaitiek{\.e}nas}}, \bibinfo {author} {\bibfnamefont {E.~B.}\ \bibnamefont
  {Hansen}}, \bibinfo {author} {\bibfnamefont {J.}~\bibnamefont {Danon}},
  \bibinfo {author} {\bibfnamefont {M.}~\bibnamefont {Leijnse}}, \bibinfo
  {author} {\bibfnamefont {K.}~\bibnamefont {Flensberg}}, \bibinfo {author}
  {\bibfnamefont {J.}~\bibnamefont {Nyg{\aa}rd}}, \bibinfo {author}
  {\bibfnamefont {P.}~\bibnamefont {Krogstrup}},\ and\ \bibinfo {author}
  {\bibfnamefont {C.~M.}\ \bibnamefont {Marcus}},\ }\bibfield  {title}
  {\bibinfo {title} {Majorana bound state in a coupled quantum-dot
  hybrid-nanowire system},\ }\href
  {http://science.sciencemag.org/content/354/6319/1557.abstract?casa_token=iM4dNJjIvEwAAAAA:EZj32k4K_Wj6ZicTGaO0AGQQlMuRZZ8wypAaqXRZyZgY66JCXd0w_rc_dC1Y2SO26oa2JO66xzFy}
  {\bibfield  {journal} {\bibinfo  {journal} {Science}\ }\textbf {\bibinfo
  {volume} {354}},\ \bibinfo {pages} {1557} (\bibinfo {year}
  {2016})}\BibitemShut {NoStop}%
\bibitem [{\citenamefont {Nichele}\ \emph {et~al.}(2017)\citenamefont
  {Nichele}, \citenamefont {Drachmann}, \citenamefont {Whiticar}, \citenamefont
  {O'Farrell}, \citenamefont {Suominen}, \citenamefont {Fornieri},
  \citenamefont {Wang}, \citenamefont {Gardner}, \citenamefont {Thomas},
  \citenamefont {Hatke}, \citenamefont {Krogstrup}, \citenamefont {Manfra},
  \citenamefont {Flensberg},\ and\ \citenamefont
  {Marcus}}]{nichele2017scaling}%
  \BibitemOpen
  \bibfield  {author} {\bibinfo {author} {\bibfnamefont {F.}~\bibnamefont
  {Nichele}}, \bibinfo {author} {\bibfnamefont {A.~C.~C.}\ \bibnamefont
  {Drachmann}}, \bibinfo {author} {\bibfnamefont {A.~M.}\ \bibnamefont
  {Whiticar}}, \bibinfo {author} {\bibfnamefont {E.~C.~T.}\ \bibnamefont
  {O'Farrell}}, \bibinfo {author} {\bibfnamefont {H.~J.}\ \bibnamefont
  {Suominen}}, \bibinfo {author} {\bibfnamefont {A.}~\bibnamefont {Fornieri}},
  \bibinfo {author} {\bibfnamefont {T.}~\bibnamefont {Wang}}, \bibinfo {author}
  {\bibfnamefont {G.~C.}\ \bibnamefont {Gardner}}, \bibinfo {author}
  {\bibfnamefont {C.}~\bibnamefont {Thomas}}, \bibinfo {author} {\bibfnamefont
  {A.~T.}\ \bibnamefont {Hatke}}, \bibinfo {author} {\bibfnamefont
  {P.}~\bibnamefont {Krogstrup}}, \bibinfo {author} {\bibfnamefont {M.~J.}\
  \bibnamefont {Manfra}}, \bibinfo {author} {\bibfnamefont {K.}~\bibnamefont
  {Flensberg}},\ and\ \bibinfo {author} {\bibfnamefont {C.~M.}\ \bibnamefont
  {Marcus}},\ }\bibfield  {title} {\bibinfo {title} {Scaling of {{Majorana
  Zero}}-{{Bias Conductance Peaks}}},\ }\href
  {https://doi.org/10.1103/PhysRevLett.119.136803} {\bibfield  {journal}
  {\bibinfo  {journal} {Phys. Rev. Lett.}\ }\textbf {\bibinfo {volume} {119}},\
  \bibinfo {pages} {136803} (\bibinfo {year} {2017})}\BibitemShut {NoStop}%
\bibitem [{\citenamefont {Zhang}\ \emph {et~al.}(2017)\citenamefont {Zhang},
  \citenamefont {G{\"u}l}, \citenamefont {{Conesa-Boj}}, \citenamefont {Nowak},
  \citenamefont {Wimmer}, \citenamefont {Zuo}, \citenamefont {Mourik},
  \citenamefont {{de Vries}}, \citenamefont {{van Veen}}, \citenamefont {{de
  Moor}}, \citenamefont {Bommer}, \citenamefont {{van Woerkom}}, \citenamefont
  {Car}, \citenamefont {Plissard}, \citenamefont {Bakkers}, \citenamefont
  {{Quintero-P{\'e}rez}}, \citenamefont {Cassidy}, \citenamefont {Koelling},
  \citenamefont {Goswami}, \citenamefont {Watanabe}, \citenamefont
  {Taniguchi},\ and\ \citenamefont {Kouwenhoven}}]{zhang2017ballistic}%
  \BibitemOpen
  \bibfield  {author} {\bibinfo {author} {\bibfnamefont {H.}~\bibnamefont
  {Zhang}}, \bibinfo {author} {\bibfnamefont {{\"O}.}~\bibnamefont {G{\"u}l}},
  \bibinfo {author} {\bibfnamefont {S.}~\bibnamefont {{Conesa-Boj}}}, \bibinfo
  {author} {\bibfnamefont {M.~P.}\ \bibnamefont {Nowak}}, \bibinfo {author}
  {\bibfnamefont {M.}~\bibnamefont {Wimmer}}, \bibinfo {author} {\bibfnamefont
  {K.}~\bibnamefont {Zuo}}, \bibinfo {author} {\bibfnamefont {V.}~\bibnamefont
  {Mourik}}, \bibinfo {author} {\bibfnamefont {F.~K.}\ \bibnamefont {{de
  Vries}}}, \bibinfo {author} {\bibfnamefont {J.}~\bibnamefont {{van Veen}}},
  \bibinfo {author} {\bibfnamefont {M.~W.~A.}\ \bibnamefont {{de Moor}}},
  \bibinfo {author} {\bibfnamefont {J.~D.~S.}\ \bibnamefont {Bommer}}, \bibinfo
  {author} {\bibfnamefont {D.~J.}\ \bibnamefont {{van Woerkom}}}, \bibinfo
  {author} {\bibfnamefont {D.}~\bibnamefont {Car}}, \bibinfo {author}
  {\bibfnamefont {S.~R.}\ \bibnamefont {Plissard}}, \bibinfo {author}
  {\bibfnamefont {E.~P. A.~M.}\ \bibnamefont {Bakkers}}, \bibinfo {author}
  {\bibfnamefont {M.}~\bibnamefont {{Quintero-P{\'e}rez}}}, \bibinfo {author}
  {\bibfnamefont {M.~C.}\ \bibnamefont {Cassidy}}, \bibinfo {author}
  {\bibfnamefont {S.}~\bibnamefont {Koelling}}, \bibinfo {author}
  {\bibfnamefont {S.}~\bibnamefont {Goswami}}, \bibinfo {author} {\bibfnamefont
  {K.}~\bibnamefont {Watanabe}}, \bibinfo {author} {\bibfnamefont
  {T.}~\bibnamefont {Taniguchi}},\ and\ \bibinfo {author} {\bibfnamefont
  {L.~P.}\ \bibnamefont {Kouwenhoven}},\ }\bibfield  {title} {\bibinfo {title}
  {Ballistic superconductivity in semiconductor nanowires},\ }\href
  {https://doi.org/10.1038/ncomms16025} {\bibfield  {journal} {\bibinfo
  {journal} {Nature Communications}\ }\textbf {\bibinfo {volume} {8}},\
  \bibinfo {pages} {16025} (\bibinfo {year} {2017})}\BibitemShut {NoStop}%
\bibitem [{\citenamefont {Vaitiek{\.e}nas}\ \emph {et~al.}(2018)\citenamefont
  {Vaitiek{\.e}nas}, \citenamefont {Deng}, \citenamefont {Nyg{\aa}rd},
  \citenamefont {Krogstrup},\ and\ \citenamefont
  {Marcus}}]{vaitiekenas2018effective}%
  \BibitemOpen
  \bibfield  {author} {\bibinfo {author} {\bibfnamefont {S.}~\bibnamefont
  {Vaitiek{\.e}nas}}, \bibinfo {author} {\bibfnamefont {M.-T.}\ \bibnamefont
  {Deng}}, \bibinfo {author} {\bibfnamefont {J.}~\bibnamefont {Nyg{\aa}rd}},
  \bibinfo {author} {\bibfnamefont {P.}~\bibnamefont {Krogstrup}},\ and\
  \bibinfo {author} {\bibfnamefont {C.~M.}\ \bibnamefont {Marcus}},\ }\bibfield
   {title} {\bibinfo {title} {Effective g {{Factor}} of {{Subgap States}} in
  {{Hybrid Nanowires}}},\ }\href
  {https://doi.org/10.1103/PhysRevLett.121.037703} {\bibfield  {journal}
  {\bibinfo  {journal} {Phys. Rev. Lett.}\ }\textbf {\bibinfo {volume} {121}},\
  \bibinfo {pages} {037703} (\bibinfo {year} {2018})}\BibitemShut {NoStop}%
\bibitem [{\citenamefont {de~Moor}\ \emph {et~al.}(2018)\citenamefont
  {de~Moor}, \citenamefont {Bommer}, \citenamefont {Xu}, \citenamefont
  {Winkler}, \citenamefont {Antipov}, \citenamefont {Bargerbos}, \citenamefont
  {Wang}, \citenamefont {van Loo}, \citenamefont {het Veld}, \citenamefont
  {Gazibegovic}, \citenamefont {Car}, \citenamefont {Logan}, \citenamefont
  {Pendharkar}, \citenamefont {Lee}, \citenamefont {Bakkers}, \citenamefont
  {Palmstr{\o}m}, \citenamefont {Lutchyn}, \citenamefont {Kouwenhoven},\ and\
  \citenamefont {Zhang}}]{moor2018electric}%
  \BibitemOpen
  \bibfield  {author} {\bibinfo {author} {\bibfnamefont {M.~W.~A.}\
  \bibnamefont {de~Moor}}, \bibinfo {author} {\bibfnamefont {J.~D.~S.}\
  \bibnamefont {Bommer}}, \bibinfo {author} {\bibfnamefont {D.}~\bibnamefont
  {Xu}}, \bibinfo {author} {\bibfnamefont {G.~W.}\ \bibnamefont {Winkler}},
  \bibinfo {author} {\bibfnamefont {A.~E.}\ \bibnamefont {Antipov}}, \bibinfo
  {author} {\bibfnamefont {A.}~\bibnamefont {Bargerbos}}, \bibinfo {author}
  {\bibfnamefont {G.}~\bibnamefont {Wang}}, \bibinfo {author} {\bibfnamefont
  {N.}~\bibnamefont {van Loo}}, \bibinfo {author} {\bibfnamefont {R.~L. M.~O.}\
  \bibnamefont {het Veld}}, \bibinfo {author} {\bibfnamefont {S.}~\bibnamefont
  {Gazibegovic}}, \bibinfo {author} {\bibfnamefont {D.}~\bibnamefont {Car}},
  \bibinfo {author} {\bibfnamefont {J.~A.}\ \bibnamefont {Logan}}, \bibinfo
  {author} {\bibfnamefont {M.}~\bibnamefont {Pendharkar}}, \bibinfo {author}
  {\bibfnamefont {J.~S.}\ \bibnamefont {Lee}}, \bibinfo {author} {\bibfnamefont
  {E.~P. A.~M.}\ \bibnamefont {Bakkers}}, \bibinfo {author} {\bibfnamefont
  {C.~J.}\ \bibnamefont {Palmstr{\o}m}}, \bibinfo {author} {\bibfnamefont
  {R.~M.}\ \bibnamefont {Lutchyn}}, \bibinfo {author} {\bibfnamefont {L.~P.}\
  \bibnamefont {Kouwenhoven}},\ and\ \bibinfo {author} {\bibfnamefont
  {H.}~\bibnamefont {Zhang}},\ }\bibfield  {title} {\bibinfo {title} {Electric
  field tunable superconductor-semiconductor coupling in {{Majorana}}
  nanowires},\ }\href {https://doi.org/10.1088/1367-2630/aae61d} {\bibfield
  {journal} {\bibinfo  {journal} {New J. Phys.}\ }\textbf {\bibinfo {volume}
  {20}},\ \bibinfo {pages} {103049} (\bibinfo {year} {2018})}\BibitemShut
  {NoStop}%
\bibitem [{\citenamefont {Zhang}\ \emph {et~al.}(2021)\citenamefont {Zhang},
  \citenamefont {{de Moor}}, \citenamefont {Bommer}, \citenamefont {Xu},
  \citenamefont {Wang}, \citenamefont {{van Loo}}, \citenamefont {Liu},
  \citenamefont {Gazibegovic}, \citenamefont {Logan}, \citenamefont {Car},
  \citenamefont {het Veld}, \citenamefont {{van Veldhoven}}, \citenamefont
  {Koelling}, \citenamefont {Verheijen}, \citenamefont {Pendharkar},
  \citenamefont {Pennachio}, \citenamefont {Shojaei}, \citenamefont {Lee},
  \citenamefont {Palmstr{\o}m}, \citenamefont {Bakkers}, \citenamefont
  {Sarma},\ and\ \citenamefont {Kouwenhoven}}]{zhang2021large}%
  \BibitemOpen
  \bibfield  {author} {\bibinfo {author} {\bibfnamefont {H.}~\bibnamefont
  {Zhang}}, \bibinfo {author} {\bibfnamefont {M.~W.~A.}\ \bibnamefont {{de
  Moor}}}, \bibinfo {author} {\bibfnamefont {J.~D.~S.}\ \bibnamefont {Bommer}},
  \bibinfo {author} {\bibfnamefont {D.}~\bibnamefont {Xu}}, \bibinfo {author}
  {\bibfnamefont {G.}~\bibnamefont {Wang}}, \bibinfo {author} {\bibfnamefont
  {N.}~\bibnamefont {{van Loo}}}, \bibinfo {author} {\bibfnamefont {C.-X.}\
  \bibnamefont {Liu}}, \bibinfo {author} {\bibfnamefont {S.}~\bibnamefont
  {Gazibegovic}}, \bibinfo {author} {\bibfnamefont {J.~A.}\ \bibnamefont
  {Logan}}, \bibinfo {author} {\bibfnamefont {D.}~\bibnamefont {Car}}, \bibinfo
  {author} {\bibfnamefont {R.~L. M.~O.}\ \bibnamefont {het Veld}}, \bibinfo
  {author} {\bibfnamefont {P.~J.}\ \bibnamefont {{van Veldhoven}}}, \bibinfo
  {author} {\bibfnamefont {S.}~\bibnamefont {Koelling}}, \bibinfo {author}
  {\bibfnamefont {M.~A.}\ \bibnamefont {Verheijen}}, \bibinfo {author}
  {\bibfnamefont {M.}~\bibnamefont {Pendharkar}}, \bibinfo {author}
  {\bibfnamefont {D.~J.}\ \bibnamefont {Pennachio}}, \bibinfo {author}
  {\bibfnamefont {B.}~\bibnamefont {Shojaei}}, \bibinfo {author} {\bibfnamefont
  {J.~S.}\ \bibnamefont {Lee}}, \bibinfo {author} {\bibfnamefont {C.~J.}\
  \bibnamefont {Palmstr{\o}m}}, \bibinfo {author} {\bibfnamefont {E.~P. A.~M.}\
  \bibnamefont {Bakkers}}, \bibinfo {author} {\bibfnamefont {S.~D.}\
  \bibnamefont {Sarma}},\ and\ \bibinfo {author} {\bibfnamefont {L.~P.}\
  \bibnamefont {Kouwenhoven}},\ }\bibfield  {title} {\bibinfo {title} {Large
  zero-bias peaks in {{InSb}}-{{Al}} hybrid semiconductor-superconductor
  nanowire devices},\ }\href {http://arxiv.org/abs/2101.11456} {\bibfield
  {journal} {\bibinfo  {journal} {arXiv:2101.11456}\ } (\bibinfo {year}
  {2021})}\BibitemShut {NoStop}%
\bibitem [{\citenamefont {Zhang}\ \emph {et~al.}(2018)\citenamefont {Zhang},
  \citenamefont {Liu}, \citenamefont {Gazibegovic}, \citenamefont {Xu},
  \citenamefont {Logan}, \citenamefont {Wang}, \citenamefont {{van Loo}},
  \citenamefont {Bommer}, \citenamefont {{de Moor}}, \citenamefont {Car},
  \citenamefont {{Op het Veld}}, \citenamefont {{van Veldhoven}}, \citenamefont
  {Koelling}, \citenamefont {Verheijen}, \citenamefont {Pendharkar},
  \citenamefont {Pennachio}, \citenamefont {Shojaei}, \citenamefont {Lee},
  \citenamefont {Palmstr{\o}m}, \citenamefont {Bakkers}, \citenamefont
  {Sarma},\ and\ \citenamefont {Kouwenhoven}}]{zhang2018quantizeda}%
  \BibitemOpen
  \bibfield  {author} {\bibinfo {author} {\bibfnamefont {H.}~\bibnamefont
  {Zhang}}, \bibinfo {author} {\bibfnamefont {C.-X.}\ \bibnamefont {Liu}},
  \bibinfo {author} {\bibfnamefont {S.}~\bibnamefont {Gazibegovic}}, \bibinfo
  {author} {\bibfnamefont {D.}~\bibnamefont {Xu}}, \bibinfo {author}
  {\bibfnamefont {J.~A.}\ \bibnamefont {Logan}}, \bibinfo {author}
  {\bibfnamefont {G.}~\bibnamefont {Wang}}, \bibinfo {author} {\bibfnamefont
  {N.}~\bibnamefont {{van Loo}}}, \bibinfo {author} {\bibfnamefont {J.~D.~S.}\
  \bibnamefont {Bommer}}, \bibinfo {author} {\bibfnamefont {M.~W.~A.}\
  \bibnamefont {{de Moor}}}, \bibinfo {author} {\bibfnamefont {D.}~\bibnamefont
  {Car}}, \bibinfo {author} {\bibfnamefont {R.~L.~M.}\ \bibnamefont {{Op het
  Veld}}}, \bibinfo {author} {\bibfnamefont {P.~J.}\ \bibnamefont {{van
  Veldhoven}}}, \bibinfo {author} {\bibfnamefont {S.}~\bibnamefont {Koelling}},
  \bibinfo {author} {\bibfnamefont {M.~A.}\ \bibnamefont {Verheijen}}, \bibinfo
  {author} {\bibfnamefont {M.}~\bibnamefont {Pendharkar}}, \bibinfo {author}
  {\bibfnamefont {D.~J.}\ \bibnamefont {Pennachio}}, \bibinfo {author}
  {\bibfnamefont {B.}~\bibnamefont {Shojaei}}, \bibinfo {author} {\bibfnamefont
  {J.~S.}\ \bibnamefont {Lee}}, \bibinfo {author} {\bibfnamefont {C.~J.}\
  \bibnamefont {Palmstr{\o}m}}, \bibinfo {author} {\bibfnamefont {E.~P. A.~M.}\
  \bibnamefont {Bakkers}}, \bibinfo {author} {\bibfnamefont {S.~D.}\
  \bibnamefont {Sarma}},\ and\ \bibinfo {author} {\bibfnamefont {L.~P.}\
  \bibnamefont {Kouwenhoven}},\ }\bibfield  {title} {\bibinfo {title}
  {Quantized majorana conductance},\ }\href
  {https://doi.org/10.1038/nature26142} {\bibfield  {journal} {\bibinfo
  {journal} {[Retracted] Nature}\ }\textbf {\bibinfo {volume} {556}},\ \bibinfo
  {pages} {74} (\bibinfo {year} {2018})},\ \Eprint
  {https://arxiv.org/abs/1710.10701} {arXiv:1710.10701} \BibitemShut {NoStop}%
\bibitem [{\citenamefont {Bommer}\ \emph {et~al.}(2019)\citenamefont {Bommer},
  \citenamefont {Zhang}, \citenamefont {G{\"u}l}, \citenamefont {Nijholt},
  \citenamefont {Wimmer}, \citenamefont {Rybakov}, \citenamefont {Garaud},
  \citenamefont {Rodic}, \citenamefont {Babaev}, \citenamefont {Troyer},
  \citenamefont {Car}, \citenamefont {Plissard}, \citenamefont {Bakkers},
  \citenamefont {Watanabe}, \citenamefont {Taniguchi},\ and\ \citenamefont
  {Kouwenhoven}}]{bommer2019spinorbit}%
  \BibitemOpen
  \bibfield  {author} {\bibinfo {author} {\bibfnamefont {J.~D.~S.}\
  \bibnamefont {Bommer}}, \bibinfo {author} {\bibfnamefont {H.}~\bibnamefont
  {Zhang}}, \bibinfo {author} {\bibfnamefont {{\"O}.}~\bibnamefont {G{\"u}l}},
  \bibinfo {author} {\bibfnamefont {B.}~\bibnamefont {Nijholt}}, \bibinfo
  {author} {\bibfnamefont {M.}~\bibnamefont {Wimmer}}, \bibinfo {author}
  {\bibfnamefont {F.~N.}\ \bibnamefont {Rybakov}}, \bibinfo {author}
  {\bibfnamefont {J.}~\bibnamefont {Garaud}}, \bibinfo {author} {\bibfnamefont
  {D.}~\bibnamefont {Rodic}}, \bibinfo {author} {\bibfnamefont
  {E.}~\bibnamefont {Babaev}}, \bibinfo {author} {\bibfnamefont
  {M.}~\bibnamefont {Troyer}}, \bibinfo {author} {\bibfnamefont
  {D.}~\bibnamefont {Car}}, \bibinfo {author} {\bibfnamefont {S.~R.}\
  \bibnamefont {Plissard}}, \bibinfo {author} {\bibfnamefont {E.~P. A.~M.}\
  \bibnamefont {Bakkers}}, \bibinfo {author} {\bibfnamefont {K.}~\bibnamefont
  {Watanabe}}, \bibinfo {author} {\bibfnamefont {T.}~\bibnamefont
  {Taniguchi}},\ and\ \bibinfo {author} {\bibfnamefont {L.~P.}\ \bibnamefont
  {Kouwenhoven}},\ }\bibfield  {title} {\bibinfo {title} {Spin-{{Orbit
  Protection}} of {{Induced Superconductivity}} in {{Majorana Nanowires}}},\
  }\href {https://doi.org/10.1103/PhysRevLett.122.187702} {\bibfield  {journal}
  {\bibinfo  {journal} {Phys. Rev. Lett.}\ }\textbf {\bibinfo {volume} {122}},\
  \bibinfo {pages} {187702} (\bibinfo {year} {2019})}\BibitemShut {NoStop}%
\bibitem [{\citenamefont {Grivnin}\ \emph {et~al.}(2019)\citenamefont
  {Grivnin}, \citenamefont {Bor}, \citenamefont {Heiblum}, \citenamefont
  {Oreg},\ and\ \citenamefont {Shtrikman}}]{grivnin2019concomitant}%
  \BibitemOpen
  \bibfield  {author} {\bibinfo {author} {\bibfnamefont {A.}~\bibnamefont
  {Grivnin}}, \bibinfo {author} {\bibfnamefont {E.}~\bibnamefont {Bor}},
  \bibinfo {author} {\bibfnamefont {M.}~\bibnamefont {Heiblum}}, \bibinfo
  {author} {\bibfnamefont {Y.}~\bibnamefont {Oreg}},\ and\ \bibinfo {author}
  {\bibfnamefont {H.}~\bibnamefont {Shtrikman}},\ }\bibfield  {title} {\bibinfo
  {title} {Concomitant opening of a bulk-gap with an emerging possible
  {{Majorana}} zero mode},\ }\href {https://doi.org/10.1038/s41467-019-09771-0}
  {\bibfield  {journal} {\bibinfo  {journal} {Nat Commun}\ }\textbf {\bibinfo
  {volume} {10}},\ \bibinfo {pages} {1940} (\bibinfo {year}
  {2019})}\BibitemShut {NoStop}%
\bibitem [{\citenamefont {Anselmetti}\ \emph {et~al.}(2019)\citenamefont
  {Anselmetti}, \citenamefont {Martinez}, \citenamefont {M{\'e}nard},
  \citenamefont {Puglia}, \citenamefont {Malinowski}, \citenamefont {Lee},
  \citenamefont {Choi}, \citenamefont {Pendharkar}, \citenamefont
  {Palmstr{\o}m}, \citenamefont {Marcus}, \citenamefont {Casparis},\ and\
  \citenamefont {Higginbotham}}]{anselmetti2019endtoend}%
  \BibitemOpen
  \bibfield  {author} {\bibinfo {author} {\bibfnamefont {G.~L.~R.}\
  \bibnamefont {Anselmetti}}, \bibinfo {author} {\bibfnamefont {E.~A.}\
  \bibnamefont {Martinez}}, \bibinfo {author} {\bibfnamefont {G.~C.}\
  \bibnamefont {M{\'e}nard}}, \bibinfo {author} {\bibfnamefont
  {D.}~\bibnamefont {Puglia}}, \bibinfo {author} {\bibfnamefont {F.~K.}\
  \bibnamefont {Malinowski}}, \bibinfo {author} {\bibfnamefont {J.~S.}\
  \bibnamefont {Lee}}, \bibinfo {author} {\bibfnamefont {S.}~\bibnamefont
  {Choi}}, \bibinfo {author} {\bibfnamefont {M.}~\bibnamefont {Pendharkar}},
  \bibinfo {author} {\bibfnamefont {C.~J.}\ \bibnamefont {Palmstr{\o}m}},
  \bibinfo {author} {\bibfnamefont {C.~M.}\ \bibnamefont {Marcus}}, \bibinfo
  {author} {\bibfnamefont {L.}~\bibnamefont {Casparis}},\ and\ \bibinfo
  {author} {\bibfnamefont {A.~P.}\ \bibnamefont {Higginbotham}},\ }\bibfield
  {title} {\bibinfo {title} {End-to-end correlated subgap states in hybrid
  nanowires},\ }\href {https://doi.org/10.1103/PhysRevB.100.205412} {\bibfield
  {journal} {\bibinfo  {journal} {Phys. Rev. B}\ }\textbf {\bibinfo {volume}
  {100}},\ \bibinfo {pages} {205412} (\bibinfo {year} {2019})}\BibitemShut
  {NoStop}%
\bibitem [{\citenamefont {M{\'e}nard}\ \emph {et~al.}(2020)\citenamefont
  {M{\'e}nard}, \citenamefont {Anselmetti}, \citenamefont {Martinez},
  \citenamefont {Puglia}, \citenamefont {Malinowski}, \citenamefont {Lee},
  \citenamefont {Choi}, \citenamefont {Pendharkar}, \citenamefont
  {Palmstr{\o}m}, \citenamefont {Flensberg}, \citenamefont {Marcus},
  \citenamefont {Casparis},\ and\ \citenamefont
  {Higginbotham}}]{menard2020conductancematrix}%
  \BibitemOpen
  \bibfield  {author} {\bibinfo {author} {\bibfnamefont {G.~C.}\ \bibnamefont
  {M{\'e}nard}}, \bibinfo {author} {\bibfnamefont {G.~L.~R.}\ \bibnamefont
  {Anselmetti}}, \bibinfo {author} {\bibfnamefont {E.~A.}\ \bibnamefont
  {Martinez}}, \bibinfo {author} {\bibfnamefont {D.}~\bibnamefont {Puglia}},
  \bibinfo {author} {\bibfnamefont {F.~K.}\ \bibnamefont {Malinowski}},
  \bibinfo {author} {\bibfnamefont {J.~S.}\ \bibnamefont {Lee}}, \bibinfo
  {author} {\bibfnamefont {S.}~\bibnamefont {Choi}}, \bibinfo {author}
  {\bibfnamefont {M.}~\bibnamefont {Pendharkar}}, \bibinfo {author}
  {\bibfnamefont {C.~J.}\ \bibnamefont {Palmstr{\o}m}}, \bibinfo {author}
  {\bibfnamefont {K.}~\bibnamefont {Flensberg}}, \bibinfo {author}
  {\bibfnamefont {C.~M.}\ \bibnamefont {Marcus}}, \bibinfo {author}
  {\bibfnamefont {L.}~\bibnamefont {Casparis}},\ and\ \bibinfo {author}
  {\bibfnamefont {A.~P.}\ \bibnamefont {Higginbotham}},\ }\bibfield  {title}
  {\bibinfo {title} {Conductance-{{Matrix Symmetries}} of a
  {{Three}}-{{Terminal Hybrid Device}}},\ }\href
  {https://doi.org/10.1103/PhysRevLett.124.036802} {\bibfield  {journal}
  {\bibinfo  {journal} {Phys. Rev. Lett.}\ }\textbf {\bibinfo {volume} {124}},\
  \bibinfo {pages} {036802} (\bibinfo {year} {2020})}\BibitemShut {NoStop}%
\bibitem [{\citenamefont {Puglia}\ \emph {et~al.}(2021)\citenamefont {Puglia},
  \citenamefont {Martinez}, \citenamefont {M{\'e}nard}, \citenamefont
  {P{\"o}schl}, \citenamefont {Gronin}, \citenamefont {Gardner}, \citenamefont
  {Kallaher}, \citenamefont {Manfra}, \citenamefont {Marcus}, \citenamefont
  {Higginbotham},\ and\ \citenamefont {Casparis}}]{puglia2021closing}%
  \BibitemOpen
  \bibfield  {author} {\bibinfo {author} {\bibfnamefont {D.}~\bibnamefont
  {Puglia}}, \bibinfo {author} {\bibfnamefont {E.~A.}\ \bibnamefont
  {Martinez}}, \bibinfo {author} {\bibfnamefont {G.~C.}\ \bibnamefont
  {M{\'e}nard}}, \bibinfo {author} {\bibfnamefont {A.}~\bibnamefont
  {P{\"o}schl}}, \bibinfo {author} {\bibfnamefont {S.}~\bibnamefont {Gronin}},
  \bibinfo {author} {\bibfnamefont {G.~C.}\ \bibnamefont {Gardner}}, \bibinfo
  {author} {\bibfnamefont {R.}~\bibnamefont {Kallaher}}, \bibinfo {author}
  {\bibfnamefont {M.~J.}\ \bibnamefont {Manfra}}, \bibinfo {author}
  {\bibfnamefont {C.~M.}\ \bibnamefont {Marcus}}, \bibinfo {author}
  {\bibfnamefont {A.~P.}\ \bibnamefont {Higginbotham}},\ and\ \bibinfo {author}
  {\bibfnamefont {L.}~\bibnamefont {Casparis}},\ }\bibfield  {title} {\bibinfo
  {title} {Closing of the induced gap in a hybrid superconductor-semiconductor
  nanowire},\ }\href {https://doi.org/10.1103/PhysRevB.103.235201} {\bibfield
  {journal} {\bibinfo  {journal} {Phys. Rev. B}\ }\textbf {\bibinfo {volume}
  {103}},\ \bibinfo {pages} {235201} (\bibinfo {year} {2021})}\BibitemShut
  {NoStop}%
\bibitem [{\citenamefont {Sengupta}\ \emph {et~al.}(2001)\citenamefont
  {Sengupta}, \citenamefont {{\v Z}uti{\'c}}, \citenamefont {Kwon},
  \citenamefont {Yakovenko},\ and\ \citenamefont
  {Das~Sarma}}]{sengupta2001midgap}%
  \BibitemOpen
  \bibfield  {author} {\bibinfo {author} {\bibfnamefont {K.}~\bibnamefont
  {Sengupta}}, \bibinfo {author} {\bibfnamefont {I.}~\bibnamefont {{\v
  Z}uti{\'c}}}, \bibinfo {author} {\bibfnamefont {H.-J.}\ \bibnamefont {Kwon}},
  \bibinfo {author} {\bibfnamefont {V.~M.}\ \bibnamefont {Yakovenko}},\ and\
  \bibinfo {author} {\bibfnamefont {S.}~\bibnamefont {Das~Sarma}},\ }\bibfield
  {title} {\bibinfo {title} {Midgap edge states and pairing symmetry of
  quasi-one-dimensional organic superconductors},\ }\href
  {https://doi.org/10.1103/PhysRevB.63.144531} {\bibfield  {journal} {\bibinfo
  {journal} {Phys. Rev. B}\ }\textbf {\bibinfo {volume} {63}},\ \bibinfo
  {pages} {144531} (\bibinfo {year} {2001})}\BibitemShut {NoStop}%
\bibitem [{\citenamefont {Law}\ \emph {et~al.}(2009)\citenamefont {Law},
  \citenamefont {Lee},\ and\ \citenamefont {Ng}}]{law2009majorana}%
  \BibitemOpen
  \bibfield  {author} {\bibinfo {author} {\bibfnamefont {K.~T.}\ \bibnamefont
  {Law}}, \bibinfo {author} {\bibfnamefont {P.~A.}\ \bibnamefont {Lee}},\ and\
  \bibinfo {author} {\bibfnamefont {T.~K.}\ \bibnamefont {Ng}},\ }\bibfield
  {title} {\bibinfo {title} {Majorana {{Fermion Induced Resonant Andreev
  Reflection}}},\ }\href {https://doi.org/10.1103/PhysRevLett.103.237001}
  {\bibfield  {journal} {\bibinfo  {journal} {Phys. Rev. Lett.}\ }\textbf
  {\bibinfo {volume} {103}},\ \bibinfo {pages} {237001} (\bibinfo {year}
  {2009})}\BibitemShut {NoStop}%
\bibitem [{\citenamefont {Flensberg}(2010)}]{flensberg2010tunneling}%
  \BibitemOpen
  \bibfield  {author} {\bibinfo {author} {\bibfnamefont {K.}~\bibnamefont
  {Flensberg}},\ }\bibfield  {title} {\bibinfo {title} {Tunneling
  characteristics of a chain of {{Majorana}} bound states},\ }\href
  {https://doi.org/10.1103/PhysRevB.82.180516} {\bibfield  {journal} {\bibinfo
  {journal} {Phys. Rev. B}\ }\textbf {\bibinfo {volume} {82}},\ \bibinfo
  {pages} {180516} (\bibinfo {year} {2010})}\BibitemShut {NoStop}%
\bibitem [{\citenamefont {Wimmer}\ \emph {et~al.}(2011)\citenamefont {Wimmer},
  \citenamefont {Akhmerov}, \citenamefont {Dahlhaus},\ and\ \citenamefont
  {Beenakker}}]{wimmer2011quantum}%
  \BibitemOpen
  \bibfield  {author} {\bibinfo {author} {\bibfnamefont {M.}~\bibnamefont
  {Wimmer}}, \bibinfo {author} {\bibfnamefont {A.~R.}\ \bibnamefont
  {Akhmerov}}, \bibinfo {author} {\bibfnamefont {J.~P.}\ \bibnamefont
  {Dahlhaus}},\ and\ \bibinfo {author} {\bibfnamefont {C.~W.~J.}\ \bibnamefont
  {Beenakker}},\ }\bibfield  {title} {\bibinfo {title} {Quantum point contact
  as a probe of a topological superconductor},\ }\href
  {https://doi.org/10.1088/1367-2630/13/5/053016} {\bibfield  {journal}
  {\bibinfo  {journal} {New J. Phys.}\ }\textbf {\bibinfo {volume} {13}},\
  \bibinfo {pages} {053016} (\bibinfo {year} {2011})}\BibitemShut {NoStop}%
\bibitem [{\citenamefont {Das~Sarma}\ \emph {et~al.}(2012)\citenamefont
  {Das~Sarma}, \citenamefont {Sau},\ and\ \citenamefont
  {Stanescu}}]{dassarma2012splitting}%
  \BibitemOpen
  \bibfield  {author} {\bibinfo {author} {\bibfnamefont {S.}~\bibnamefont
  {Das~Sarma}}, \bibinfo {author} {\bibfnamefont {J.~D.}\ \bibnamefont {Sau}},\
  and\ \bibinfo {author} {\bibfnamefont {T.~D.}\ \bibnamefont {Stanescu}},\
  }\bibfield  {title} {\bibinfo {title} {Splitting of the zero-bias conductance
  peak as smoking gun evidence for the existence of the {{Majorana}} mode in a
  superconductor-semiconductor nanowire},\ }\href
  {https://doi.org/10.1103/PhysRevB.86.220506} {\bibfield  {journal} {\bibinfo
  {journal} {Phys. Rev. B}\ }\textbf {\bibinfo {volume} {86}},\ \bibinfo
  {pages} {220506} (\bibinfo {year} {2012})}\BibitemShut {NoStop}%
\bibitem [{\citenamefont {Lai}\ \emph {et~al.}(2019)\citenamefont {Lai},
  \citenamefont {Sau},\ and\ \citenamefont {Das~Sarma}}]{lai2019presence}%
  \BibitemOpen
  \bibfield  {author} {\bibinfo {author} {\bibfnamefont {Y.-H.}\ \bibnamefont
  {Lai}}, \bibinfo {author} {\bibfnamefont {J.~D.}\ \bibnamefont {Sau}},\ and\
  \bibinfo {author} {\bibfnamefont {S.}~\bibnamefont {Das~Sarma}},\ }\bibfield
  {title} {\bibinfo {title} {Presence versus absence of end-to-end nonlocal
  conductance correlations in {{Majorana}} nanowires: {{Majorana}} bound states
  versus {{Andreev}} bound states},\ }\href
  {https://doi.org/10.1103/PhysRevB.100.045302} {\bibfield  {journal} {\bibinfo
   {journal} {Phys. Rev. B}\ }\textbf {\bibinfo {volume} {100}},\ \bibinfo
  {pages} {045302} (\bibinfo {year} {2019})}\BibitemShut {NoStop}%
\bibitem [{\citenamefont {Pan}\ \emph {et~al.}(2021{\natexlab{a}})\citenamefont
  {Pan}, \citenamefont {Sau},\ and\ \citenamefont
  {Das~Sarma}}]{pan2021threeterminal}%
  \BibitemOpen
  \bibfield  {author} {\bibinfo {author} {\bibfnamefont {H.}~\bibnamefont
  {Pan}}, \bibinfo {author} {\bibfnamefont {J.~D.}\ \bibnamefont {Sau}},\ and\
  \bibinfo {author} {\bibfnamefont {S.}~\bibnamefont {Das~Sarma}},\ }\bibfield
  {title} {\bibinfo {title} {Three-terminal nonlocal conductance in
  {{Majorana}} nanowires: {{Distinguishing}} topological and trivial in
  realistic systems with disorder and inhomogeneous potential},\ }\href
  {https://doi.org/10.1103/PhysRevB.103.014513} {\bibfield  {journal} {\bibinfo
   {journal} {Phys. Rev. B}\ }\textbf {\bibinfo {volume} {103}},\ \bibinfo
  {pages} {014513} (\bibinfo {year} {2021}{\natexlab{a}})}\BibitemShut
  {NoStop}%
\bibitem [{\citenamefont {Rosdahl}\ \emph {et~al.}(2018)\citenamefont
  {Rosdahl}, \citenamefont {Vuik}, \citenamefont {Kjaergaard},\ and\
  \citenamefont {Akhmerov}}]{rosdahl2018andreev}%
  \BibitemOpen
  \bibfield  {author} {\bibinfo {author} {\bibfnamefont {T.~{\"O}.}\
  \bibnamefont {Rosdahl}}, \bibinfo {author} {\bibfnamefont {A.}~\bibnamefont
  {Vuik}}, \bibinfo {author} {\bibfnamefont {M.}~\bibnamefont {Kjaergaard}},\
  and\ \bibinfo {author} {\bibfnamefont {A.~R.}\ \bibnamefont {Akhmerov}},\
  }\bibfield  {title} {\bibinfo {title} {Andreev rectifier: {{A}} nonlocal
  conductance signature of topological phase transitions},\ }\href
  {https://doi.org/10.1103/PhysRevB.97.045421} {\bibfield  {journal} {\bibinfo
  {journal} {Phys. Rev. B}\ }\textbf {\bibinfo {volume} {97}},\ \bibinfo
  {pages} {045421} (\bibinfo {year} {2018})}\BibitemShut {NoStop}%
\bibitem [{\citenamefont {Pan}\ and\ \citenamefont
  {Das~Sarma}(2020)}]{pan2020physical}%
  \BibitemOpen
  \bibfield  {author} {\bibinfo {author} {\bibfnamefont {H.}~\bibnamefont
  {Pan}}\ and\ \bibinfo {author} {\bibfnamefont {S.}~\bibnamefont
  {Das~Sarma}},\ }\bibfield  {title} {\bibinfo {title} {Physical mechanisms for
  zero-bias conductance peaks in {{Majorana}} nanowires},\ }\href
  {https://doi.org/10.1103/PhysRevResearch.2.013377} {\bibfield  {journal}
  {\bibinfo  {journal} {Phys. Rev. Research}\ }\textbf {\bibinfo {volume}
  {2}},\ \bibinfo {pages} {013377} (\bibinfo {year} {2020})}\BibitemShut
  {NoStop}%
\bibitem [{\citenamefont {Kells}\ \emph {et~al.}(2012)\citenamefont {Kells},
  \citenamefont {Meidan},\ and\ \citenamefont
  {Brouwer}}]{kells2012nearzeroenergy}%
  \BibitemOpen
  \bibfield  {author} {\bibinfo {author} {\bibfnamefont {G.}~\bibnamefont
  {Kells}}, \bibinfo {author} {\bibfnamefont {D.}~\bibnamefont {Meidan}},\ and\
  \bibinfo {author} {\bibfnamefont {P.~W.}\ \bibnamefont {Brouwer}},\
  }\bibfield  {title} {\bibinfo {title} {Near-zero-energy end states in
  topologically trivial spin-orbit coupled superconducting nanowires with a
  smooth confinement},\ }\href {https://doi.org/10.1103/PhysRevB.86.100503}
  {\bibfield  {journal} {\bibinfo  {journal} {Phys. Rev. B}\ }\textbf {\bibinfo
  {volume} {86}},\ \bibinfo {pages} {100503} (\bibinfo {year}
  {2012})}\BibitemShut {NoStop}%
\bibitem [{\citenamefont {Prada}\ \emph {et~al.}(2012)\citenamefont {Prada},
  \citenamefont {{San-Jose}},\ and\ \citenamefont
  {Aguado}}]{prada2012transport}%
  \BibitemOpen
  \bibfield  {author} {\bibinfo {author} {\bibfnamefont {E.}~\bibnamefont
  {Prada}}, \bibinfo {author} {\bibfnamefont {P.}~\bibnamefont {{San-Jose}}},\
  and\ \bibinfo {author} {\bibfnamefont {R.}~\bibnamefont {Aguado}},\
  }\bibfield  {title} {\bibinfo {title} {Transport spectroscopy of {$NS$}
  nanowire junctions with {{Majorana}} fermions},\ }\href
  {https://doi.org/10.1103/PhysRevB.86.180503} {\bibfield  {journal} {\bibinfo
  {journal} {Phys. Rev. B}\ }\textbf {\bibinfo {volume} {86}},\ \bibinfo
  {pages} {180503} (\bibinfo {year} {2012})}\BibitemShut {NoStop}%
\bibitem [{\citenamefont {Liu}\ \emph {et~al.}(2017)\citenamefont {Liu},
  \citenamefont {Sau}, \citenamefont {Stanescu},\ and\ \citenamefont
  {Das~Sarma}}]{liu2017andreev}%
  \BibitemOpen
  \bibfield  {author} {\bibinfo {author} {\bibfnamefont {C.-X.}\ \bibnamefont
  {Liu}}, \bibinfo {author} {\bibfnamefont {J.~D.}\ \bibnamefont {Sau}},
  \bibinfo {author} {\bibfnamefont {T.~D.}\ \bibnamefont {Stanescu}},\ and\
  \bibinfo {author} {\bibfnamefont {S.}~\bibnamefont {Das~Sarma}},\ }\bibfield
  {title} {\bibinfo {title} {Andreev bound states versus {{Majorana}} bound
  states in quantum dot-nanowire-superconductor hybrid structures: {{Trivial}}
  versus topological zero-bias conductance peaks},\ }\href
  {https://doi.org/10.1103/PhysRevB.96.075161} {\bibfield  {journal} {\bibinfo
  {journal} {Phys. Rev. B}\ }\textbf {\bibinfo {volume} {96}},\ \bibinfo
  {pages} {075161} (\bibinfo {year} {2017})}\BibitemShut {NoStop}%
\bibitem [{\citenamefont {Stanescu}\ and\ \citenamefont
  {Tewari}(2019)}]{stanescu2019robust}%
  \BibitemOpen
  \bibfield  {author} {\bibinfo {author} {\bibfnamefont {T.~D.}\ \bibnamefont
  {Stanescu}}\ and\ \bibinfo {author} {\bibfnamefont {S.}~\bibnamefont
  {Tewari}},\ }\bibfield  {title} {\bibinfo {title} {Robust low-energy
  {{Andreev}} bound states in semiconductor-superconductor structures:
  {{Importance}} of partial separation of component {{Majorana}} bound
  states},\ }\href {https://doi.org/10.1103/PhysRevB.100.155429} {\bibfield
  {journal} {\bibinfo  {journal} {Phys. Rev. B}\ }\textbf {\bibinfo {volume}
  {100}},\ \bibinfo {pages} {155429} (\bibinfo {year} {2019})}\BibitemShut
  {NoStop}%
\bibitem [{\citenamefont {Moore}\ \emph
  {et~al.}(2018{\natexlab{a}})\citenamefont {Moore}, \citenamefont {Zeng},
  \citenamefont {Stanescu},\ and\ \citenamefont {Tewari}}]{moore2018quantized}%
  \BibitemOpen
  \bibfield  {author} {\bibinfo {author} {\bibfnamefont {C.}~\bibnamefont
  {Moore}}, \bibinfo {author} {\bibfnamefont {C.}~\bibnamefont {Zeng}},
  \bibinfo {author} {\bibfnamefont {T.~D.}\ \bibnamefont {Stanescu}},\ and\
  \bibinfo {author} {\bibfnamefont {S.}~\bibnamefont {Tewari}},\ }\bibfield
  {title} {\bibinfo {title} {Quantized zero-bias conductance plateau in
  semiconductor-superconductor heterostructures without topological
  {{Majorana}} zero modes},\ }\href
  {https://doi.org/10.1103/PhysRevB.98.155314} {\bibfield  {journal} {\bibinfo
  {journal} {Phys. Rev. B}\ }\textbf {\bibinfo {volume} {98}},\ \bibinfo
  {pages} {155314} (\bibinfo {year} {2018}{\natexlab{a}})}\BibitemShut
  {NoStop}%
\bibitem [{\citenamefont {Moore}\ \emph
  {et~al.}(2018{\natexlab{b}})\citenamefont {Moore}, \citenamefont {Stanescu},\
  and\ \citenamefont {Tewari}}]{moore2018twoterminal}%
  \BibitemOpen
  \bibfield  {author} {\bibinfo {author} {\bibfnamefont {C.}~\bibnamefont
  {Moore}}, \bibinfo {author} {\bibfnamefont {T.~D.}\ \bibnamefont
  {Stanescu}},\ and\ \bibinfo {author} {\bibfnamefont {S.}~\bibnamefont
  {Tewari}},\ }\bibfield  {title} {\bibinfo {title} {Two-terminal charge
  tunneling: {{Disentangling Majorana}} zero modes from partially separated
  {{Andreev}} bound states in semiconductor-superconductor heterostructures},\
  }\href {https://doi.org/10.1103/PhysRevB.97.165302} {\bibfield  {journal}
  {\bibinfo  {journal} {Phys. Rev. B}\ }\textbf {\bibinfo {volume} {97}},\
  \bibinfo {pages} {165302} (\bibinfo {year} {2018}{\natexlab{b}})}\BibitemShut
  {NoStop}%
\bibitem [{\citenamefont {Mi}\ \emph {et~al.}(2014)\citenamefont {Mi},
  \citenamefont {Pikulin}, \citenamefont {Marciani},\ and\ \citenamefont
  {Beenakker}}]{mi2014xshaped}%
  \BibitemOpen
  \bibfield  {author} {\bibinfo {author} {\bibfnamefont {S.}~\bibnamefont
  {Mi}}, \bibinfo {author} {\bibfnamefont {D.~I.}\ \bibnamefont {Pikulin}},
  \bibinfo {author} {\bibfnamefont {M.}~\bibnamefont {Marciani}},\ and\
  \bibinfo {author} {\bibfnamefont {C.~W.~J.}\ \bibnamefont {Beenakker}},\
  }\bibfield  {title} {\bibinfo {title} {X-shaped and {{Y}}-shaped {{Andreev}}
  resonance profiles in a superconducting quantum dot},\ }\href
  {https://doi.org/10.1134/S1063776114120176} {\bibfield  {journal} {\bibinfo
  {journal} {J. Exp. Theor. Phys.}\ }\textbf {\bibinfo {volume} {119}},\
  \bibinfo {pages} {1018} (\bibinfo {year} {2014})}\BibitemShut {NoStop}%
\bibitem [{\citenamefont {Sau}\ and\ \citenamefont
  {Das~Sarma}(2013)}]{sau2013density}%
  \BibitemOpen
  \bibfield  {author} {\bibinfo {author} {\bibfnamefont {J.~D.}\ \bibnamefont
  {Sau}}\ and\ \bibinfo {author} {\bibfnamefont {S.}~\bibnamefont
  {Das~Sarma}},\ }\bibfield  {title} {\bibinfo {title} {Density of states of
  disordered topological superconductor-semiconductor hybrid nanowires},\
  }\href {https://doi.org/10.1103/PhysRevB.88.064506} {\bibfield  {journal}
  {\bibinfo  {journal} {Phys. Rev. B}\ }\textbf {\bibinfo {volume} {88}},\
  \bibinfo {pages} {064506} (\bibinfo {year} {2013})}\BibitemShut {NoStop}%
\bibitem [{\citenamefont {Pikulin}\ \emph {et~al.}(2012)\citenamefont
  {Pikulin}, \citenamefont {Dahlhaus}, \citenamefont {Wimmer}, \citenamefont
  {Schomerus},\ and\ \citenamefont {Beenakker}}]{pikulin2012zerovoltage}%
  \BibitemOpen
  \bibfield  {author} {\bibinfo {author} {\bibfnamefont {D.~I.}\ \bibnamefont
  {Pikulin}}, \bibinfo {author} {\bibfnamefont {J.~P.}\ \bibnamefont
  {Dahlhaus}}, \bibinfo {author} {\bibfnamefont {M.}~\bibnamefont {Wimmer}},
  \bibinfo {author} {\bibfnamefont {H.}~\bibnamefont {Schomerus}},\ and\
  \bibinfo {author} {\bibfnamefont {C.~W.~J.}\ \bibnamefont {Beenakker}},\
  }\bibfield  {title} {\bibinfo {title} {A zero-voltage conductance peak from
  weak antilocalization in a {{Majorana}} nanowire},\ }\href
  {https://doi.org/10.1088/1367-2630/14/12/125011} {\bibfield  {journal}
  {\bibinfo  {journal} {New J. Phys.}\ }\textbf {\bibinfo {volume} {14}},\
  \bibinfo {pages} {125011} (\bibinfo {year} {2012})}\BibitemShut {NoStop}%
\bibitem [{\citenamefont {Bagrets}\ and\ \citenamefont
  {Altland}(2012)}]{bagrets2012class}%
  \BibitemOpen
  \bibfield  {author} {\bibinfo {author} {\bibfnamefont {D.}~\bibnamefont
  {Bagrets}}\ and\ \bibinfo {author} {\bibfnamefont {A.}~\bibnamefont
  {Altland}},\ }\bibfield  {title} {\bibinfo {title} {Class {$D$} {{Spectral
  Peak}} in {{Majorana Quantum Wires}}},\ }\href
  {https://doi.org/10.1103/PhysRevLett.109.227005} {\bibfield  {journal}
  {\bibinfo  {journal} {Phys. Rev. Lett.}\ }\textbf {\bibinfo {volume} {109}},\
  \bibinfo {pages} {227005} (\bibinfo {year} {2012})}\BibitemShut {NoStop}%
\bibitem [{\citenamefont {Liu}\ \emph {et~al.}(2012)\citenamefont {Liu},
  \citenamefont {Potter}, \citenamefont {Law},\ and\ \citenamefont
  {Lee}}]{liu2012zerobias}%
  \BibitemOpen
  \bibfield  {author} {\bibinfo {author} {\bibfnamefont {J.}~\bibnamefont
  {Liu}}, \bibinfo {author} {\bibfnamefont {A.~C.}\ \bibnamefont {Potter}},
  \bibinfo {author} {\bibfnamefont {K.~T.}\ \bibnamefont {Law}},\ and\ \bibinfo
  {author} {\bibfnamefont {P.~A.}\ \bibnamefont {Lee}},\ }\bibfield  {title}
  {\bibinfo {title} {Zero-{{Bias Peaks}} in the {{Tunneling Conductance}} of
  {{Spin}}-{{Orbit}}-{{Coupled Superconducting Wires}} with and without
  {{Majorana End}}-{{States}}},\ }\href
  {https://doi.org/10.1103/PhysRevLett.109.267002} {\bibfield  {journal}
  {\bibinfo  {journal} {Phys. Rev. Lett.}\ }\textbf {\bibinfo {volume} {109}},\
  \bibinfo {pages} {267002} (\bibinfo {year} {2012})}\BibitemShut {NoStop}%
\bibitem [{\citenamefont {Pan}\ \emph {et~al.}(2020)\citenamefont {Pan},
  \citenamefont {Cole}, \citenamefont {Sau},\ and\ \citenamefont
  {Das~Sarma}}]{pan2020generic}%
  \BibitemOpen
  \bibfield  {author} {\bibinfo {author} {\bibfnamefont {H.}~\bibnamefont
  {Pan}}, \bibinfo {author} {\bibfnamefont {W.~S.}\ \bibnamefont {Cole}},
  \bibinfo {author} {\bibfnamefont {J.~D.}\ \bibnamefont {Sau}},\ and\ \bibinfo
  {author} {\bibfnamefont {S.}~\bibnamefont {Das~Sarma}},\ }\bibfield  {title}
  {\bibinfo {title} {Generic quantized zero-bias conductance peaks in
  superconductor-semiconductor hybrid structures},\ }\href
  {https://doi.org/10.1103/PhysRevB.101.024506} {\bibfield  {journal} {\bibinfo
   {journal} {Phys. Rev. B}\ }\textbf {\bibinfo {volume} {101}},\ \bibinfo
  {pages} {024506} (\bibinfo {year} {2020})}\BibitemShut {NoStop}%
\bibitem [{cro()}]{crossover_SM}%
  \BibitemOpen
  \href@noop {} {\bibinfo {title} {See the {{Supplementary Information}} for
  animations}}\BibitemShut {NoStop}%
\bibitem [{\citenamefont {Woods}\ \emph {et~al.}(2021)\citenamefont {Woods},
  \citenamefont {Sarma},\ and\ \citenamefont {Stanescu}}]{woods2021charge}%
  \BibitemOpen
  \bibfield  {author} {\bibinfo {author} {\bibfnamefont {B.~D.}\ \bibnamefont
  {Woods}}, \bibinfo {author} {\bibfnamefont {S.~D.}\ \bibnamefont {Sarma}},\
  and\ \bibinfo {author} {\bibfnamefont {T.~D.}\ \bibnamefont {Stanescu}},\
  }\bibfield  {title} {\bibinfo {title} {Charge impurity effects in hybrid
  {{Majorana}} nanowires},\ }\href {http://arxiv.org/abs/2103.06880} {\bibfield
   {journal} {\bibinfo  {journal} {arXiv:2103.06880}\ } (\bibinfo {year}
  {2021})}\BibitemShut {NoStop}%
\bibitem [{\citenamefont {Woods}\ \emph {et~al.}(2020)\citenamefont {Woods},
  \citenamefont {Das~Sarma},\ and\ \citenamefont
  {Stanescu}}]{woods2020subband}%
  \BibitemOpen
  \bibfield  {author} {\bibinfo {author} {\bibfnamefont {B.~D.}\ \bibnamefont
  {Woods}}, \bibinfo {author} {\bibfnamefont {S.}~\bibnamefont {Das~Sarma}},\
  and\ \bibinfo {author} {\bibfnamefont {T.~D.}\ \bibnamefont {Stanescu}},\
  }\bibfield  {title} {\bibinfo {title} {Subband occupation in
  semiconductor-superconductor nanowires},\ }\href
  {https://doi.org/10.1103/PhysRevB.101.045405} {\bibfield  {journal} {\bibinfo
   {journal} {Phys. Rev. B}\ }\textbf {\bibinfo {volume} {101}},\ \bibinfo
  {pages} {045405} (\bibinfo {year} {2020})}\BibitemShut {NoStop}%
\bibitem [{\citenamefont {Liu}\ \emph {et~al.}(2019)\citenamefont {Liu},
  \citenamefont {Sau}, \citenamefont {Stanescu},\ and\ \citenamefont
  {Das~Sarma}}]{liu2019conductance}%
  \BibitemOpen
  \bibfield  {author} {\bibinfo {author} {\bibfnamefont {C.-X.}\ \bibnamefont
  {Liu}}, \bibinfo {author} {\bibfnamefont {J.~D.}\ \bibnamefont {Sau}},
  \bibinfo {author} {\bibfnamefont {T.~D.}\ \bibnamefont {Stanescu}},\ and\
  \bibinfo {author} {\bibfnamefont {S.}~\bibnamefont {Das~Sarma}},\ }\bibfield
  {title} {\bibinfo {title} {Conductance smearing and anisotropic suppression
  of induced superconductivity in a {{Majorana}} nanowire},\ }\href
  {https://doi.org/10.1103/PhysRevB.99.024510} {\bibfield  {journal} {\bibinfo
  {journal} {Phys. Rev. B}\ }\textbf {\bibinfo {volume} {99}},\ \bibinfo
  {pages} {024510} (\bibinfo {year} {2019})}\BibitemShut {NoStop}%
\bibitem [{\citenamefont {Pan}\ \emph {et~al.}(2021{\natexlab{b}})\citenamefont
  {Pan}, \citenamefont {Liu}, \citenamefont {Wimmer},\ and\ \citenamefont
  {Das~Sarma}}]{pan2021quantized}%
  \BibitemOpen
  \bibfield  {author} {\bibinfo {author} {\bibfnamefont {H.}~\bibnamefont
  {Pan}}, \bibinfo {author} {\bibfnamefont {C.-X.}\ \bibnamefont {Liu}},
  \bibinfo {author} {\bibfnamefont {M.}~\bibnamefont {Wimmer}},\ and\ \bibinfo
  {author} {\bibfnamefont {S.}~\bibnamefont {Das~Sarma}},\ }\bibfield  {title}
  {\bibinfo {title} {Quantized and unquantized zero-bias tunneling conductance
  peaks in {{Majorana}} nanowires: {{Conductance}} below and above
  {$2{e}^{2}/h$}},\ }\href {https://doi.org/10.1103/PhysRevB.103.214502}
  {\bibfield  {journal} {\bibinfo  {journal} {Phys. Rev. B}\ }\textbf {\bibinfo
  {volume} {103}},\ \bibinfo {pages} {214502} (\bibinfo {year}
  {2021}{\natexlab{b}})}\BibitemShut {NoStop}%
\bibitem [{\citenamefont {Stanescu}\ \emph {et~al.}(2010)\citenamefont
  {Stanescu}, \citenamefont {Sau}, \citenamefont {Lutchyn},\ and\ \citenamefont
  {Das~Sarma}}]{stanescu2010proximity}%
  \BibitemOpen
  \bibfield  {author} {\bibinfo {author} {\bibfnamefont {T.~D.}\ \bibnamefont
  {Stanescu}}, \bibinfo {author} {\bibfnamefont {J.~D.}\ \bibnamefont {Sau}},
  \bibinfo {author} {\bibfnamefont {R.~M.}\ \bibnamefont {Lutchyn}},\ and\
  \bibinfo {author} {\bibfnamefont {S.}~\bibnamefont {Das~Sarma}},\ }\bibfield
  {title} {\bibinfo {title} {Proximity effect at the
  superconductor--topological insulator interface},\ }\href
  {https://doi.org/10.1103/PhysRevB.81.241310} {\bibfield  {journal} {\bibinfo
  {journal} {Phys. Rev. B}\ }\textbf {\bibinfo {volume} {81}},\ \bibinfo
  {pages} {241310} (\bibinfo {year} {2010})}\BibitemShut {NoStop}%
\bibitem [{\citenamefont {Stanescu}\ and\ \citenamefont
  {Das~Sarma}(2017)}]{stanescu2017proximityinduced}%
  \BibitemOpen
  \bibfield  {author} {\bibinfo {author} {\bibfnamefont {T.~D.}\ \bibnamefont
  {Stanescu}}\ and\ \bibinfo {author} {\bibfnamefont {S.}~\bibnamefont
  {Das~Sarma}},\ }\bibfield  {title} {\bibinfo {title} {Proximity-induced
  low-energy renormalization in hybrid semiconductor-superconductor
  {{Majorana}} structures},\ }\href
  {https://doi.org/10.1103/PhysRevB.96.014510} {\bibfield  {journal} {\bibinfo
  {journal} {Phys. Rev. B}\ }\textbf {\bibinfo {volume} {96}},\ \bibinfo
  {pages} {014510} (\bibinfo {year} {2017})}\BibitemShut {NoStop}%
\bibitem [{\citenamefont {Lutchyn}\ \emph {et~al.}(2018)\citenamefont
  {Lutchyn}, \citenamefont {Bakkers}, \citenamefont {Kouwenhoven},
  \citenamefont {Krogstrup}, \citenamefont {Marcus},\ and\ \citenamefont
  {Oreg}}]{lutchyn2018majorana}%
  \BibitemOpen
  \bibfield  {author} {\bibinfo {author} {\bibfnamefont {R.~M.}\ \bibnamefont
  {Lutchyn}}, \bibinfo {author} {\bibfnamefont {E.~P. A.~M.}\ \bibnamefont
  {Bakkers}}, \bibinfo {author} {\bibfnamefont {L.~P.}\ \bibnamefont
  {Kouwenhoven}}, \bibinfo {author} {\bibfnamefont {P.}~\bibnamefont
  {Krogstrup}}, \bibinfo {author} {\bibfnamefont {C.~M.}\ \bibnamefont
  {Marcus}},\ and\ \bibinfo {author} {\bibfnamefont {Y.}~\bibnamefont {Oreg}},\
  }\bibfield  {title} {\bibinfo {title} {Majorana zero modes in
  superconductor\textendash semiconductor heterostructures},\ }\href
  {https://doi.org/10.1038/s41578-018-0003-1} {\bibfield  {journal} {\bibinfo
  {journal} {Nature Reviews Materials}\ }\textbf {\bibinfo {volume} {3}},\
  \bibinfo {pages} {52} (\bibinfo {year} {2018})}\BibitemShut {NoStop}%
\bibitem [{\citenamefont {Annett}(2004)}]{annett2004superconductivity}%
  \BibitemOpen
  \bibfield  {author} {\bibinfo {author} {\bibfnamefont {J.~F.}\ \bibnamefont
  {Annett}},\ }\href@noop {} {\emph {\bibinfo {title} {Superconductivity,
  Superfluids, and Condensates}}},\ Oxford Master Series in Condensed Matter
  Physics\ (\bibinfo  {publisher} {{Oxford University Press}},\ \bibinfo
  {address} {{Oxford ; New York}},\ \bibinfo {year} {2004})\BibitemShut
  {NoStop}%
\bibitem [{\citenamefont {Setiawan}\ \emph {et~al.}(2017)\citenamefont
  {Setiawan}, \citenamefont {Liu}, \citenamefont {Sau},\ and\ \citenamefont
  {Das~Sarma}}]{setiawan2017electron}%
  \BibitemOpen
  \bibfield  {author} {\bibinfo {author} {\bibfnamefont {F.}~\bibnamefont
  {Setiawan}}, \bibinfo {author} {\bibfnamefont {C.-X.}\ \bibnamefont {Liu}},
  \bibinfo {author} {\bibfnamefont {J.~D.}\ \bibnamefont {Sau}},\ and\ \bibinfo
  {author} {\bibfnamefont {S.}~\bibnamefont {Das~Sarma}},\ }\bibfield  {title}
  {\bibinfo {title} {Electron temperature and tunnel coupling dependence of
  zero-bias and almost-zero-bias conductance peaks in {{Majorana}} nanowires},\
  }\href {https://doi.org/10.1103/PhysRevB.96.184520} {\bibfield  {journal}
  {\bibinfo  {journal} {Phys. Rev. B}\ }\textbf {\bibinfo {volume} {96}},\
  \bibinfo {pages} {184520} (\bibinfo {year} {2017})}\BibitemShut {NoStop}%
\bibitem [{\citenamefont {Liu}\ \emph {et~al.}(2018)\citenamefont {Liu},
  \citenamefont {Sau},\ and\ \citenamefont
  {Das~Sarma}}]{liu2018distinguishing}%
  \BibitemOpen
  \bibfield  {author} {\bibinfo {author} {\bibfnamefont {C.-X.}\ \bibnamefont
  {Liu}}, \bibinfo {author} {\bibfnamefont {J.~D.}\ \bibnamefont {Sau}},\ and\
  \bibinfo {author} {\bibfnamefont {S.}~\bibnamefont {Das~Sarma}},\ }\bibfield
  {title} {\bibinfo {title} {Distinguishing topological {{Majorana}} bound
  states from trivial {{Andreev}} bound states: {{Proposed}} tests through
  differential tunneling conductance spectroscopy},\ }\href
  {https://doi.org/10.1103/PhysRevB.97.214502} {\bibfield  {journal} {\bibinfo
  {journal} {Phys. Rev. B}\ }\textbf {\bibinfo {volume} {97}},\ \bibinfo
  {pages} {214502} (\bibinfo {year} {2018})}\BibitemShut {NoStop}%
\bibitem [{\citenamefont {Beenakker}(1997)}]{beenakker1997randommatrix}%
  \BibitemOpen
  \bibfield  {author} {\bibinfo {author} {\bibfnamefont {C.~W.~J.}\
  \bibnamefont {Beenakker}},\ }\bibfield  {title} {\bibinfo {title}
  {Random-matrix theory of quantum transport},\ }\href
  {https://doi.org/10.1103/RevModPhys.69.731} {\bibfield  {journal} {\bibinfo
  {journal} {Rev. Mod. Phys.}\ }\textbf {\bibinfo {volume} {69}},\ \bibinfo
  {pages} {731} (\bibinfo {year} {1997})}\BibitemShut {NoStop}%
\bibitem [{\citenamefont {Guhr}\ \emph {et~al.}(1998)\citenamefont {Guhr},
  \citenamefont {{M{\"u}ller{\textendash}Groeling}},\ and\ \citenamefont
  {Weidenm{\"u}ller}}]{guhr1998randommatrix}%
  \BibitemOpen
  \bibfield  {author} {\bibinfo {author} {\bibfnamefont {T.}~\bibnamefont
  {Guhr}}, \bibinfo {author} {\bibfnamefont {A.}~\bibnamefont
  {{M{\"u}ller{\textendash}Groeling}}},\ and\ \bibinfo {author} {\bibfnamefont
  {H.~A.}\ \bibnamefont {Weidenm{\"u}ller}},\ }\bibfield  {title} {\bibinfo
  {title} {Random-matrix theories in quantum physics: Common concepts},\ }\href
  {https://doi.org/10.1016/S0370-1573(97)00088-4} {\bibfield  {journal}
  {\bibinfo  {journal} {Physics Reports}\ }\textbf {\bibinfo {volume} {299}},\
  \bibinfo {pages} {189} (\bibinfo {year} {1998})}\BibitemShut {NoStop}%
\bibitem [{\citenamefont {Brouwer}\ \emph {et~al.}(1999)\citenamefont
  {Brouwer}, \citenamefont {Frahm},\ and\ \citenamefont
  {Beenakker}}]{brouwer1999distribution}%
  \BibitemOpen
  \bibfield  {author} {\bibinfo {author} {\bibfnamefont {P.~W.}\ \bibnamefont
  {Brouwer}}, \bibinfo {author} {\bibfnamefont {K.~M.}\ \bibnamefont {Frahm}},\
  and\ \bibinfo {author} {\bibfnamefont {C.~W.~J.}\ \bibnamefont {Beenakker}},\
  }\bibfield  {title} {\bibinfo {title} {Distribution of the quantum mechanical
  time-delay matrix for a chaotic cavity},\ }\href
  {https://doi.org/10.1088/0959-7174/9/2/303} {\bibfield  {journal} {\bibinfo
  {journal} {Waves in Random Media}\ }\textbf {\bibinfo {volume} {9}},\
  \bibinfo {pages} {91} (\bibinfo {year} {1999})}\BibitemShut {NoStop}%
\bibitem [{\citenamefont {Beenakker}(2015)}]{beenakker2015randommatrix}%
  \BibitemOpen
  \bibfield  {author} {\bibinfo {author} {\bibfnamefont {C.~W.~J.}\
  \bibnamefont {Beenakker}},\ }\bibfield  {title} {\bibinfo {title}
  {Random-matrix theory of {{Majorana}} fermions and topological
  superconductors},\ }\href {https://doi.org/10.1103/RevModPhys.87.1037}
  {\bibfield  {journal} {\bibinfo  {journal} {Rev. Mod. Phys.}\ }\textbf
  {\bibinfo {volume} {87}},\ \bibinfo {pages} {1037} (\bibinfo {year}
  {2015})}\BibitemShut {NoStop}%
\bibitem [{\citenamefont {Das~Sarma}\ \emph {et~al.}(2016)\citenamefont
  {Das~Sarma}, \citenamefont {Nag},\ and\ \citenamefont
  {Sau}}]{dassarma2016how}%
  \BibitemOpen
  \bibfield  {author} {\bibinfo {author} {\bibfnamefont {S.}~\bibnamefont
  {Das~Sarma}}, \bibinfo {author} {\bibfnamefont {A.}~\bibnamefont {Nag}},\
  and\ \bibinfo {author} {\bibfnamefont {J.~D.}\ \bibnamefont {Sau}},\
  }\bibfield  {title} {\bibinfo {title} {How to infer non-{{Abelian}}
  statistics and topological visibility from tunneling conductance properties
  of realistic {{Majorana}} nanowires},\ }\href
  {https://doi.org/10.1103/PhysRevB.94.035143} {\bibfield  {journal} {\bibinfo
  {journal} {Phys. Rev. B}\ }\textbf {\bibinfo {volume} {94}},\ \bibinfo
  {pages} {035143} (\bibinfo {year} {2016})}\BibitemShut {NoStop}%
\bibitem [{\citenamefont {Blonder}\ \emph {et~al.}(1982)\citenamefont
  {Blonder}, \citenamefont {Tinkham},\ and\ \citenamefont
  {Klapwijk}}]{blonder1982transition}%
  \BibitemOpen
  \bibfield  {author} {\bibinfo {author} {\bibfnamefont {G.~E.}\ \bibnamefont
  {Blonder}}, \bibinfo {author} {\bibfnamefont {M.}~\bibnamefont {Tinkham}},\
  and\ \bibinfo {author} {\bibfnamefont {T.~M.}\ \bibnamefont {Klapwijk}},\
  }\bibfield  {title} {\bibinfo {title} {Transition from metallic to tunneling
  regimes in superconducting microconstrictions: {{Excess}} current, charge
  imbalance, and supercurrent conversion},\ }\href
  {https://doi.org/10.1103/PhysRevB.25.4515} {\bibfield  {journal} {\bibinfo
  {journal} {Phys. Rev. B}\ }\textbf {\bibinfo {volume} {25}},\ \bibinfo
  {pages} {4515} (\bibinfo {year} {1982})}\BibitemShut {NoStop}%
\bibitem [{\citenamefont {Datta}(1995)}]{datta1995electronic}%
  \BibitemOpen
  \bibfield  {author} {\bibinfo {author} {\bibfnamefont {S.}~\bibnamefont
  {Datta}},\ }\href {https://doi.org/10.1017/CBO9780511805776} {\emph {\bibinfo
  {title} {Electronic {{Transport}} in {{Mesoscopic Systems}}}}},\ Cambridge
  {{Studies}} in {{Semiconductor Physics}} and {{Microelectronic Engineering}}\
  (\bibinfo  {publisher} {{Cambridge University Press}},\ \bibinfo {address}
  {{Cambridge}},\ \bibinfo {year} {1995})\BibitemShut {NoStop}%
\bibitem [{\citenamefont {Anantram}\ and\ \citenamefont
  {Datta}(1996)}]{anantram1996current}%
  \BibitemOpen
  \bibfield  {author} {\bibinfo {author} {\bibfnamefont {M.~P.}\ \bibnamefont
  {Anantram}}\ and\ \bibinfo {author} {\bibfnamefont {S.}~\bibnamefont
  {Datta}},\ }\bibfield  {title} {\bibinfo {title} {Current fluctuations in
  mesoscopic systems with {{Andreev}} scattering},\ }\href
  {https://doi.org/10.1103/PhysRevB.53.16390} {\bibfield  {journal} {\bibinfo
  {journal} {Phys. Rev. B}\ }\textbf {\bibinfo {volume} {53}},\ \bibinfo
  {pages} {16390} (\bibinfo {year} {1996})}\BibitemShut {NoStop}%
\bibitem [{\citenamefont {Groth}\ \emph {et~al.}(2014)\citenamefont {Groth},
  \citenamefont {Wimmer}, \citenamefont {Akhmerov},\ and\ \citenamefont
  {Waintal}}]{groth2014kwant}%
  \BibitemOpen
  \bibfield  {author} {\bibinfo {author} {\bibfnamefont {C.~W.}\ \bibnamefont
  {Groth}}, \bibinfo {author} {\bibfnamefont {M.}~\bibnamefont {Wimmer}},
  \bibinfo {author} {\bibfnamefont {A.~R.}\ \bibnamefont {Akhmerov}},\ and\
  \bibinfo {author} {\bibfnamefont {X.}~\bibnamefont {Waintal}},\ }\bibfield
  {title} {\bibinfo {title} {Kwant: A software package for quantum transport},\
  }\href {http://iopscience.iop.org/article/10.1088/1367-2630/16/6/063065/meta}
  {\bibfield  {journal} {\bibinfo  {journal} {New Journal of Physics}\ }\textbf
  {\bibinfo {volume} {16}},\ \bibinfo {pages} {063065} (\bibinfo {year}
  {2014})}\BibitemShut {NoStop}%
\bibitem [{\citenamefont {Setiawan}\ \emph {et~al.}(2015)\citenamefont
  {Setiawan}, \citenamefont {Brydon}, \citenamefont {Sau},\ and\ \citenamefont
  {Das~Sarma}}]{setiawan2015conductance}%
  \BibitemOpen
  \bibfield  {author} {\bibinfo {author} {\bibfnamefont {F.}~\bibnamefont
  {Setiawan}}, \bibinfo {author} {\bibfnamefont {P.~M.~R.}\ \bibnamefont
  {Brydon}}, \bibinfo {author} {\bibfnamefont {J.~D.}\ \bibnamefont {Sau}},\
  and\ \bibinfo {author} {\bibfnamefont {S.}~\bibnamefont {Das~Sarma}},\
  }\bibfield  {title} {\bibinfo {title} {Conductance spectroscopy of
  topological superconductor wire junctions},\ }\href
  {https://doi.org/10.1103/PhysRevB.91.214513} {\bibfield  {journal} {\bibinfo
  {journal} {Phys. Rev. B}\ }\textbf {\bibinfo {volume} {91}},\ \bibinfo
  {pages} {214513} (\bibinfo {year} {2015})}\BibitemShut {NoStop}%
\bibitem [{\citenamefont {Vuik}\ \emph {et~al.}(2019)\citenamefont {Vuik},
  \citenamefont {Nijholt}, \citenamefont {Akhmerov},\ and\ \citenamefont
  {Wimmer}}]{vuik2019reproducing}%
  \BibitemOpen
  \bibfield  {author} {\bibinfo {author} {\bibfnamefont {A.}~\bibnamefont
  {Vuik}}, \bibinfo {author} {\bibfnamefont {B.}~\bibnamefont {Nijholt}},
  \bibinfo {author} {\bibfnamefont {A.}~\bibnamefont {Akhmerov}},\ and\
  \bibinfo {author} {\bibfnamefont {M.}~\bibnamefont {Wimmer}},\ }\bibfield
  {title} {\bibinfo {title} {Reproducing topological properties with
  quasi-{{Majorana}} states},\ }\href
  {https://doi.org/10.21468/SciPostPhys.7.5.061} {\bibfield  {journal}
  {\bibinfo  {journal} {SciPost Physics}\ }\textbf {\bibinfo {volume} {7}},\
  \bibinfo {pages} {061} (\bibinfo {year} {2019})}\BibitemShut {NoStop}%
\bibitem [{\citenamefont {Das~Sarma}\ and\ \citenamefont
  {Pan}(2021)}]{dassarma2021disorderinduced}%
  \BibitemOpen
  \bibfield  {author} {\bibinfo {author} {\bibfnamefont {S.}~\bibnamefont
  {Das~Sarma}}\ and\ \bibinfo {author} {\bibfnamefont {H.}~\bibnamefont
  {Pan}},\ }\bibfield  {title} {\bibinfo {title} {Disorder-induced zero-bias
  peaks in {{Majorana}} nanowires},\ }\href
  {https://doi.org/10.1103/PhysRevB.103.195158} {\bibfield  {journal} {\bibinfo
   {journal} {Phys. Rev. B}\ }\textbf {\bibinfo {volume} {103}},\ \bibinfo
  {pages} {195158} (\bibinfo {year} {2021})}\BibitemShut {NoStop}%
\bibitem [{hpc()}]{hpcc}%
  \BibitemOpen
  \href {https://hpcc.umd.edu/} {\bibinfo {title}
  {{{https://hpcc.umd.edu/}}}}\BibitemShut {NoStop}%
\end{thebibliography}%
\appendix
\section{The crossover from ugly to ugly ZBCPs with another disorder realizations}\label{app:A}

\begin{figure}[htbp]
	\centering
	\includegraphics[width=3.4in]{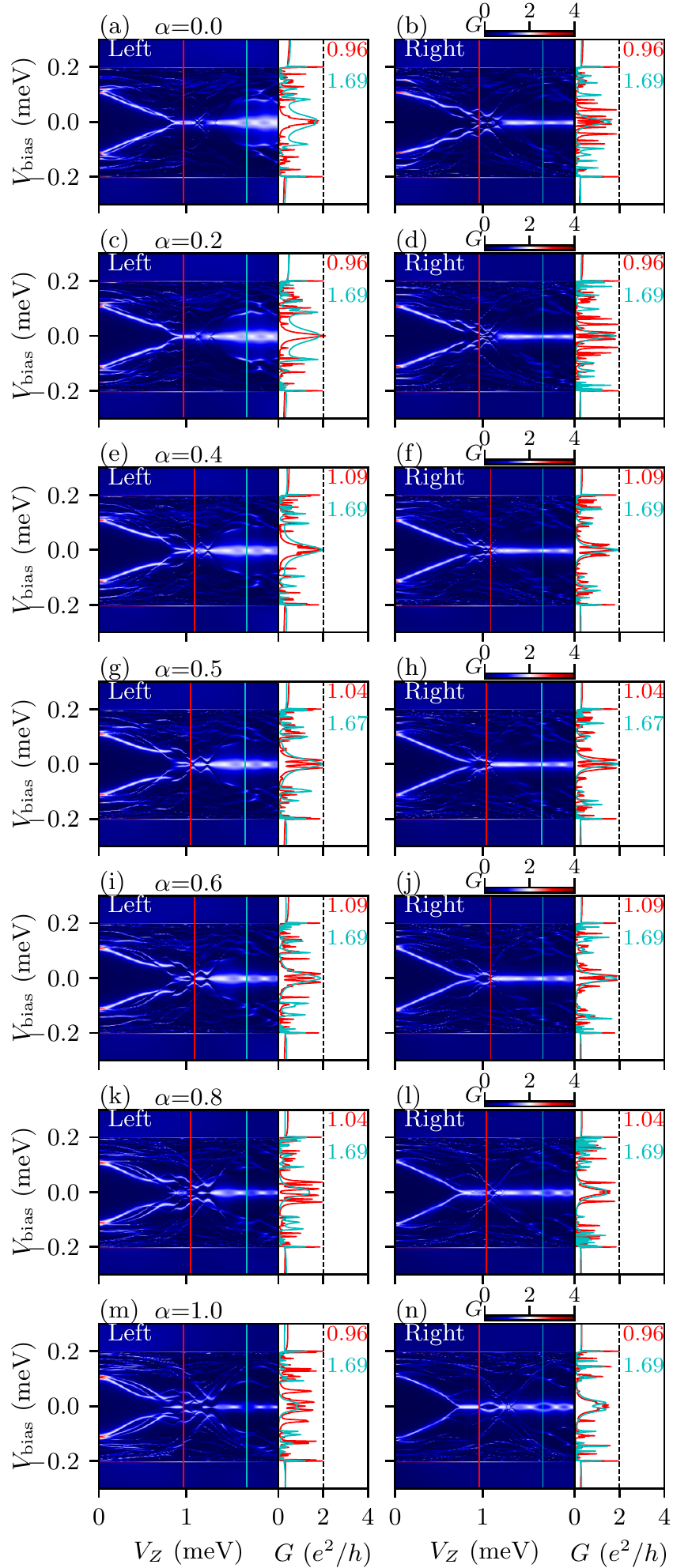}
	\caption{The tunnel conductance measured from the left end (in the left panels) and right end (in the right panels) of the nanowire in the crossover between two ugly ZBCPs using the simple linear interpolation of Eq.~\eqref{eq:ugly2ugly}.
	The method to interpolate here is identical to Fig.~\ref{fig:ugly_ugly_linear} but realizations of potential disorder in (a),(b) ($ \alpha=0 $) and (m),(n) ($ \alpha=1 $) are different. }
	\label{fig:ugly_ugly_linear_2}
\end{figure}

\begin{figure}[htbp]
	\centering
	\includegraphics[width=3.4in]{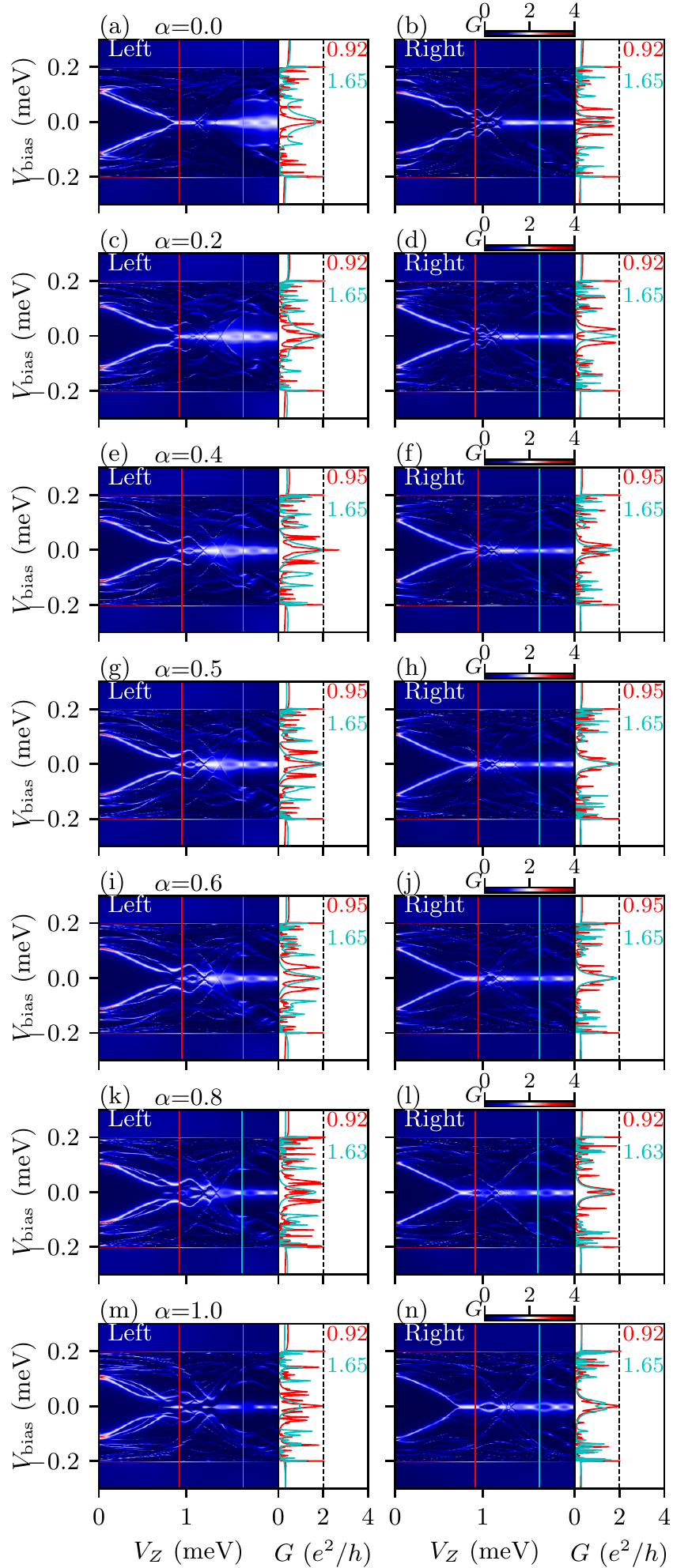}
	\caption{The tunnel conductance measured from the left end (in the left panels) and right end (in the right panels) of the nanowire in the crossover between two ugly ZBCPs using the variance-conserving interpolation of Eq.~\eqref{eq:ugly2ugly_sqrt}.
	The method to interpolate here is identical to Fig.~\ref{fig:ugly_ugly_var} but realizations of the potential disorder in (a),(b) ($ \alpha=0 $) and (m),(n) ($ \alpha=1 $) are different.}
	\label{fig:ugly_ugly_var_2}
\end{figure}

To manifest that the disorder realizations in Figs.~\ref{fig:ugly_ugly_linear} and \ref{fig:ugly_ugly_var} are not unique, we additionally provide another example in the presence of another disorder realizations. Figures~\ref{fig:ugly_ugly_linear_2} and \ref{fig:ugly_ugly_var_2} show the crossover using the linear and variance-conserving interpolation, respectively. Their corresponding LDOSs are shown in Figs.~\ref{fig:ugly_ugly_linear_2_LDOS} and~\ref{fig:ugly_ugly_var_2_LDOS}.

\clearpage
\section{Local density of states}\label{app:B}
\begin{figure*}[htbp]
	\centering
	\includegraphics[width=6.8in]{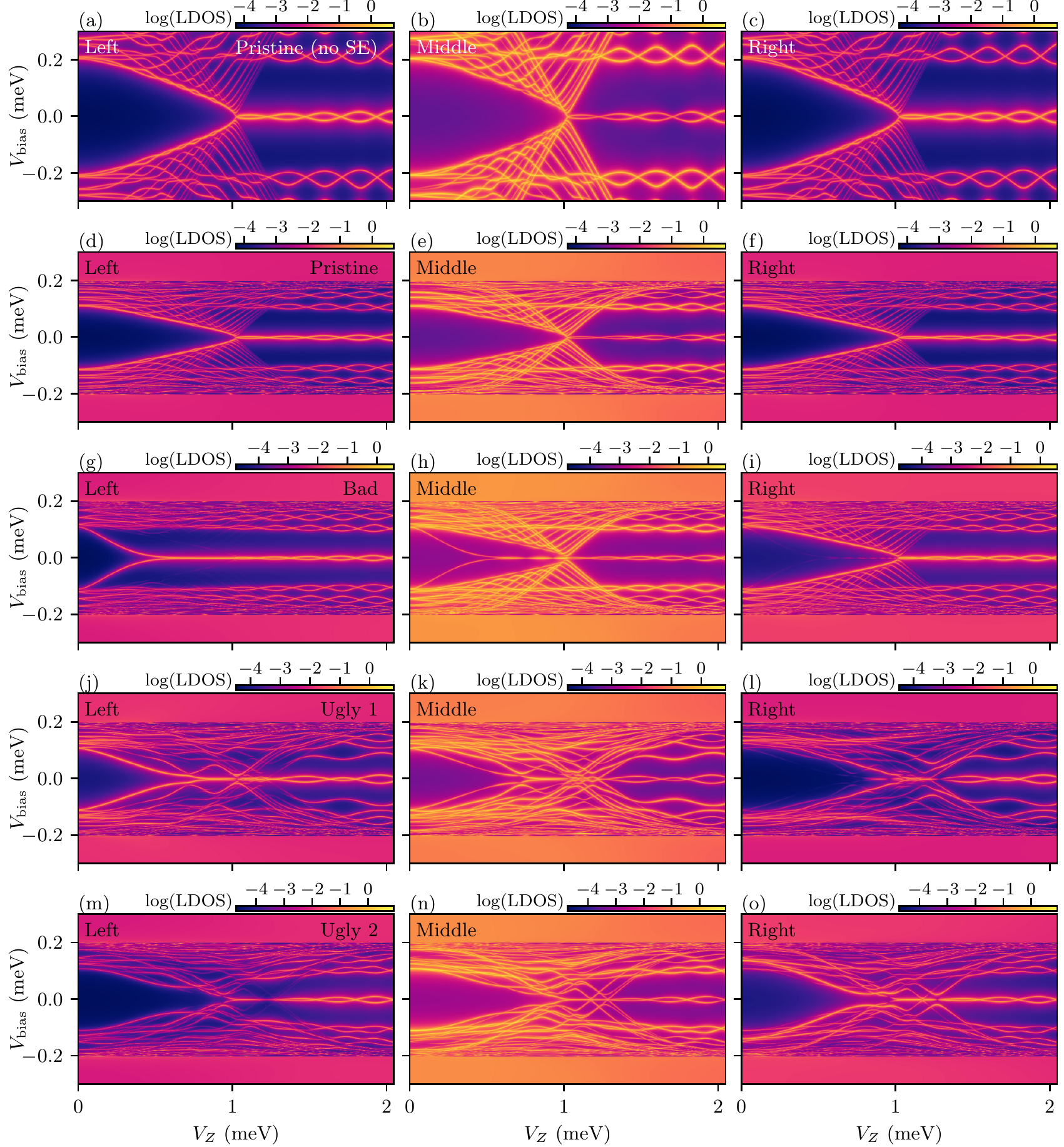}
	\caption{The LDOSs (corresponding to Fig.~\ref{fig:static}) of different static cases (from the top to the bottom: pristine without self-energy, pristine, bad, the first ugly, and the second ugly) at the left end (left panels), in the middle of the wire (middle panels), and at the right end (right panels).
	}
	\label{fig:static_LDOS}
\end{figure*}

\begin{figure*}[htbp]
	\centering
	\includegraphics[width=6.8in]{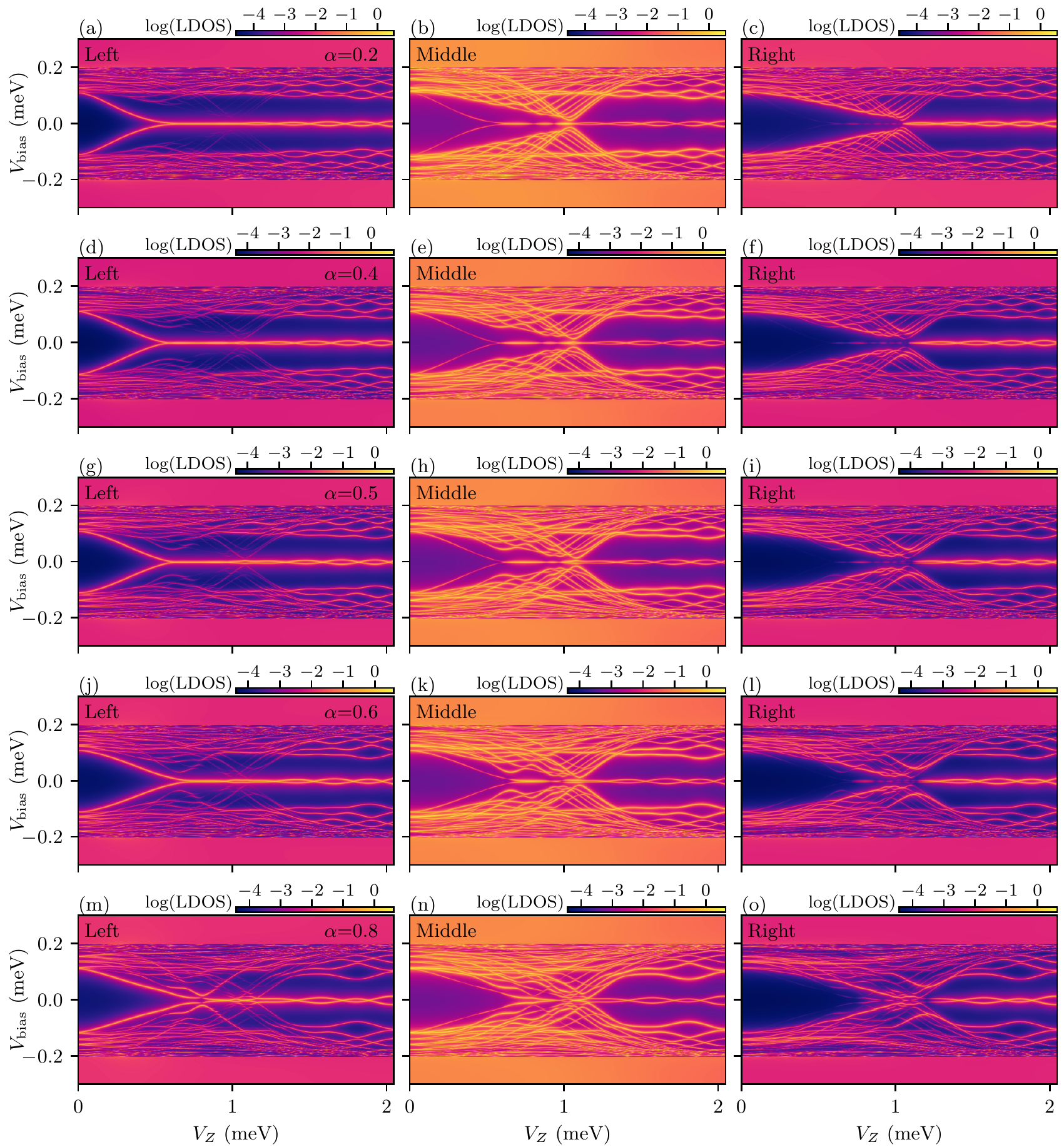}
	\caption{The LDOSs (corresponding to Fig.~\ref{fig:bad_ugly}) of a wire in the crossover between the bad ZBCP (the first row) and the ugly ZBCP (the last row) at the left end (left panels), in the middle of the wire (middle panels), and at the right end (right panels). 
	}
	\label{fig:bad_ugly_LDOS}
\end{figure*}

\begin{figure*}[htbp]
	\centering
	\includegraphics[width=6.8in]{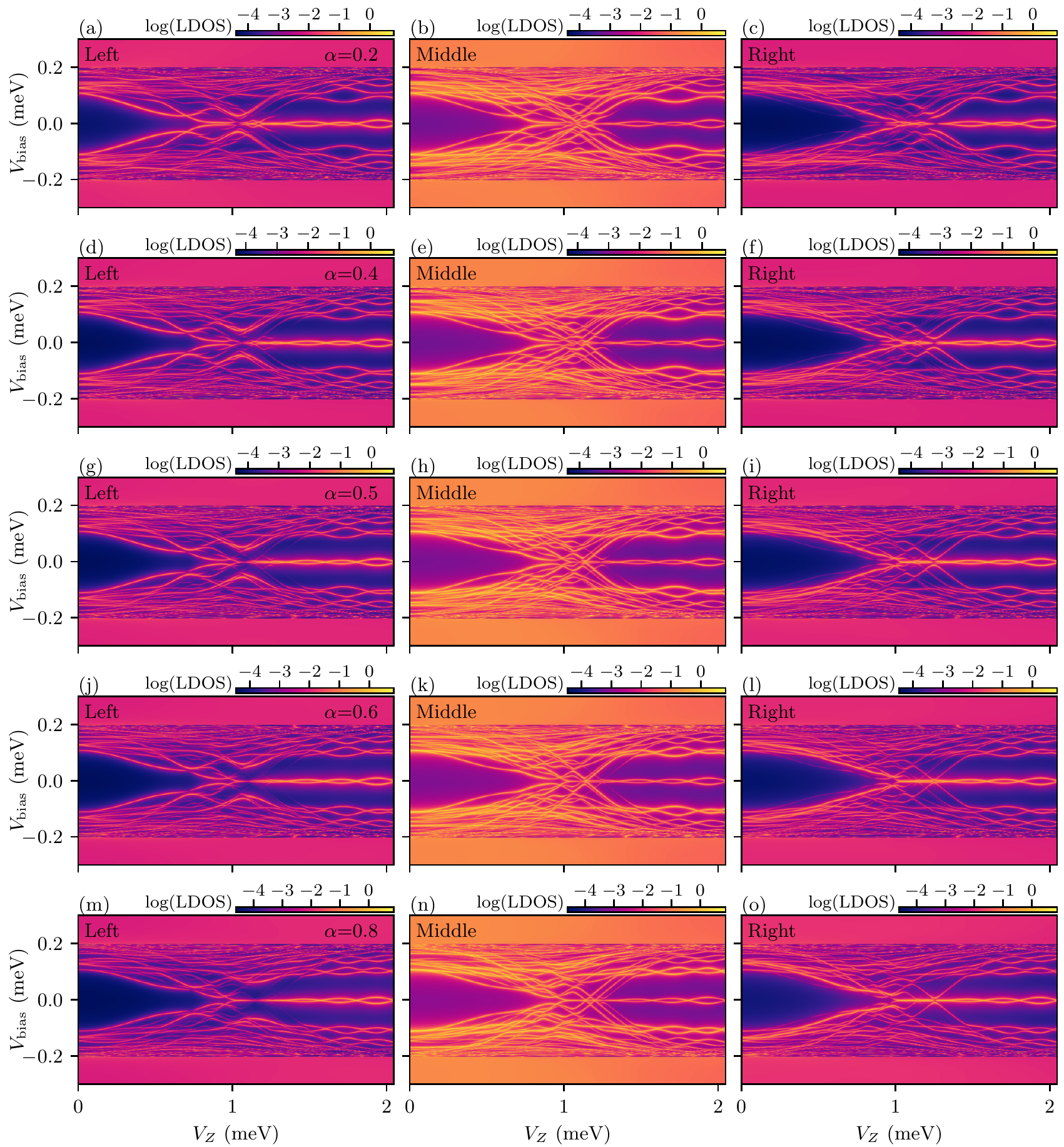}
	\caption{The LDOSs (corresponding to Fig.~\ref{fig:ugly_ugly_linear}) of a wire in the crossover between one ugly ZBCP (the first row) and another ugly ZBCP (the last row) using the linear interpolation at the left end (left panels), in the middle of the wire (middle panels), and at the right end (right panels). 
	}
	\label{fig:ugly_ugly_linear_LDOS}
\end{figure*}

\begin{figure*}[htbp]
	\centering
	\includegraphics[width=6.8in]{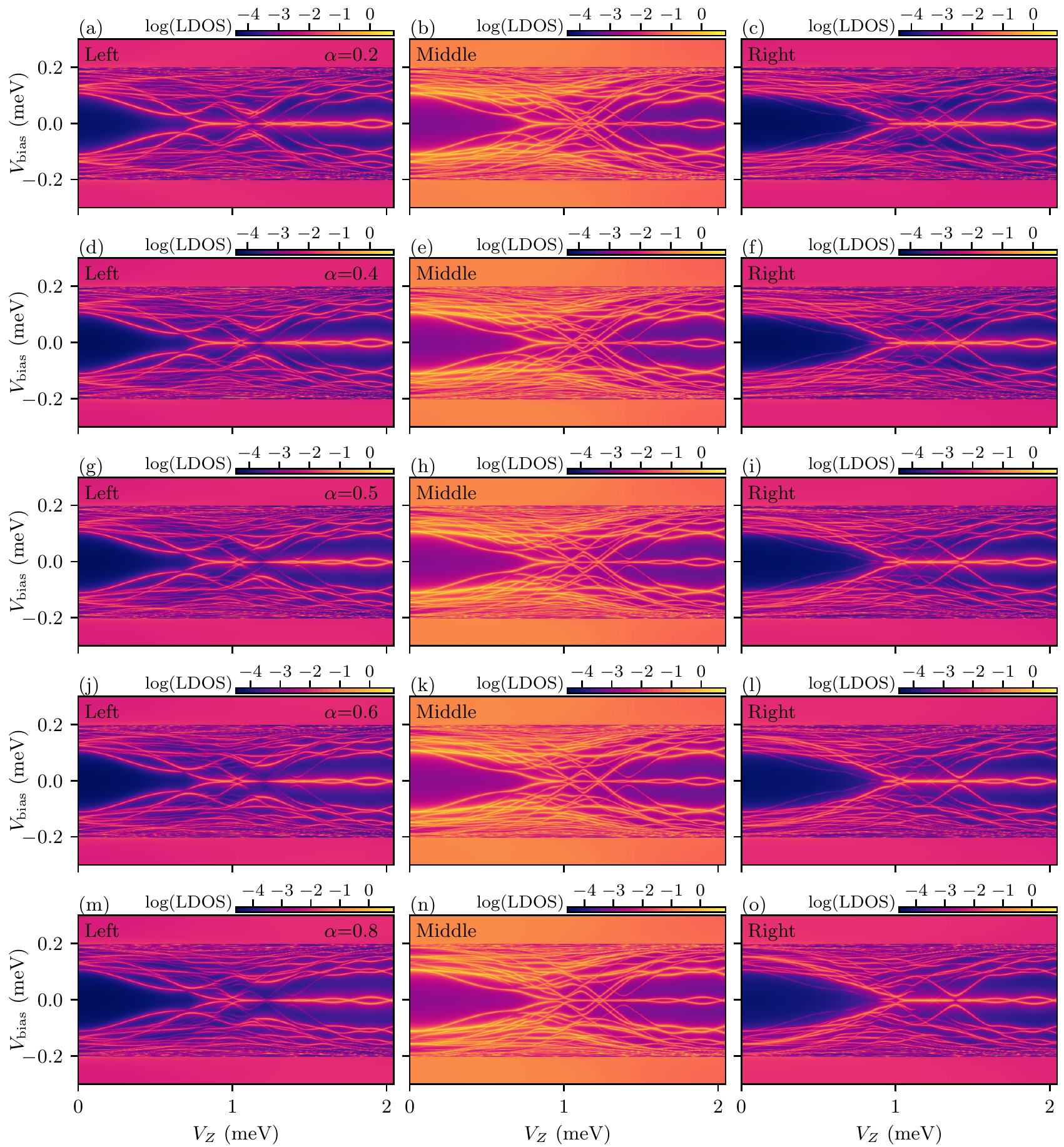}
	\caption{The LDOSs (corresponding to Fig.~\ref{fig:ugly_ugly_var}) of a wire in the crossover between one ugly ZBCP (the first row) and another ugly ZBCP (the last row) using the variance-conserving interpolation at the left end (left panels), in the middle of the wire (middle panels), and at the right end (right panels). 
	}
	\label{fig:ugly_ugly_var_LDOS}
\end{figure*}

\begin{figure*}[htbp]
	\centering
	\includegraphics[width=6.8in]{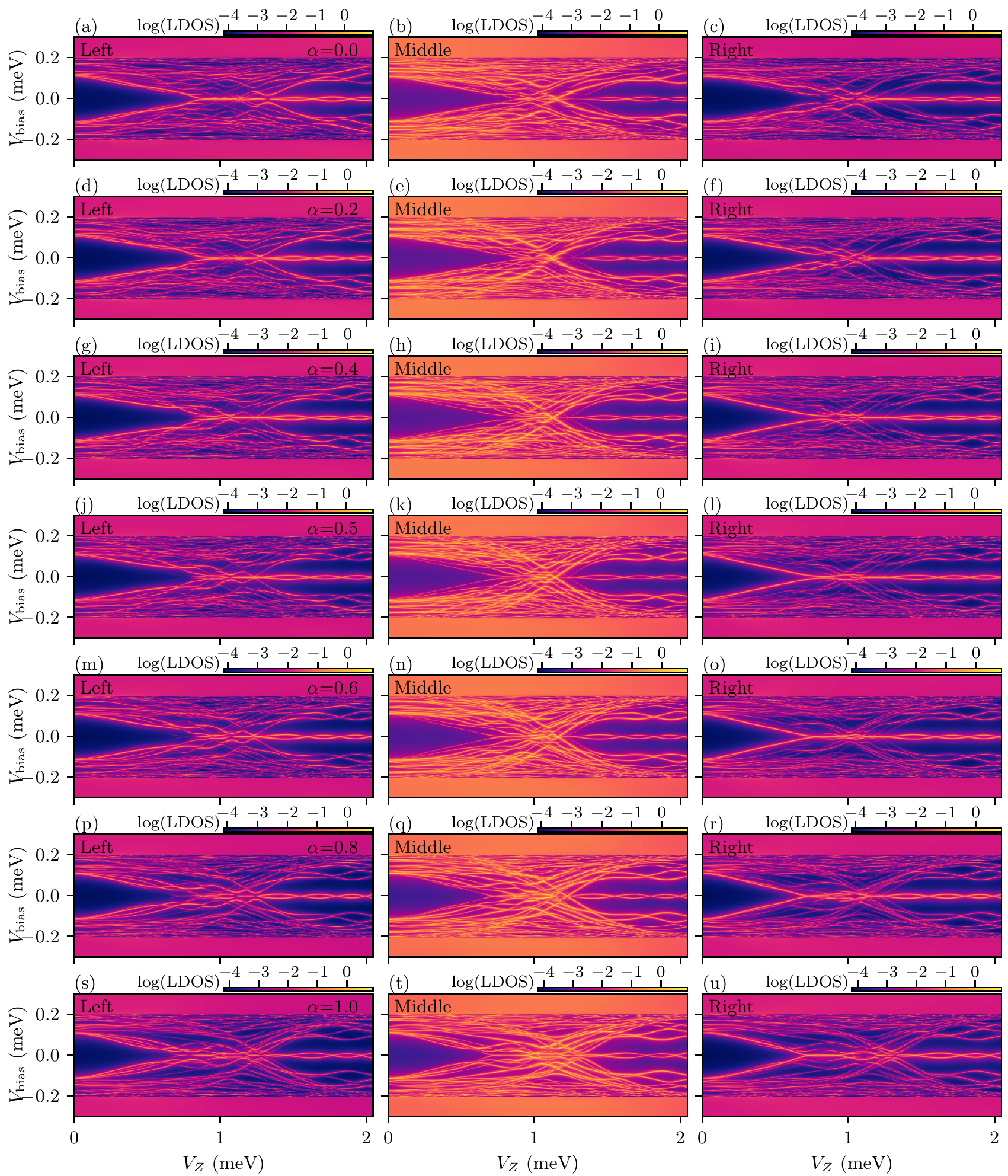}
	\caption{The LDOSs (corresponding to Fig.~\ref{fig:ugly_ugly_linear_2}) of a wire in the crossover between one ugly ZBCP (the first row) and another ugly ZBCP (the last row) using the simple linear interpolation at the left end (left panels), in the middle of the wire (middle panels), and at the right end (right panels).
	}
	\label{fig:ugly_ugly_linear_2_LDOS}
\end{figure*}

\begin{figure*}[htbp]
	\centering
	\includegraphics[width=6.8in]{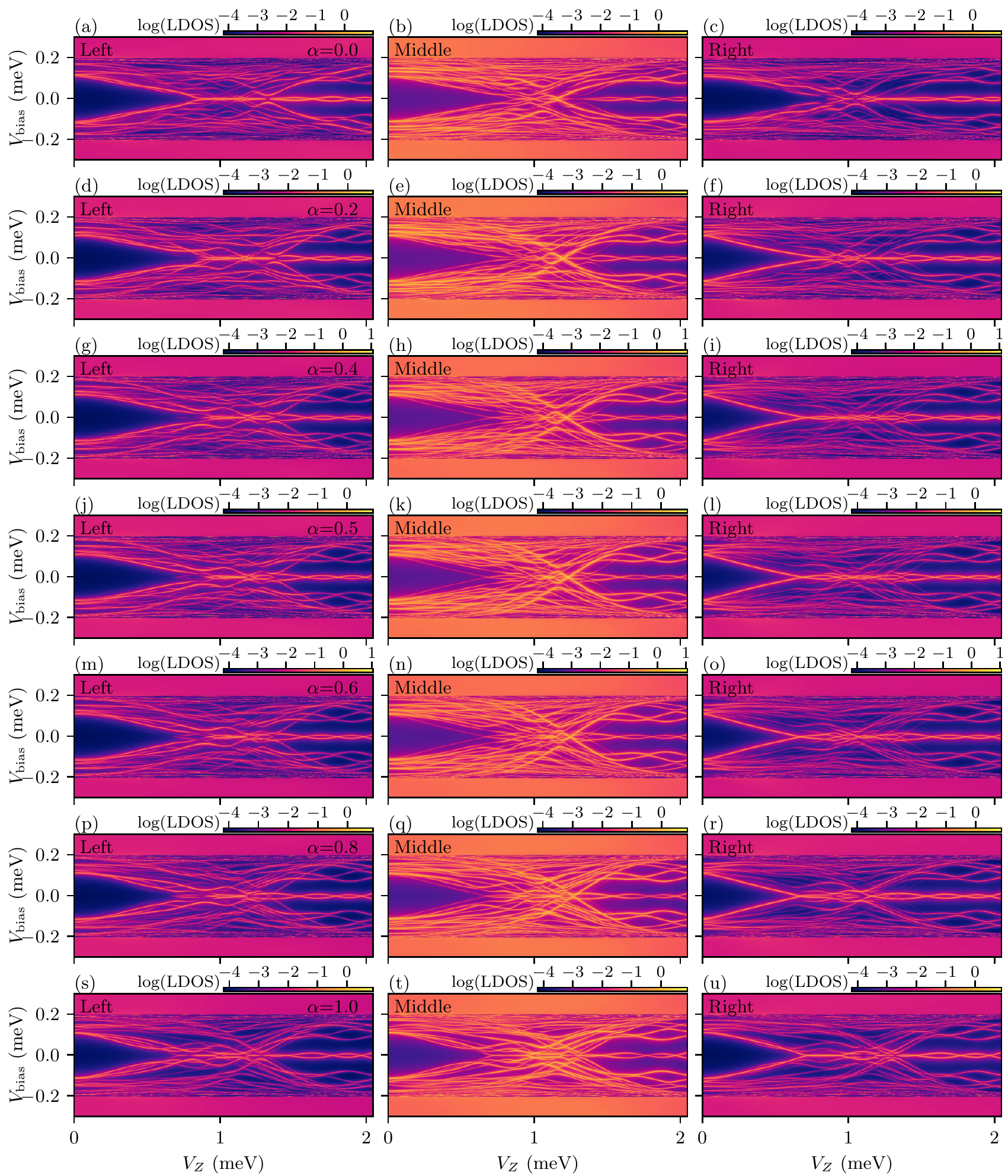}
	\caption{The LDOSs (corresponding to Fig.~\ref{fig:ugly_ugly_var_2}) of a wire in the crossover between one ugly ZBCP (the first row) and another ugly ZBCP (the last row) using the variance-conserving interpolation at the left end (left panels), in the middle of the wire (middle panels), and at the right end (right panels). 
	}
	\label{fig:ugly_ugly_var_2_LDOS}
\end{figure*}

In this appendix, we present in Figs.~\ref{fig:static_LDOS} to \ref{fig:ugly_ugly_var_2_LDOS} the LDOSs corresponding to the tunnel conductance (Figs.~\ref{fig:static} to~\ref{fig:ugly_ugly_var}) in the main text and Appendix~\ref{app:A} (Figs.~\ref{fig:ugly_ugly_linear_2} and~\ref{fig:ugly_ugly_var_2}) to show that the tunnel conductance and LDOS results are consistent. Therefore, to avoid showing redundant figures, we have also verified that the other tunnel conductance spectra, which do not have a corresponding LDOS presented here, are also consistent with their LDOS.

\end{document}